\documentclass[pdflatex,sn-mathphys-num]{sn-jnl}
\usepackage{lmodern}
\usepackage{graphicx}%
\usepackage{multirow}%
\usepackage{amsmath,amssymb,amsfonts}%
\usepackage{amsthm}%
\usepackage{mathrsfs}%
\usepackage[title]{appendix}%
\usepackage{xcolor}%
\usepackage{textcomp}%
\usepackage{manyfoot}%
\usepackage{booktabs}%
\usepackage{algorithm}%
\usepackage{algorithmicx}%
\usepackage{algpseudocode}%
\usepackage{listings}%

\newcommand{\bbf}{\mathbf{b}}

\newcommand{\Gbf}{\mathbf{G}}

\newcommand{\tbf}{\mathbf{t}}

\newcommand{\wbf}{\mathbf{w}}

\newcommand{\wstar}{\wbf^\star}

\newtheorem{definition}{Definition}%
\theoremstyle{plain}
\theoremstyle{definition}
\newtheorem{exmp}{Example}[section]
\usepackage{placeins}

\begin{document}
\title{A practical identifiability criterion leveraging weak-form parameter estimation}
\author*[1]{\fnm{Nora} \sur{Heitzman-Breen}}\email{nora.heitzman-breen@colorado.edu}
\author[1,2]{\fnm{Vanja} \sur{Dukic}}\email{vanja.dukic@colorado.edu}
\author[1,2]{\fnm{David M.} \sur{Bortz}}\email{david.bortz@colorado.edu}

\affil[1]{\orgdiv{Department of Applied Mathematics}, \orgname{University of Colorado}, \orgaddress{\city{Boulder}, \state{CO}, \postcode{80309-0526}, \country{USA}}}
\affil[2]{Joint senior authors}

\abstract{
In this work, we define a practical identifiability criterion, $(e,q)$-identifiability, based on a parameter $e$, reflecting the noise in observed variables, and a parameter $q$, reflecting the mean-square error of the parameter estimator. This criterion is better able to encompass changes in the quality of the parameter estimate\textcolor{black}{(s)} due to increased noise in the data (compared to existing criteria based solely on average relative errors). \textcolor{black}{We illustrate the usefulness of the criteria in several challenging identifiability studies, involving parameter estimation in partially observed systems.}
Furthermore, we leverage a weak-form equation error-based method of parameter estimation for systems with unobserved variables to assess practical identifiability far more quickly in comparison to output error-based parameter estimation. We do so by generating weak-form input-output equations using differential algebra techniques, as previously proposed by \textit{Boulier et al} \cite{BoulierKorporalLemaireEtAl2014ComputerAlgebrainScientificComputing}, and then applying Weak form Estimation of Nonlinear Dynamics (WENDy) to obtain parameter estimates. This method is computationally efficient and robust to noise, as demonstrated through two classical biological modeling examples. 

}




\maketitle

\section{Introduction}

Parameter estimation is central to mathematical modeling in the biological sciences.  However, despite the increasing sophistication of computational \textcolor{black}{and statistical } techniques over the years, there are many examples of researchers using the same model and obtaining widely varying parameter estimates \textcolor{black}{for the same biological phenomena}  \cite{NguyenKlawonnMikolajczykEtAl2016PLoSONE, GoncalvesMentreLemenuel-DiotEtAl2020AAPSJ, StepaniantsHastewellSkinnerEtAl2023arXiv230404818, SimpsonMaclaren2024BullMathBiol,ZitzmannKeRibeiroEtAl2024PLoSComputBiola}. \textcolor{black}{It is well-recognized that parameter estimation in dynamical systems remains a challenging task \cite{McGoffMukherjeePillai2015StatistSurva}.  While some of this variation is due to causes such as quality and quantity of data and uncertainty in initial conditions, the primary reason for that difficulty lies in the relationship between the model structure and data.}  Analyzing model identifiability, i.e., the ability to recover unique parameters of a system from observations, is \textcolor{black}{thus } critical to developing robust and replicable \textcolor{black}{inference in} biological systems. 

\emph{Structural identifiability} addresses whether model parameters can be uniquely determined given the observation of some output variables of a system. There are multiple approaches to assess the structural identifiability of a model including but not limited to a Taylor series expansion \cite{WielandHauberRosenblattEtAl2021CurrentOpinioninSystemsBiology, ChisBangaBalsa-Canto2011IFACProceedingsVolumes, MiaoWuXue2014JournaloftheAmericanStatisticalAssociation}, the implicit function theorem \cite{ChisBangaBalsa-Canto2011IFACProceedingsVolumes, MiaoWuXue2014JournaloftheAmericanStatisticalAssociation}, generalizations of observability criteria \cite{StigterMolenaar2015Automatica, VanWilligenburgStigterMolenaar2022NonlinearDyn, VillaverdeBarreiroPapachristodoulou2016PLoSComputBiol}, and differential algebra\footnote{A subfield of algebra created by Ritt and his student Kolchin to study the existence of solutions to differential equations (\cite{Ritt1950,Kolchin1985}).} approaches \cite{LjungGlad1994Automatica,MeshkatEisenbergDiStefano2009MathematicalBiosciences,BelluSaccomaniAudolyEtAl2007ComputerMethodsandProgramsinBiomedicine,DongGoodbrakeHarringtonEtAl2023SIAMJApplAlgebraGeometry}. 

However, in practice, the quantity, frequency, and timing of available data have been shown to have large impacts on the reliability of parameter estimates. For examples in models of acute viral infections, \textcolor{black}{we direct the reader to} \cite{NguyenKlawonnMikolajczykEtAl2016PLoSONE,CiupeTuncer2022SciRep,Heitzman-BreenLiyanageDuggalEtAl2024RSocOpenSci}. Furthermore, choices in the numerical discretizations (needed to approximate solutions) can also impact parameter estimates \cite{DukicBortz2018InverseProblSciEng, NardiniBortz2019InverseProbl}. \emph{Practical identifiability}\footnote{\textcolor{black}{This has also been referred to as statistical identifiability.}} refers to whether parameters can be uniquely estimated given the available observations and the chosen parameter estimation method, including numerics, and is \textcolor{black}{directly related to} uncertainty quantification of model parameters. \textcolor{black}{Practical identifiability methods can be applied to study estimator properties for a model with a specific observed dataset (\textit{a posteriori}), or to study general properties of a model in different hypothetical data scenarios, i.e., noise and collection frequency (\textit{a priori}). The latter can provide important insights to guide experimental design, power analyses, and data collection.}

\textcolor{black}{While simple models can benefit from traditional statistical methods, such as the Fisher Information Matrix, most practical identifiability techniques for complex models rely on repeated simulations, which can become computationally expensive.} \textcolor{black}{In contrast,} weak-form based parameter estimation, which is typically much faster than output error-based methods, can address the challenge incurred by repeating the parameter estimation. The idea of using the weak form in system identification can be traced back to the Equations-of-Motion method described in \cite{Shinbrot1954NACATN3288}, in which an equation (in the strong form) is multiplied by a compactly supported test function\footnote{Shinbrot calls them \emph{method functions}} $\phi$ and then the equation is integrated. As the derivatives of the test functions are also assumed to have compact support, integration by parts removes the need to evaluate the derivatives of the data.  Shinbrot proposed that the $\phi$'s be trigonometric functions with different frequencies, while Loeb and Cahen \cite{LoebCahen1963Automatisme} later proposed piecewise polynomials (denoting their approach as the \emph{Modulating Functions Method}). Recently, our group has developed a \textcolor{black}{suite of }weak form-based parameter estimation and equation learning methods \textcolor{black}{(WENDy and WSINDy)} in the spirit of Shinbrot, Loeb and Cahen, but with mathematically motivated test functions \cite{MessengerBortz2021JComputPhys, MessengerBortz2021MultiscaleModelSimul, BortzMessengerDukic2023BullMathBiol}.  This choice of test functions improves both the computational efficiency of performing parameter estimation and is highly robust to noise in the data. These weak form-based methods have also been applied to biological and environmental systems \cite{MessengerWheelerLiuEtAl2022JRSocInterface, MessengerDwyerDukic2024JRSocInterface, MinorMessengerDukicEtAl2025JGeophysResMachLearnComput}. Morevoer, we have also performed anlayses establishing asymptotic properties of WSINDy \cite{MessengerBortz2025IMAJNumerAnal} as well as bias and coverage properties of WENDy \cite{ChawlaBortzDukic2025HandbookofVisualExperimentalandComputationalMathematicsBridgesthroughData}. For an overview of this novel class of methods, we direct the interested reader to \cite{MessengerTranDukicEtAl2024SIAMNews,BortzMessengerTran2024NumericalAnalysisMeetsMachineLearning}.

One challenge in using \textcolor{black}{WENDy type weak-form estimation methods is that, thus far,} all state variables \textcolor{black}{had to} be measured. However, in biological systems, data is rarely observed for all model compartments. For example, epidemiological models are often calibrated to only a subset of model compartments, such as mortality or hospitalization data. Accordingly, \textcolor{black}{ in this study, we consider }differential elimination methods to collapse a system of low-order equations to an equivalent lower-dimensional high-order system with only observable variables. \textcolor{black}{This technique has been applied in} \cite{Denis-VidalJoly-BlanchardNoiret2003NumericalAlgorithms, VerdiereDenis-VidalJoly-BlanchardEtAl2005IntJApplMathComputSci, BoulierKorporalLemaireEtAl2014ComputerAlgebrainScientificComputing, VerdiereZhuDenis-Vidal2018JournalofComputationalandAppliedMathematics}, and\textcolor{black}{,} in some cases\textcolor{black}{,} these systems have been cast into either integral or weak forms \cite{BoulierKorporalLemaireEtAl2014ComputerAlgebrainScientificComputing, VerdiereZhuDenis-Vidal2018JournalofComputationalandAppliedMathematics,WongvanichHannSirisena2015MathematicalBiosciences}.
However, previous applications of these elimination methods have not yielded a robust weak-form parameter estimation process.

In this work, we propose a process for efficiently assessing \emph{a priori} model identifiability by leveraging weak-form parameter estimation. First, we apply differential elimination methods to generate an input-output equation, \textcolor{black}{for the states which will be observed.} At this step, we can also assess the structural identifiability of the model, which is a prerequisite to practical identifiability. Second, we cast the input-output equation into its weak form. Third, \textcolor{black}{ we perform weak-form parameter estimation on repeated simulations of noisy data.} \textcolor{black}{Note, unlike output error based methods, }this method is feasible even for large sample sizes and high levels of measurement error when the Weak form Estimation of Nonlinear Dynamics (WENDy) is used for parameter estimation, due to the method's robustness to noise and computational efficiency. Finally, we assess practical identifiability from these simulations using a newly defined criterion, \textcolor{black}{which we refer to as} $(e,q)-$identifiability. 


In Sections \ref{subsec:struct_ident} and \ref{subsec:pract_ident}, we will give a brief overview of input-output-based methods to determine structural identifiability and methods used to determine practical identifiability. Then, in Section \ref{subsec:eq-ident}, we define $(e,q)-$\emph{identifiability}, a practical identifiability criterion based on the noise in observed variables and the mean-square error for the \textcolor{black}{corresponding} parameter estimator. Next, in \textcolor{black}{Sections \ref{subsec:weak_gen} and \ref{subsec:wendy1} we describe the generation of weak-form input-output equations and weak-form parameter estimation. The structural identifiability and limitations of the weak-form input-output parameter estimation are discussed in \ref{subsec:struct_weakform}. The weak-form input-ouput parameter estimation is applied \textcolor{black}{to} two canonical biological examples in Section \ref{subsec:weakIO-examples}.} Finally, in Section \ref{sec:pract_ident} we apply our definition of $(e,q)-$identifiability to the examples, compare this criterion to the average relative error, and lastly compare the computational efficiency of the weak-form parameter estimation to an output error method.

\section{Background}\label{sec:Background}

In Sections \ref{subsec:struct_ident} and \ref{subsec:pract_ident}, we provide background information on structural and practical identifiability. Readers already familiar with these topics are encouraged to skip to Section \ref{subsec:eq-ident}, in which we introduce $(e,q)-$\emph{identifiability}, a practical identifiability criterion based on relating the noise in observed variables and the mean-square error for the parameter estimator.



\subsection{Structural Identifiability}\label{subsec:struct_ident}

The goal of studying the structural identifiability of a system of differential equations is to determine whether there are unique model parameters that give rise to the dynamics of the observable states. This property is a prerequisite to recovering parameters of interest from partially observed systems. Below, we provide a brief overview of the input-output based method of determining structural identifiability, along with some standard definitions.

Consider the system of ordinary differential equations $\dot{U}=\Theta(U,w)$, where $U$ is a vector of the state variables of the system and $w$ is a vector of model parameters. If the \textcolor{black}{noiseless} observations (output of the system) are given by $y(t)=\Omega(U,w)$, then structural identifiability can be defined as follows.

\begin{definition}
Let $w$ and $\widehat{w}$ be distinct model parameter vectors. A model is said to be \underline{globally structurally identifiable} when $$\Omega(U,w)=\Omega(U,\widehat{w}) \quad\Rightarrow\quad w\equiv\widehat{w}.$$ A model is said to be \underline{locally structurally identifiable} if for any $w$ within an open neighborhood of some point $\widehat{w}$ in the parameter space, $$\Omega(U,w)=\Omega(U,\widehat{w}) \quad\Rightarrow\quad w\equiv\widehat{w}.$$
\textcolor{black}{Otherwise, a model is said to be \underline{structurally unidentifiable.}}\label{def:ident}
\end{definition}

One way of assessing the structural identifiability of a system of differential equations is by studying the input-output equation. Input-output equations have been used to study the structural identifiability of numerous biological systems, for examples, see \cite{EisenbergRobertsonTien2013JournalofTheoreticalBiology,ChowellDahalLiyanageEtAl2023JMathBiol,Heitzman-BreenLiyanageDuggalEtAl2024RSocOpenSci,LiyanageHeitzman-BreenTuncerEtAl2024MBEa,LiyanageChowellPogudinEtAl2025Viruses}. An input-output equation is a differential equation that depends only on the inputs, outputs, their respective derivatives, and the model parameters. Given an input-output equation for a system of differential equations, global structural identifiability can be restated as given below.

\begin{definition}
A model is said to be \underline{globally structurally identifiable} if coefficient map of the corresponding input-output equation, $c(w)$ is injective.\footnote{Local structural identifiability can be similarly restated.}\footnote{\textcolor{black}{For this definition to be equivalent to Definition \ref{def:ident}, there is an criterion on the Wronskian of the differential monomials from the input-output equation. For more details, see \cite{OvchinnikovPogudinThompson2023AAECC}.}}\label{def:io_ident}
\end{definition}

Input-output equations can be found from a system of differential equations using differential elimination algorithms, including the Ritt, the Ritt-Kolchin \cite{Kolchin1985}, and the Rosenfeld-Groebner \cite{Boulier2007GrobnerBasesinSymbolicAnalysis} algorithms\textcolor{black}{, or with a more recently developed, projection-based algorithm \cite{DongGoodbrakeHarringtonEtAl2023SIAMJApplAlgebraGeometry}.} There are many computational tools available to obtain input-output equations, including \cite{BelluSaccomaniAudolyEtAl2007ComputerMethodsandProgramsinBiomedicine, MeshkatKuoDiStefano2014PLoSONE,HongOvchinnikovPogudinEtAl2019Bioinformatics}. In this study, we use the \texttt{DifferentialAlgebra v4} package \cite{BoulierThiery2024} in Python to generate input-output equations, which applies a Rosenfeld-Groebner algorithm. \textcolor{black}{Note, an input-output equation may not exist or may not be computationally feasible to discover; however, addressing the weak-form estimation in such cases is a subject of future work. This paper focuses on a weak-form method for the case that an input-output equation, as referred to in Definition \ref{def:io_ident}, can be found.}

\subsection{Practical Identifiability}\label{subsec:pract_ident}

In practice, observations of the state variable are often sparse and subject to measurement error. Practical identifiability refers to whether parameters can be uniquely estimated given the available observations and chosen estimation method. Below, we provide a brief overview of current methods to determine practical identifiability and propose a criterion for practical identifiability based on the mean-square error observed for the parameter estimator.

There are many proposed methods to assess the practical identifiability of a system, and there is no commonly agreed-upon single criterion to determine practical identifiability. The Fisher Information Matrix (FIM) can be used to approximate the parameter covariance matrix and as a measure of local practical identifiability \cite{JacquezGreif1985MathematicalBiosciences,LandawDiStefano1984AmericanJournalofPhysiology-RegulatoryIntegrativeandComparativePhysiology}, however, \textcolor{black}{this criterion is difficult to calculate in more complex systems where likelihood modes may be difficult to find.} The FIM method has been used to assess the practical identifiability in biological systems \cite{EisenbergRobertsonTien2013JournalofTheoreticalBiology,GoncalvesMentreLemenuel-DiotEtAl2020AAPSJ}. \textcolor{black}{In these more complicated systems, }Profile Likelihood based methods test practical identifiability through successive refittings in the parameter space, which give insight into the shape of the likelihood and whether it is uniquely maximized\footnote{\textcolor{black}{Depending on the choice of estimator or proxy used for the likelihood, this problem may be restated as a minimization.}} \cite{MurphyVanDerVaart2000JournaloftheAmericanStatisticalAssociation}. Profile likelihood methods have been used to assess practical identifiability in biological systems as well \cite{KreutzRaueKaschekEtAl2013TheFEBSJournal,LiyanageHeitzman-BreenTuncerEtAl2024MBEa}.

\textcolor{black}{Simulation}-based methods of assessing practical identifiability involve repeated estimation performed using a prescribed error model for the observed data. From these repeated estimations, one criterion that has been proposed to assess practical identifiability is the average relative error. Examples of \textcolor{black}{applications of} this criterion applied to biological systems include \cite{TuncerMarcthevaLaBarreEtAl2018BullMathBiol,Heitzman-BreenLiyanageDuggalEtAl2024RSocOpenSci}. \textcolor{black}{Simulation-}based methods can be applied to any choice of objective function, but they also require numerous repetitions of the chosen parameter estimation method, which can become challenging to implement if the estimation method is not computationally efficient.

\subsection{$(e,q)$-identifiability}\label{subsec:eq-ident}

We are interested in using a \textcolor{black}{simulation-}based method of assessing practical identifiability. By choosing such a method, we can avoid making assumptions about the local \textcolor{black}{approximation} of the system, which is required to find the Fisher Information Matrix. \textcolor{black}{Likewise}, we do not need to restrict our choice of objective function, \textcolor{black}{ as in} Profile Likelihood Analysis. While \textcolor{black}{simulation-}based methods can pose a computational challenge, recent advances in weak-form parameter estimation can be leveraged to perform the many required repetitions of the estimation quickly. 

We propose a practical identifiability criterion that depends on the mean-squared error \textcolor{black}{on the chosen estimator}. Below we define the \textcolor{black}{\underline{($e,q$)-identifiability}}, which can be computed for systems and data using either output error or equation error optimization; \textcolor{black}{ however, in this paper we use equation error optimization.}

First, we define the error structure of the model observations.

\begin{definition}\label{def:obs}
Let $y(t)$ be \underline{observations of the model} $\dot{U}=\Theta(U,w)$ with some associated measurement noise so that
$$y(t)=\textcolor{black}{\mathcal{Y}(\Omega(U\textcolor{black}{(t)},w), \epsilon)}$$
where $\epsilon\sim\mathcal{F}$ and $\mathcal{F}$ is an arbitrary distribution with variance $\sigma^2$. \textcolor{black}{For example, in the case of additive noise, $y(t)=\Omega(U\textcolor{black}{(t)},w)+\epsilon$, and, in the case of multiplicative noise, $y(t)=\epsilon\Omega(U\textcolor{black}{(t)},w)$.}
\end{definition}

\textcolor{black}{Note that $\Omega$ is a vector of the observation of states in the absence of noise as described in Section \ref{subsec:struct_ident}. The examples in this study and the following definitions are written in terms of a single observation state. While $(e,q)-$ identifiability can be applied for more than one observation state, we leave choosing an appropriate scaling of multiple observation states for future work.} 

Next, we define a scalar that describes the size of the measurement error relative to the size of the data measurements.

\begin{definition}\label{def:e}
Let $e$ denote the \underline{\textcolor{black}{observation} error ratio} such that \textcolor{black}{$e= \sigma / \text{RMS}(\Omega)$}, \textcolor{black}{where $\sigma^2$ is variance of the error distribution, as defined above,} and RMS is the root mean square, i.e., \textcolor{black}{$\text{RMS}(\Omega)=
{\frac{1}{\sqrt{T_m-T_0}}\int_{T_0}^{T_M} \Omega(t)^2dt}$. Here, we assume that observations are made at time points $\{T_0,...,T_M\}$. }  
\end{definition}

Next, we define a scalar that describes the size of the error in parameter estimates relative to the size of the parameter.

\begin{definition}\label{def:q}
    Let $q$ denote the \underline{\textcolor{black}{estimator} error ratio} such that \textcolor{black}{$q=\sqrt{M_i}/|w_i|$}, where $M_{i}$ is the maximum tolerated \textcolor{black}{mean-squared} error in the parameter estimator, $\widehat{w}_i$. \textcolor{black}{Here, $\text{MSE}(\widehat{w})=\mathbb{E}[(\widehat{w}_k-w)^2]=\text{Var}(\widehat{w})+\text{Bias}(\widehat{w},w)^2$, and $w$ is the true value of the parameter.}
\end{definition}

Finally, we define an identifiability criterion relating these two ratios.\\

\noindent\fbox{\begin{minipage}{\textwidth}
\begin{definition}\label{def:eq-ident}
A parameter\textcolor{black}{, $w_i$,} is said to be \underline{$(e,q)$-identifiable} if, for a given \textcolor{black}{observation error} ratio, $e$, the mean-squared error (MSE) of the parameter estimator, $\widehat{w}_i$, is below \textcolor{black}{the maximum tolerated mean-squared error,} $M_i$, i.e.,
\[
MSE(\widehat{w}_i(\textcolor{black}{e}))<(\textcolor{black}{q}\textcolor{black}{w_i})^2~.
\]
Furthermore, a model is said to be \underline{$(e,q)$-identifiable} if the above holds for all $i$.
\label{def:eq_ident}
\end{definition}
\end{minipage}}
\newline

 Since the MSE can be decomposed into the variance and squared bias, $(e,q)-$\textcolor{black}{identifiability} indicates that we can obtain parameter estimates that are both near the true parameter and near each other, \textcolor{black}{ in other words both accurate and precise,} with respect to a cut-off, $(qw)^2$, that is scaled by the size of the parameter and the choice of acceptable estimate error ratio $q$. \textcolor{black}{Note, $(e,q)-$identifiability could be applied \textit{a priori} or \textit{a posteriori}. When the true parameter value is unknown, as will be the case in most {\it a posteriori} applications, and when we have an unbiased estimator, we can let $w_{i}=\mathbb{E}(\widehat{w}_{i_k})$.}

\textcolor{black}{As with most simulation-based tests of practical identifiability, our goal in using the values $e$ and $q$ is to relate the uncertainty in the parameter estimator to uncertainty in the observations, and for these values to be interpretable and flexible in application. The observation error ratio, $e$, represents a ratio of a measurement ($RMS$) of the observation magnitude and the variance of the error, which allows us to use the criterion independently of the scale of the data. The estimator error ratio, $q$, represents the ratio of parameter magnitude and mean-squared error of the parameter estimator, which allows the use of the criterion for a variety of ODE systems and estimators. For example, if two models are (5,20)-identifiable, i.e., $e=5\%$ and $q=20\%$, for some set of data, then for variance in the error of the data that is $5\%$ of the average observation, we expect all parameter MSEs to remain below $20\%$ of the magnitude of the true parameters. Thus, while these models may have different scalings, they can still be compared directly.}

\section{\textcolor{black}{Weak-form parameter estimation for systems with unobserved variables}}\label{sec:weak_IO}
In this section, we combine established differential elimination methods with recent advances in weak-form parameter estimation to accurately estimate parameters from systems with unobserved compartments. In Section \ref{subsec:weak_gen}, we describe a method to generate weak-form input-output equations by using differential elimination, then \textcolor{black}{in Section \ref{subsec:wendy1}} present a method of parameter estimation using these equations. \textcolor{black}{Next, in Section \ref{subsec:struct_weakform}, we discuss structural identifiability and limitations of the weak-form input-output parameter estimation.} Lastly, in Section \ref{subsec:weakIO-examples}, we illustrate the application of these methods to two canonical biological systems.

\subsection{\textcolor{black}{Weak-form Input-output equations}}\label{subsec:weak_gen}
In general, we convert \textcolor{black}{from}  an equation in the form $\dot{U}=\Theta(U,p)$, with \textcolor{black}{parameters $p$ and} observed states, $y(t)=\Omega(U,p)$, to an input-output equation of the form $0=F(y,w)$\textcolor{black}{, where the terms of $F(\cdot)$ are differential monomials and $w_i=f_i(p)$ are expressions of the parameters of the full ODE system}. This process involves, first, ranking the model variables and their derivatives with the observed state variable at the lowest rank. Next, differential polynomials generated from the model equations are reduced to their characteristic set according to the new ranking using a differential elimination algorithm. After the characteristic set is obtained, the lowest ranking differential polynomial(s) will be the input-output equation(s). 

\textcolor{black}{Let} the right-hand side of the input-output equation can be written in the form 
\begin{equation}
F(y)=\sum_{i=0}^n\frac{\textsf{d}^i}{\textsf{d}^it}H_i(y)\textcolor{black}{+\left(\sum_{i=0}^n\frac{\textsf{d}^i}{\textsf{d}^it}C_i(y)\right)w}.\label{eq:diff_form1}
\end{equation}\textcolor{black}{Here, $H_i$ and $C_i$ are rational expressions of $y$. Then,} it is clear that by convolving the input-output equation with a test function $\phi$ and applying integration by parts, we can generate a weak-form input-output equation of the form
\begin{equation}
\textcolor{black}{\sum_{i=0}^n\int_0^T(-1)^i(\frac{\textsf{d}^{i}}{\textsf{d}t^i}\phi(t))H_i(y)\textsf{d}t=\left(\sum_{i=0}^n\int_0^T(-1)^i(\frac{\textsf{d}^{i}}{\textsf{d}t^i}\phi(t))C_i(y)\textsf{d}t\right)w.}
\label{eq:weak_from1}\end{equation} 

However, not all nonlinear systems can be written in the form of equation \eqref{eq:diff_form1}. \textcolor{black}{Instead, consider}
\begin{equation}
F(y)=\sum_{i=0}^nK_i(y)\frac{\textsf{d}^i}{\textsf{d}^it}H_i(y)\textcolor{black}{+\left(\sum_{i=0}^nJ_i(y)\frac{\textsf{d}^i}{\textsf{d}^it}C_i(y)\right)w,}\label{eq:diff_form2}
\end{equation}
\textcolor{black}{where $K_i$ and $J_i$ are rational expression of $y$ and its derivatives $\{\dot{y}$, $\ddot{y},...\}$.} In the case that the right-hand side of the input-output equation takes the form \textcolor{black}{of equation \eqref{eq:diff_form2}}, either substitutions of variables, estimation of higher order derivatives from data, or both will be required to compute terms that are not integrable by parts.

\subsection{\textcolor{black}{Weak form Estimation of Nonlinear Dynamics (WENDy)}}\label{subsec:wendy1}

\textcolor{black}{Recently, our group has developed the Weak form Estimation of Nonlinear Dynamics (WENDy) method for parameter estimation \cite{BortzMessengerDukic2023BullMathBiol}. Briefly, in the WENDy method, a system is cast in its weak form, then data \textcolor{black}{are} substituted directly into the equation, and parameters are estimated by solving a regression problem. A more detailed description of the method can be found in \cite{BortzMessengerDukic2023BullMathBiol}. Most importantly, this method allows us to estimate parameters from systems in the form of \eqref{eq:weak_from1} directly using WENDy\footnote{\textcolor{black}{Note that in this formulation, we are limited to problems that are linear-in-parameters (not to be confused with linear problems). Recently, \cite{RummelMessengerBeckerEtAl2025arXiv250208881} has addressed problems that are nonlinear-in-parameters.}}.}

\textcolor{black}{We begin by considering the ordinary differential
equation (ODE)
$\dot{U}=\Theta(U,p)$ with row vector of $d$ observed states \begin{equation*}
    y(t):=(y_1(t),...,y_d(t)).
\end{equation*}
We assume the corresponding $d$-dimensional input-output equation $0=F(y,w)$ can be written in the form of \eqref{eq:diff_form1}.
We let the superscript $\star$ notation denote quantities based on the true (noise-free and unknown) parameter values ($p^{\star}$). We
assume that at each time point $t\in\tbf=(t_0,\dots,t_M)$ measurements of the system are observed with additive noise
\begin{equation}
y(t)=y^{\star}(t)+\varepsilon(t)\label{eq:additive noise-1}
\end{equation}
where each element of $\varepsilon(t)$ is i.i.d.~$\mathcal{N}(0,\sigma^{2})$. Using bold-face variables to represent evaluation at the timegrid $\tbf$, our observations then consist of samples $\mathbf{y}:=\mathbf{y}^{\star}+\pmb{\varepsilon}\in\mathbb{R}^{M+1\times d}$.}

\textcolor{black}{Our goal is to estimate the structurally identifiable parameter expressions $w^{\star}$ from the input-output equation representation of the ODE system. We start by considering the weak-form input-output equation 
\begin{equation}
\sum_{i=0}^n\int_0^T(-1)^{i+1}(\frac{\textsf{d}^{i}}{\textsf{d}t^i}\phi) H_i(y)\textsf{d}t=\left(\sum_{i=0}^n\int_0^T(-1)^i(\frac{\textsf{d}^{i}}{\textsf{d}t^i}\phi)C_i(y)\textsf{d}t\right)\textsf{mat}(w).\label{eq:weak_from1mat}\end{equation} 
where ``$\mathsf{mat}$'' is the matricization operation.}

\textcolor{black}{Using a set of test functions $\{\phi_k\}_{i=1}^{K}$, \eqref{eq:weak_from1mat} can be discretized at observed data $\mathbf{y}$, yielding approximated integrals 
\begin{equation}
\sum_{i=0}^n(-1)^{i+1}(\frac{\textsf{d}^{i}}{\textsf{d}t^i}\Phi_k) H_i(\mathbf{y})=\left(\sum_{i=0}^n(-1)^i(\frac{\textsf{d}^{i}}{\textsf{d}t^i}\Phi_k)C_i(\mathbf{y})\right)W.\label{eq:discrete_wk_IO}
\end{equation} Here $\frac{\textsf{d}^{i}}{\textsf{d}t^i}\Phi_k:=[\frac{\textsf{d}^{i}}{\textsf{d}t^i}\phi_k(t_0)|...|\frac{\textsf{d}^{i}}{\textsf{d}t^i}\phi_k(t_M)]\mathcal{Q}$ where $\mathcal{Q}$ is a diagonal matrix containing the Newton-Cotes quadrature weights. Let $\mathbf{C_{\phi,y}}=\sum_{i=0}^n(-1)^i(\frac{\textsf{d}^{i}}{\textsf{d}t^i}\Phi)C_i(\mathbf{y})$ and $\mathbf{H_{\phi,y}}=\sum_{i=0}^n(-1)^{i+1}(\frac{\textsf{d}^{i}}{\textsf{d}t^i}\Phi) H_i(\mathbf{y})$. By minimizing the contribution of integration error to the residual, this discretization leads to the following corresponding linear system $\Gbf\wstar\approx \bbf$ \cite{BortzMessengerDukic2023BullMathBiol}, where
\begin{align*}
\Gbf & :=[\mathbb{I}_{d}\otimes \mathbf{C_{\phi,y}}],\\
\bbf & :=\mathsf{vec}(\mathbf{H_{\phi,y}}),
\end{align*}
from which we can obtain an estimate of $w^{\star}$.}

\textcolor{black}{As presented in \cite{BortzMessengerDukic2023BullMathBiol}, the WENDy algorithm addresses the Errors-In-Variable nature of the linearized problem using iterative reweighting to approximate the correct underlying covariance structure\footnote{\textcolor{black}{The approximated covariance can also be used to calculate parameter confidence intervals. \textcolor{black}{A brief derivation of the parameter covariance approximation is given in Appendix \ref{app:wendy_covar}, and we direct readers to \cite{BortzMessengerDukic2023BullMathBiol} for further details.}}}. This reweighting is dependent on the assumed noise structure of the observed data. For Gaussian additive noise, we apply the covariance correction using \textsf{Algorithm} 2 in \cite{BortzMessengerDukic2023BullMathBiol}. Additionally, it has been shown in \cite{RummelMessengerBeckerEtAl2025arXiv250208881} that, through a strategic choice of transformations to the observations $\mathbf{y}$, it is possible to approximate the covariance structure for systems where observational noise is multiplicative and lognormal (i.e. $\mathbf{y}=\mathbf{y}^{\star}\varepsilon$ where $\log(\varepsilon(\mathbf{t}))\sim\mathcal{N}(0,\sigma^{2})$).}

\subsection{\textcolor{black}{Structural Identifiablity of Weak-form Problem}}\label{subsec:struct_weakform}

\textcolor{black}{Recall Definition \ref{def:io_ident} from Section \ref{subsec:struct_ident}, note that the terms of the input-output equation must be linearly independent on the problem's domain in order to uniquely recover model parameters. It is also important to note that the conversion of an input-output equation in the form of equation \eqref{eq:diff_form1} to a weak-form input-output equation in the form of equation \eqref{eq:weak_from1} does not preserve the linear independence of terms for arbitrary choices of test function, $\phi(t)$. In particular, the integration operation does not preserve dependence relationships. Furthermore, the use of Definition \ref{def:io_ident} to check structural identifiability is equivalent to testing the membership in a differential field. An analogous problem in the weak-form would require the generation of an integro-differential field, and is an open field of research \cite{LemaireRoussel2024ComputerAlgebrainScientificComputing}.}

\textcolor{black}{Alternatively, we propose verifying the recovery of the unique parameters, $w$, locally from the discretized problem given by equation \eqref{eq:discrete_wk_IO}. The WENDy regression problem has a unique solution when the matrix G is of full rank. In \cite{BortzMessengerDukic2023BullMathBiol}, this condition could be met for a choice of test functions such that $\text{rank}(\mathbf{\Phi})=K$ for a system without higher order derivatives of the observation states. We conjecture that for the case presented in Section \ref{subsec:wendy1}, it is necessary to require $\text{rank}(\mathbf{\Phi})=\text{rank}(\dot{\mathbf{\Phi}})=\text{rank}(\mathbf{\frac{\textsf{\textbf{d}}^{i}}{\textsf{d}t^i}}\mathbf{\Phi})=K$ to obtain a unique solution to the WENDy regression problem. We compare the expected structural identifiability derived from the strong-form input-output equation to the local condition on the rank of matrix G for two examples in Appendix C.}

\textcolor{black}{Finally, it is essential to note that the conversion to a weak-form input-output equation entails the loss of information about any initial conditions. There are examples of biological systems where initial conditions are prerequisite to structural identifiability, for example, within-host models of virus dynamics \cite{Heitzman-BreenLiyanageDuggalEtAl2024RSocOpenSci,LiyanageHeitzman-BreenTuncerEtAl2024MBEa}. In this study, we consider examples of systems where structural identifiability is not dependent on initial conditions.}

\subsection{Illustrating Examples}\label{subsec:weakIO-examples}
\begin{figure}[ht]
    \centering
\includegraphics[width=\linewidth]{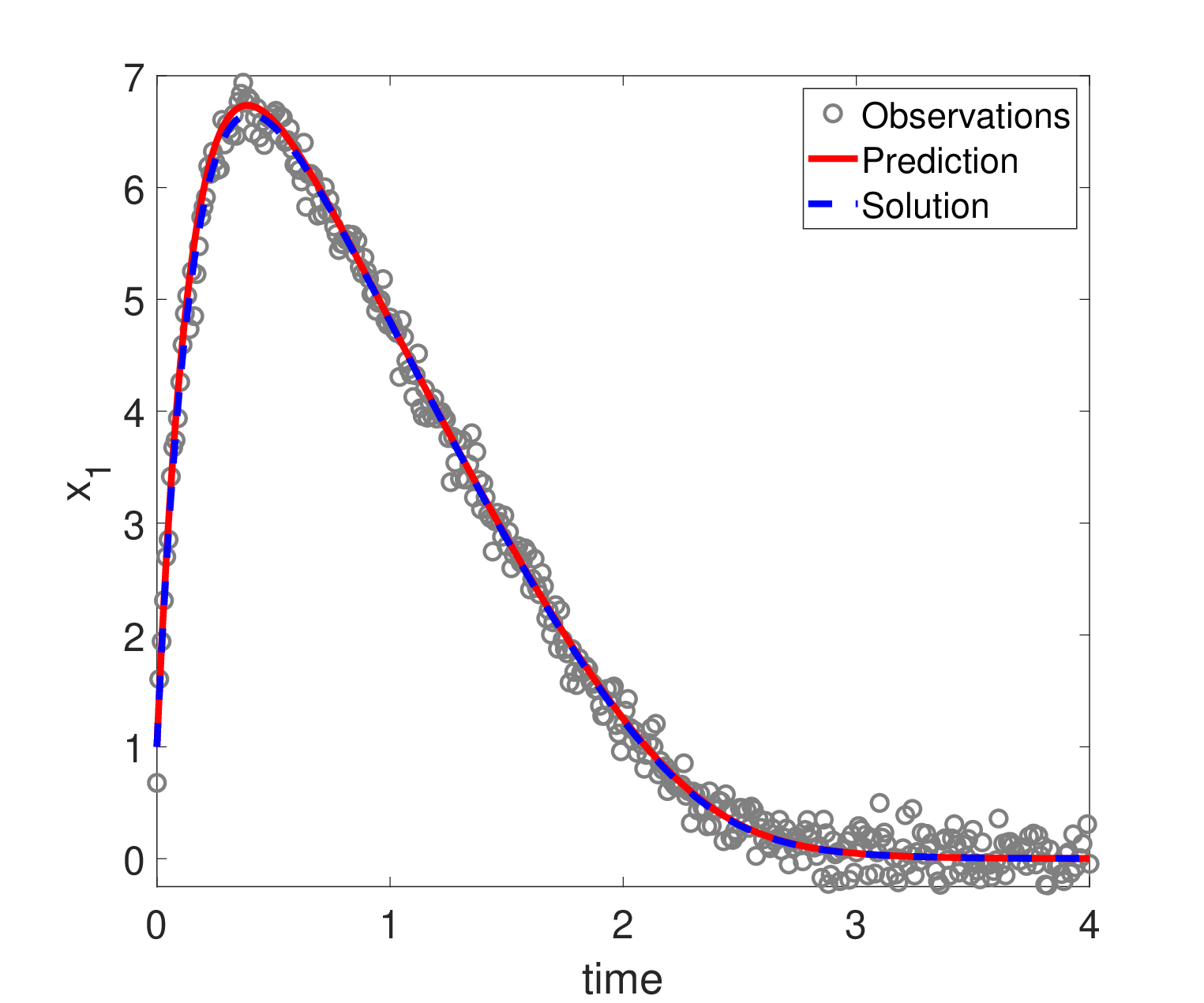}
    \caption{Using WENDy with equation \eqref{eq:weak_bld_cnc} we recover parameters for the blood diffusion model \eqref{eq:bld_cnc} with only observations in the blood compartment ($x_1(t)$). The drug concentration dynamics in the blood compartment (red) and the tissue compartment (blue) are given for model \eqref{eq:bld_cnc} with parameters $k_{12}=5$, $k_{21}=1$, and $V_e=6$. \textcolor{black}{These parameters are chosen to allow comparison to the example in \cite{BoulierKorporalLemaireEtAl2014ComputerAlgebrainScientificComputing}.} The plot depicts the relative error in estimating the true parameters using WENDy from 400 observations of the blood compartment (black dots) with $e=5\%$ \textcolor{black}{additive observation error} ratio. Note that $e=5\%$ \textcolor{black}{additive observation error} ratio is equivalent to the white noise of $\sigma=25.06\%$ for this choice of model parameters.}
    \label{fig:bld_cnc_ex}
\end{figure}

\begin{figure}[ht]
    \centering
\includegraphics[width=\linewidth]{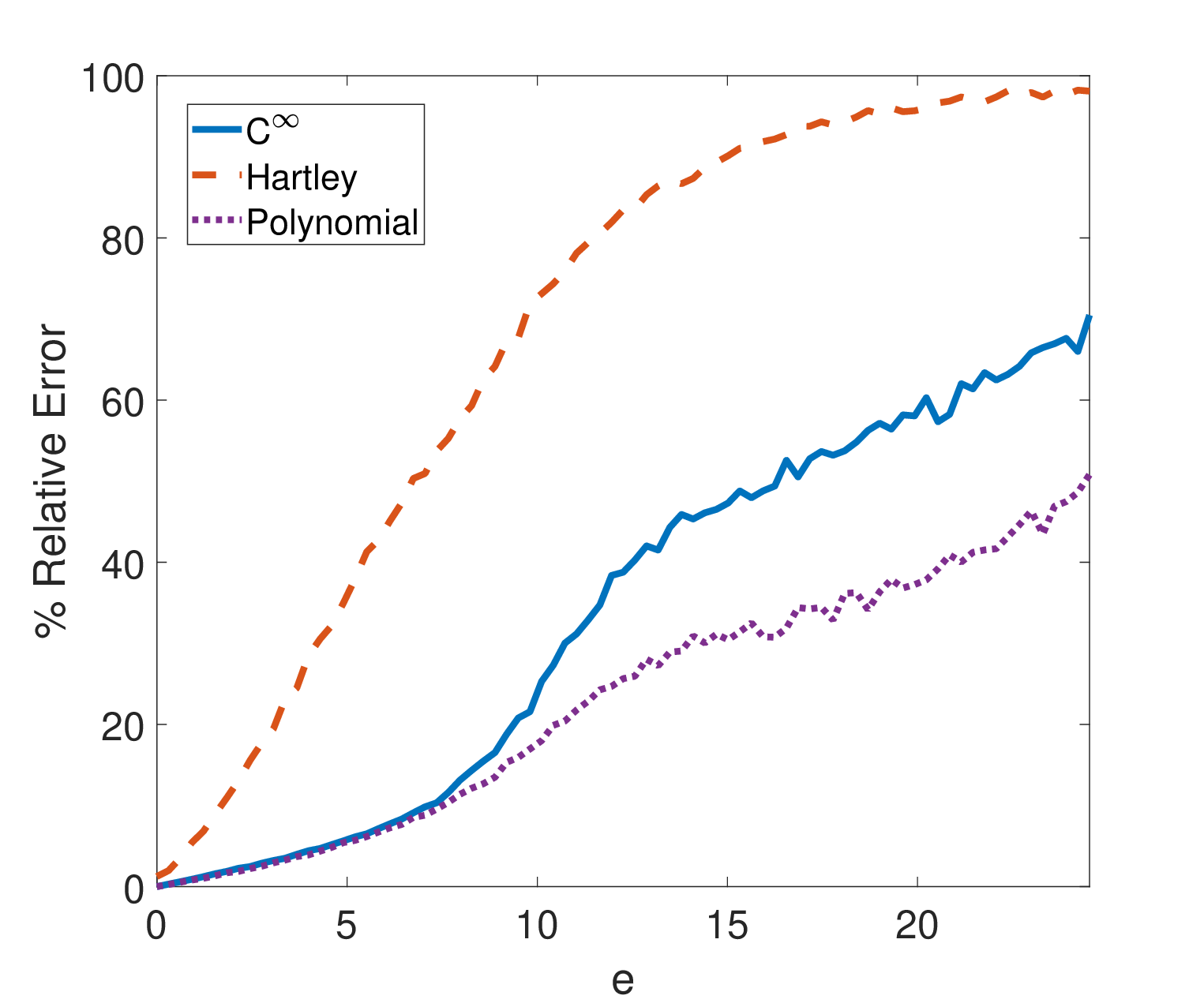}
    \caption{\textcolor{black}{For using WENDy to estimate the parameters in \eqref{eq:weak_bld_cnc}, this figure depicts the relative parameter error vs. observation error ratio $e$. The different curves represent using WENDy with different test functions ($C^\infty$ from \eqref{eq:C_inf}, Hartley from \eqref{eq:Hartley}, and Polynomial from \eqref{eq:poly}). When using either $C^\infty$ or polynomial test functions (as is standard in WENDy), the relative error remains below $25\%$ for observations with up to a $10\%$ additive error ratio, and for polynomial test functions, the relative error remains below $50\%$ for observations with up to a $24.5\%$ additive error ratio. Note that each value on these curves represents the average relative error for 1,000 simulated datasets with additive Gaussian error ratio $e\in[0\%,24.5\%]$. Also note that the error ratio is equivalent to additive white noise of the level $\sigma\in[0,0.8].$}}
    \label{fig:bld_cnc_compare}
\end{figure}

We present two example systems whose input-output equations take on the forms of equations \eqref{eq:diff_form1} and \eqref{eq:diff_form2} when a single state variable is observed.
\medskip
\begin{exmp}\label{ex:1}
We are motivated by an example \cite{BoulierKorporalLemaireEtAl2014ComputerAlgebrainScientificComputing} where a weak-form input-output equation is used to estimate parameters in the blood-tissue diffusion model. Diffusion of a drug between blood and tissue in the body is assumed to be proportional to the concentration gradient, and is represented by the following equation,
\begin{equation}
    \begin{split}
        \dot{x_1}(t)&=-k_{12}x_1(t)+k_{21}x_2(t)-\frac{V_ex_1(t)}{1+x_1(t)},\\
        \dot{x_2}(t)&=k_{12}x_1(t)-k_{21}x_2(t),
    \end{split}\label{eq:bld_cnc}
\end{equation}
where $x_1$ and $x_2$ are the concentrations of drug in the blood and tissue, respectively, $k_{12}$ is the diffusion rate from the blood into the tissue, $k_{21}$ is the diffusion rate from the tissue into the blood, and $V_e$ is the rate of drug decay in the blood. The input-output equation corresponding to \eqref{eq:bld_cnc} when only $x_1(t)$ is observed \textcolor{black}{ is given below,
\begin{equation}
\begin{split}
\ddot{x}_1(t)x_1(t)^2+2\ddot{x}_1(t)x_1(t)+\ddot{x}_1(t)+w_1\left(x_1(t)^2+x_1(t)\right)\\+w_2\left(\dot{x}_1(t)x_1(t)^2+2\dot{x}_1(t)x_1(t)\right)+w_3\dot{x}_1(t)=0,\label{eq:bldcnc_io_og}
\end{split}
\end{equation}
where $w_1=k_{21}V_e$, $w_2=k_{12}+k_{21}$, and $w_3=k_{12}+k_{21}+V_e$.} \textcolor{black}{We confirm that equation \eqref{eq:bldcnc_io_og} is globally structurally identifiable using available software \textcolor{black}{discussed} in Appendix \ref{app:weak-struct}.
}

\textcolor{black}{Equation \eqref{eq:bldcnc_io_og} can be written in the form of \eqref{eq:diff_form1} }and the weak-form input-output equation for the blood diffusion model is given below,
\begin{equation}
\begin{split}
-\int_{T_1}^{T_2}\Ddot{\phi}x_1(t)\textsf{d}t=&w_1 \int_{T_1}^{T_2}\phi\frac{x_1(t)}{x_1(t)+1}\textsf{d}t - w_2\int_{T_1}^{T_2} \Dot{\phi} \frac{x_1(t)^2}{x_1(t)+1}\textsf{d}t +w_3 \int_{T_1}^{T_2} \Dot{\phi}\frac{1}{x_1(t)+1}\textsf{d}t.
    \end{split}\label{eq:weak_bld_cnc}
\end{equation}

For this example, we consider three choices for the form of the test function, $\phi$. We consider $C^{\infty}$ bump functions, which have been used in \cite{BortzMessengerDukic2023BullMathBiol}. These functions are given by the following form,
\begin{equation}
    \phi(t,a) = \begin{cases}
        C\text{exp}(-\frac{\eta}{1-(t/a)^2}), & [-a, a]\\
        0,& \textit{elsewhere},
    \end{cases}\label{eq:C_inf}
\end{equation}
where $C$ is chosen so that $||\phi||_2=1$, $a$ is the radius of support, and $\eta$ is a shape parameter. For this example, \textcolor{black}{$a=0.6$} and $\eta=9$. 

We also consider Hartley modulating functions, which were applied to this weak-form system in \cite{BoulierKorporalLemaireEtAl2014ComputerAlgebrainScientificComputing}. Below, we give the equation for a third-order Hartley modulating function,
\begin{equation}
        \phi(t,a) = \begin{cases}
        C(\text{cas}(\frac{6\pi}{a}t)-3\text{cas}(\frac{4\pi}{a}t)+3\text{cas}(\frac{2\pi}{a}t)-1), & [-a, a]\\
        0,& \textit{elsewhere},
    \end{cases}\label{eq:Hartley}
\end{equation}
where $C$ is chosen so that $||\phi||_2=1$, $a$ is the radius of support, and $\text{cas}(t)=\cos(t)+\sin(t)$. For this example, \textcolor{black}{$a=0.8$}.

We also consider polynomial test functions, which were shown to be highly robust to noise in \cite{MessengerBortz2021MultiscaleModelSimul}. Below we give the equation for a $12^{\text{th}}$ order polynomial test function,
\begin{equation}
        \phi(t,a) = \begin{cases}
        C(t+a)^6(a-t)^6, & [-a, a]\\
        0,& \textit{elsewhere},
    \end{cases}\label{eq:poly}
\end{equation}
where $C$ is chosen so that $||\phi||_2=1$ and $a$ is the radius of support. For this example, \textcolor{black}{$a=0.52$}.

Figure \ref{fig:bld_cnc_ex} shows the drug diffusion system fit to only blood concentration\textcolor{black}{s} using WENDy with a polynomial test function, and Figure \ref{fig:bld_cnc_compare} shows the average relative error using WENDy for various additive noise ratios for three different test functions. In comparison to \cite{BoulierKorporalLemaireEtAl2014ComputerAlgebrainScientificComputing}, WENDy performs competitively, remaining under $50\%$ \textcolor{black}{for the} relative error with $80\%$ white noise in the observed variable when using a $12^{\text{th}}$ degree polynomial test function. This figure clearly demonstrates that the choice of test function greatly impacts the robustness of the parameter estimation method. \textcolor{black}{For further details on the choice of test functions and their respective radii of support, see Appendix \ref{app:testfunc}.}
\end{exmp}
\medskip

\begin{exmp}\label{ex:2}
The SIR model describes the spread of an infection in a population \textcolor{black}{with no birth, no death, nor immigration.} Susceptible individuals ($S$) become infected ($I$) at a rate $\beta$ and recover ($R$) without chance of reinfection at a rate $\alpha$. Model equations are presented in the equation below,
\begin{equation}
    \begin{split}
        \dot{S} &= -\beta SI,\\
        \dot{I} &= \beta SI -\alpha I,\\
        \dot{R} &= \alpha I.
    \end{split}\label{eq:SIR}
\end{equation}
We assume the infected compartment is the only state variable observed. We then obtain the SIR input-output equation for the observed state variable $I$,

\begin{equation}
    0=-\frac{\dot{I}^2}{I}+\beta\alpha I^2+\beta I\dot{I}+\ddot{I}.\label{eq:io_sir}
\end{equation}

Note that, as expected, both parameters $\beta$ and $\alpha$ of model \eqref{eq:SIR} are globally structurally identifiable. \textcolor{black}{We confirm that the equation \eqref{eq:io_sir} yields that the unknown model parameters are globally structurally identifiable by inspection and using the available software discussed in Appendix \ref{app:weak-struct}.}

 From equation \eqref{eq:io_sir} we can generate the weak form of model \eqref{eq:SIR}, SIR system for observed state $I$ is presented below,

\begin{equation}
    \int_0^T \ddot{\phi}I\textsf{d}t-\int_0^T \phi\frac{\dot{I}^2}{I}\textsf{d}t=\beta\int_0^T \dot{\phi}\frac{I^2}{2}\textsf{d}t-\beta\alpha\int_0^T \phi I^2\textsf{d}t.\label{eq:wf_sir}
\end{equation}

In this case, we are unable to simplify the term $\int_0^T \phi \frac{\dot{I}^2}{I}\textsf{d}t$ using integration by parts.\footnote{\textcolor{black}{Alternatively, equation \eqref{eq:wf_sir} can be written as $\int_0^T(\ddot{I}I^2-\dot{I}^2I)/I^2\textsf{d}t=\beta\int_0^T \dot{\phi}I\textsf{d}t-\beta\alpha\int_0^T \phi I\textsf{d}t.$ Applying integration by parts to the left-hand side yields $\int_0^T\ddot{\phi}\ln(I)\textsf{d}t.$ However, preliminary results using this form for WENDy parameter estimation are less accurate than the form presented in equation \eqref{eq:wf_sir_conpop} and addressing this challenge is a topic for future research.}} However, let the total population $N$ be \textcolor{black}{ a known }constant, $N=S(t)+I(t)+R(t)$, and additionally, re-write $R$ in terms of only the observed variable $I$,
\begin{equation}
    R(t)=\int_0^t\alpha I(s)\textsf{d}s.\label{eq:Rsub}
\end{equation}
Then, using equations \eqref{eq:wf_sir}-\eqref{eq:Rsub}, we define a weak-form SIR model that can be used to estimate the parameter $\beta$ below,
\textcolor{black}{\begin{equation}
\begin{split}  \int_0^T\dot{\phi}I\textsf{d}t+\alpha\int_0^T\phi I \textsf{d}t&=-\beta\int_0^T\phi(R(t)+I-S_0)I\textsf{d}t,\\
    R(t)&=\int_0^t\alpha I(s)\textsf{d}s.
\end{split}\label{eq:wf_sir_conpop}
    \end{equation}}
\textcolor{black}{We verify the structural identifiability of equation \eqref{eq:io_sir} and the local recovery of parameters in the weak form Equation \eqref{eq:wf_sir_conpop} in Appendix \ref{app:weak-struct}.}
Figure \ref{fig:sir_ex}A shows the SIR model fit to only the infected population using WENDy.
\end{exmp}

\begin{figure}[ht]
    \centering
\includegraphics[width=\linewidth]{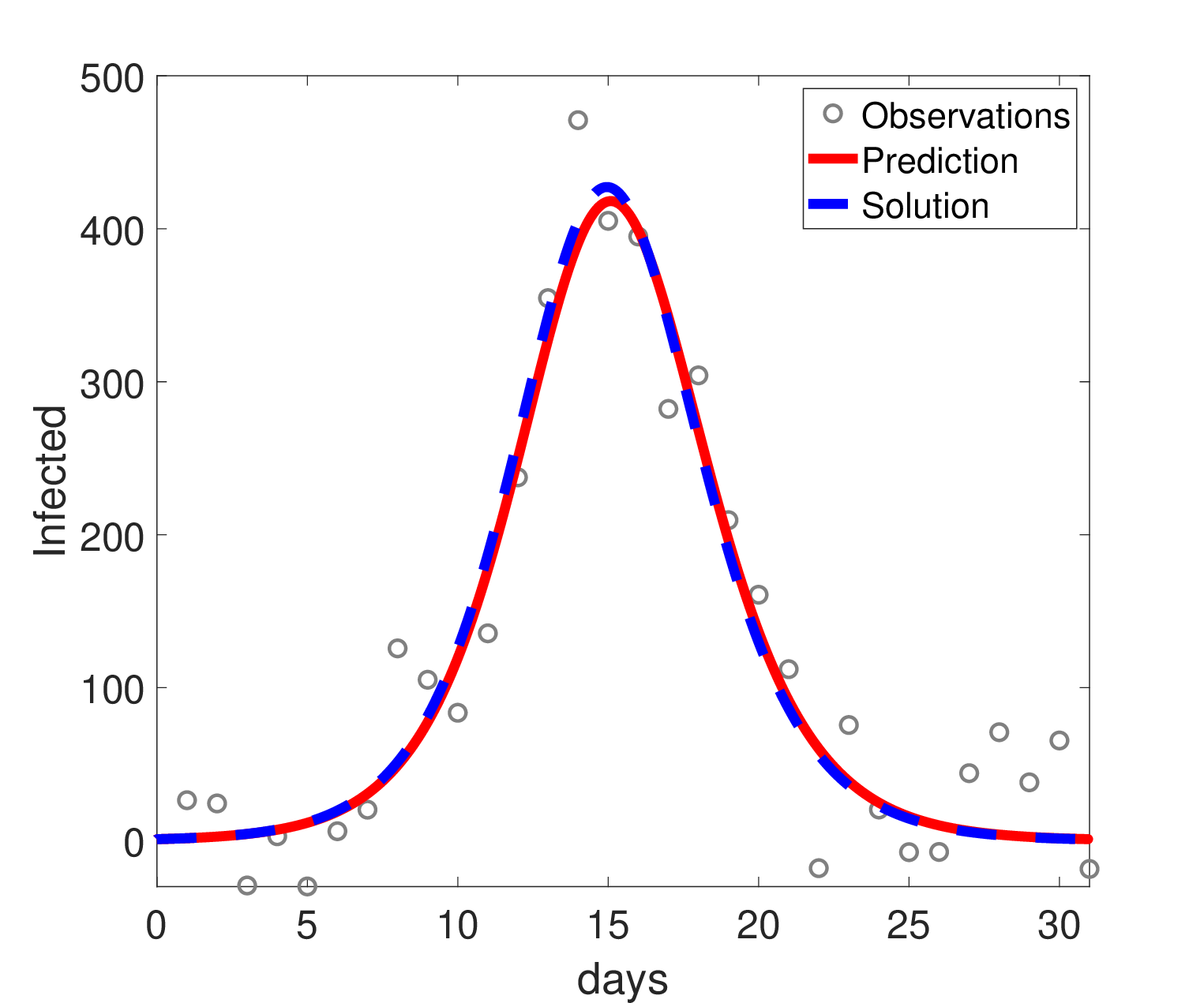}
    \caption{Using WENDy with equations \eqref{eq:wf_sir_conpop} we recover the transmission rate $\beta$ for the SIR model \eqref{eq:SIR} with only observations in the infected compartment, even with large observational noise. The infected compartment dynamics are given for model \eqref{eq:SIR} with N=10,000, $S_0=N-1$, $I_0=1$, $R_0=0$, $\beta=\frac{5.5}{N}$, $\gamma=5$. \textcolor{black}{ These parameter values correspond to an infection with a 5-day infectious period and a basic reproduction number $\mathcal{R}_0=1.1$, which could reasonably represent a seasonal influenza outbreak \cite{BiggerstaffCauchemezReedEtAl2014BMCInfectDis}.} We estimate the true value of $\beta$ using WENDy from 31 observations of the infected compartment (black dots) and recover the original dynamics (red line) observations with $e=20\%$ \textcolor{black}{additive observation error} ratio (Right).}
    \label{fig:sir_ex}
\end{figure}

\section{Examples of Practical Identifiability}\label{sec:pract_ident}

In this section, we will describe the Practical identifiability of the weak form systems of the examples generated in Section \ref{subsec:weak_gen}. In Section \ref{subsec:pract_method}, we describe the method of generating simulated datasets. In Section \ref{subsec:pract_result}, we interpret the practical identifiability of two biological examples.

\subsection{Applying $(e,q)$-identifiability}\label{subsec:pract_method}

To assess the practical identifiability of examples \ref{ex:1} and \ref{ex:2} we generated \textcolor{black}{$D=1,000$ simulated datasets} for choices of \textcolor{black}{observational error} ratios between $e=[0\%,20\%]$. We generated noisy observations using Definition \ref{def:obs} and assuming \textcolor{black}{either a Gaussian additive error distribution or a multiplicative log-normal distribution.} For each example, we assumed a limited number of data points; specifically, $40$ observations were considered for Example \ref{ex:1}, and $31$ observations were considered for Example \ref{ex:2}. Parameter estimates were generated using the WENDy method described in \textcolor{black}{Section \ref{subsec:wendy1} with test functions given in Appendix \ref{app:testfunc}.}

Additionally, we compare the performance of WENDy to a traditional output error (OE) method. We implement the OE method in MATLAB using the function $\texttt{lsqnonlin}$ with default settings. To perform the forward solves required by the OE method, we use a variable-step, variable-order solver based on the numerical differentiation formulas (NDFs) of orders 1 to 5, as implemented in $\texttt{ode15s}$ in Example \ref{ex:1} and a Dormand-Prince method as implemented in $\texttt{ode45}$ in Example \ref{ex:2}.

\subsection{Results of practical identifiability}\label{subsec:pract_result}

\begin{exmp}
We estimate the parameters $w_1,$ $w_2$, and $w_3$ from the weak-form input-output equation \eqref{eq:weak_bld_cnc} \textcolor{black}{for the blood diffusion model} using WENDy (see \textcolor{black}{Section \ref{subsec:wendy1}}), for simulated data with Gaussian \textcolor{black}{additive observation error}. We vary the  \textcolor{black}{additive observation error} ratio in the simulated data between $e\in [0\%,20\%]$ such that the standard deviation of the additive error is $\sigma=e\text{RMS}(\Omega(t))$, and at each \textcolor{black}{additive observation error} ratio, we generate 1,000 noisy datasets. The $(e,q)$-identifiability (see Section \ref{subsec:pract_ident}) for $e\in[0\%,20\%]$ and $q\in[1\%,100\%]$ of model \eqref{eq:weak_bld_cnc} is given in Figure \ref{fig:bld_cnc_pract_ident}A, where the area in blue represents where the model is $(e,q)$-identifiable. For example, the model \underline{is} (5,50)-identifiable, meaning that at a $5\%$ \textcolor{black}{additive observation error} ratio in the data, the MSE of the parameter \textcolor{black}{estimator} $\widehat{w}=[\widehat{w}_1,\widehat{w}_2,\widehat{w}_3]$ falls below $50\%\times w_i$. However, at $15\%$  \textcolor{black}{additive observation error} ratio, this property no longer holds, i.e., the model \underline{is not} (15,50)-identifiable. We also observe from the $(e,q)-$identifiability of the individual parameters that the identifiability parameter $w_3$ is lost first as noise increases. In the example above, both $w_2$ and $w_3$ are $(15,50)-$identifiable, while $w_3$ is not. As demonstrated in Figure \ref{fig:bld_cnc_pract_ident}A, as \textcolor{black}{additive observation error} increases, the MSE increases and approaches the magnitude of the true parameters. \textcolor{black}{We do not find the blood-tissue diffusion model to be generally practically identifiable, using a (10,20)-identifiable criterion.}

\begin{figure}[ht]
    \centering
    \includegraphics[width=0.5\linewidth]{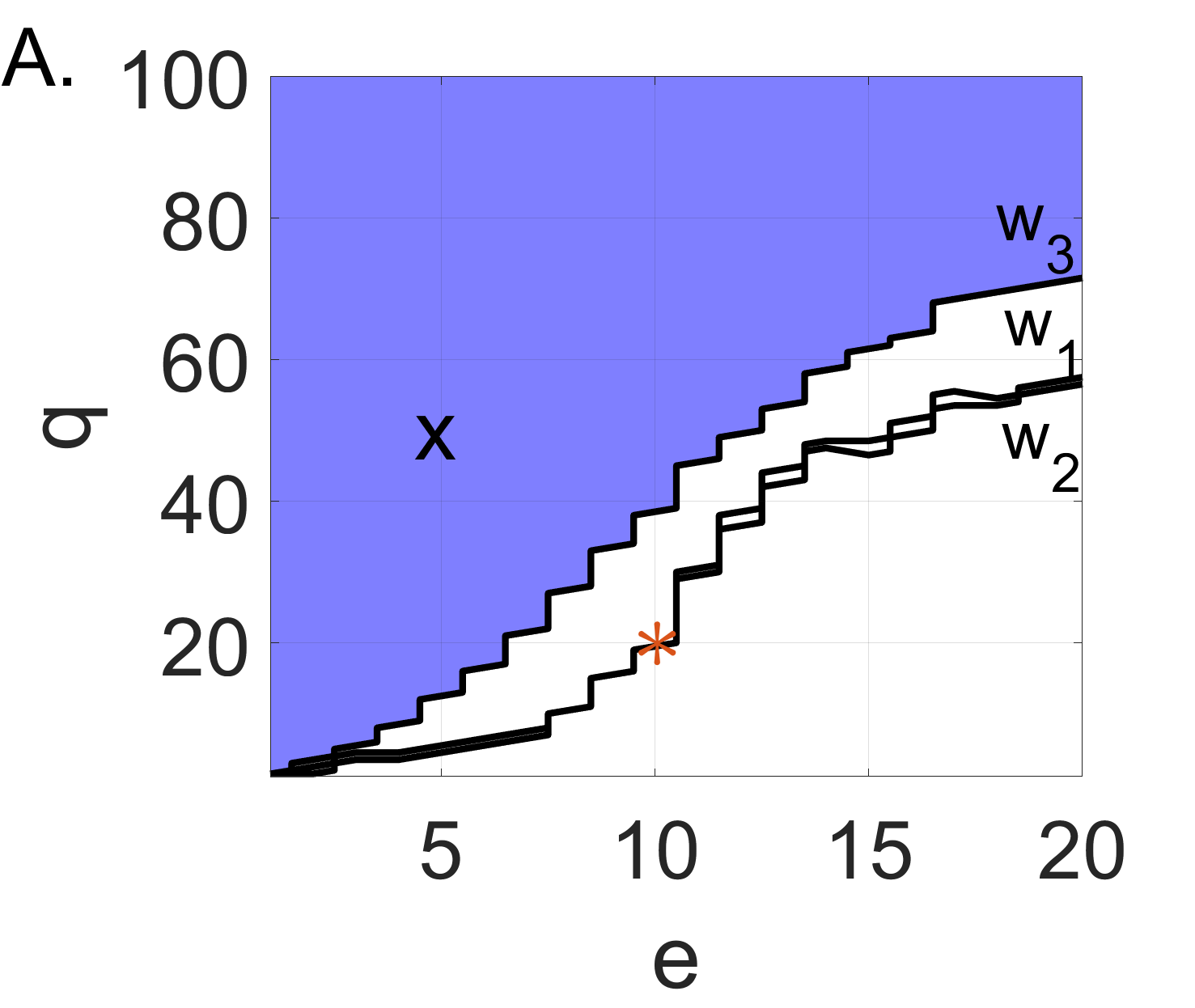}\includegraphics[width=0.5\linewidth]{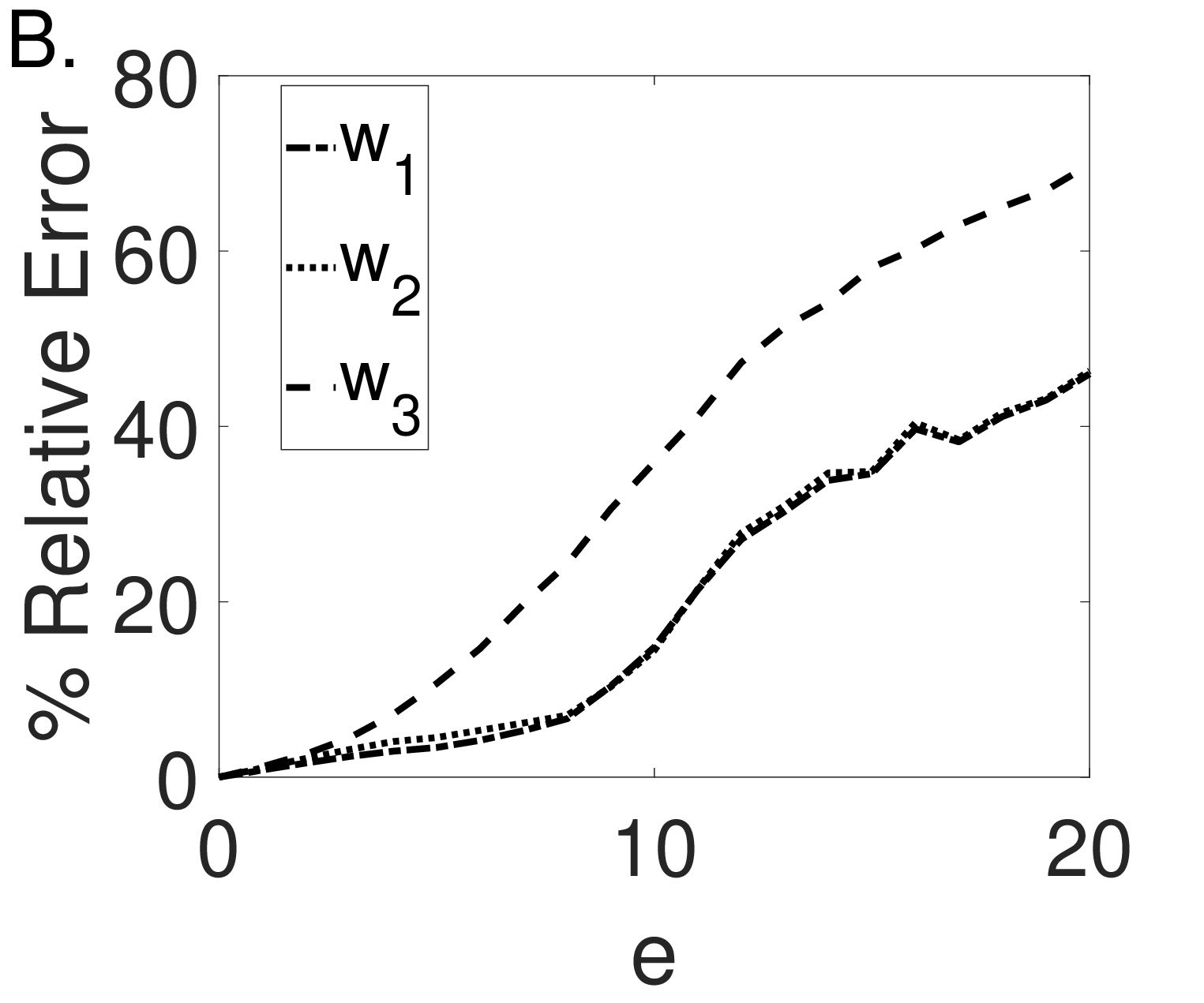}\\
\includegraphics[width=0.5\linewidth]{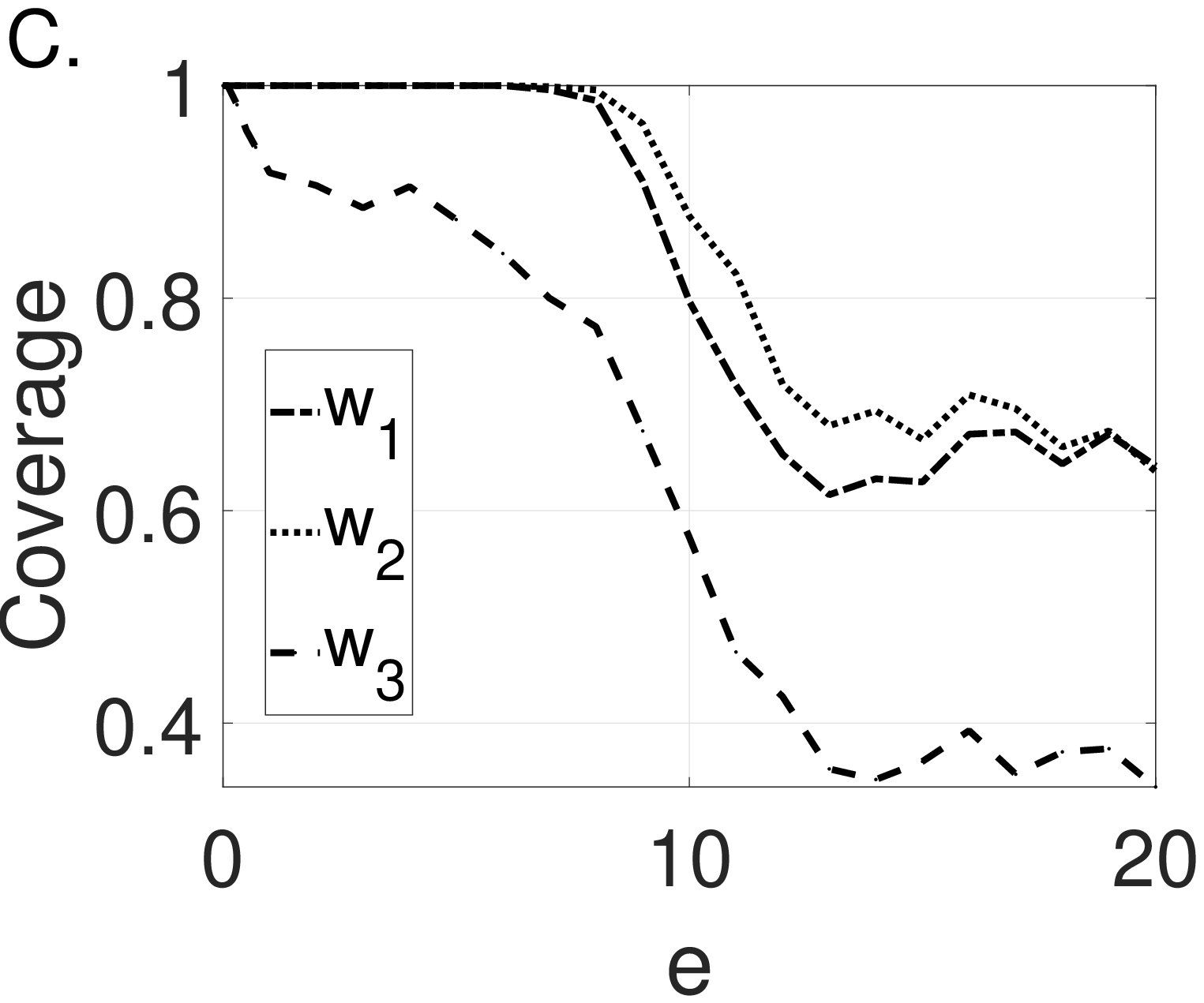}
    \caption{Using WENDy with equation \eqref{eq:weak_bld_cnc} we find the blood concentration model to be generally practically identifiable below $11\%$ scaled noise in the data. (A.) The area in blue denotes where the model is $(e,q)$-identifiable, and the area in white denotes where the model is not $(e,q)$-identifiable. The $(e,q)$-identifiability was determined from 1,000 simulations at each respective error level, e. The black $\pmb{\mathsf{X}}$ denotes the $(5,50)-$identifiability cutoff. Because this point falls within the blue region, we can say that at a $5\%$ \textcolor{black}{additive observation error} ratio in the data the MSE of $\widehat{w}_1,$ $\widehat{w}_2$, and $\widehat{w}_3$ remain below the square of $50\%$ the parameter magnitude. The red star 
\textcolor{red}{\textasteriskcentered } marks the $(10,20)$-identifiability criterion. (B.) The relative error determined from 1,000 simulations at each respective error level, e. (C.) The proportion of $95\%$ confidence intervals at \textcolor{black}{$e\in[0\%,20\%]$ } that contain the true value of $w_1,$ $w_2,$ and $w_3$.}
    \label{fig:bld_cnc_pract_ident}
\end{figure}

We additionally compare the $(e,q)$-identifiability of model \eqref{eq:weak_bld_cnc} to the average relative error and the coverage of the $95\%$ confidence intervals generated at each \textcolor{black}{additive observation error} ratio. Previous work by our group and others \cite{TuncerMarcthevaLaBarreEtAl2018BullMathBiol,Heitzman-BreenLiyanageDuggalEtAl2024RSocOpenSci} proposed (over many datasets, \textcolor{black}{$D\geq1,000$, } generated at a particular \textcolor{black}{additive observation error} ratio) that a model can be considered practically identifiable in terms of relative error, if the relative error is equal to or lower than the error introduced through additive noise. The relative error generated for 1,000 simulated datasets at each \textcolor{black}{additive observation error} level is given in Figure \ref{fig:bld_cnc_pract_ident}B, and the coverage of the $95\%$ confidence intervals for \textcolor{black}{${w}_1$, ${w}_2$, and ${w}_3$} are given in Figure \ref{fig:bld_cnc_pract_ident}C. Based on this strict relative error criterion, the blood concentration model is not practically identifiable for any error level. Even at $1\%$ additive noise, the relative error in parameter $w_3$ is $1.1\%$. However, a more relaxed condition on relative error for practical identifiability has been used in \cite{Heitzman-BreenLiyanageDuggalEtAl2024RSocOpenSci,LiyanageChowellPogudinEtAl2025Viruses}, where relative error needs only to be below $10e$. In this case, the blood concentration model is practically identifiable with respect to the relative error from $e=[0\%,20\%]$. Generally, the relative error remains near $2\times e$ for parameters $w_1$ and $w_2$ and near $3\times e$ for parameter $w_3$ for $e=[0\%,20\%]$. Additionally, above the $11\%$ \textcolor{black}{additive observation error} level, where the model is no longer (e,50)-identifiable, the coverage for all parameters $w_1$, $w_2$, and $w_3$ decreases as shown in Figure \ref{fig:bld_cnc_pract_ident}C. For parameters $w_1$ and $w_2$ at error levels below $e=6\%$ in all trials, the $95\%$ confidence interval contains the true values of $w_1$ and $w_2$. This decreases below $70\%$ at $e=11\%$ \textcolor{black}{additive observation error} ratio. All three criteria agree that parameter $w_3$ becomes non-identifiable first as noise increases.
\end{exmp}
\medskip
\begin{exmp}
We estimate the parameter $\beta$ from the modified weak-form input-output \textcolor{black}{SIR} equations \eqref{eq:wf_sir_conpop} using WENDy for simulated data with Gaussian additive error. We vary the additive error ratio in the simulated data between $e=[0\%,200\%]$ such that the standard deviation of the additive error is $\sigma=e\text{RMS}(\Omega(t))$, and at each additive error ratio, we generate 1,000 noisy datasets. The $(e,q)$-identifiability for $e=[0\%,200\%]$ and $q=[1\%,100\%]$ of model \eqref{eq:wf_sir_conpop} is given in Figure \ref{fig:sir_pract_ident}A. We find that the weak form SIR model is $(e,q)$-identifiable for all estimate error ratios above the $q=10\%$ level. For measurement error levels from $[0\%,120\%]$ the MSE always remains below $5\%\times \beta$. Therefore, we conclude the $\beta$ is generally practically identifiable for noise up to $120\%\times\text{RMS}(\Omega(t))$. 

\begin{figure}[ht]
    \centering
    \includegraphics[width=0.5\linewidth]{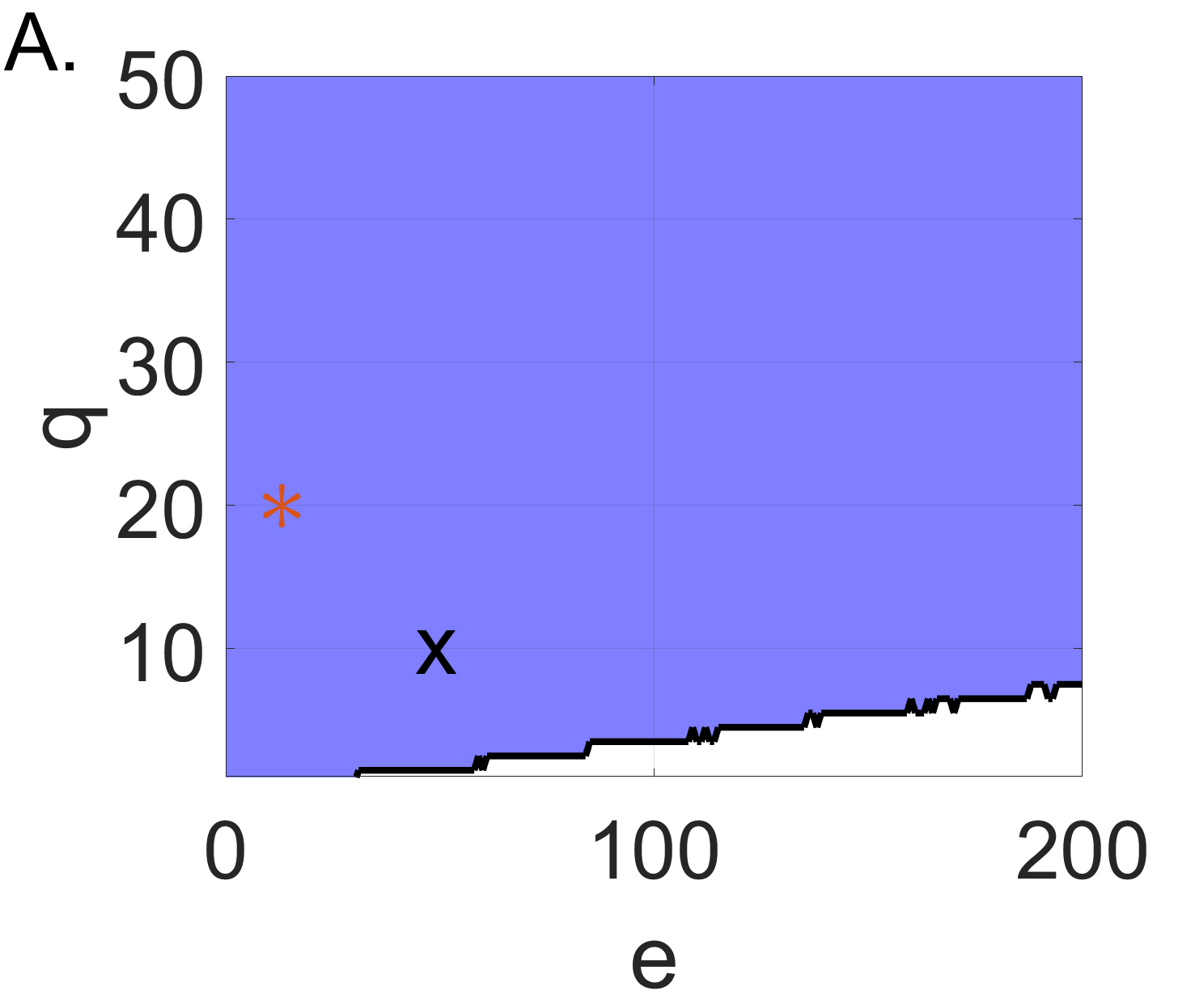}\includegraphics[width=0.5\linewidth]{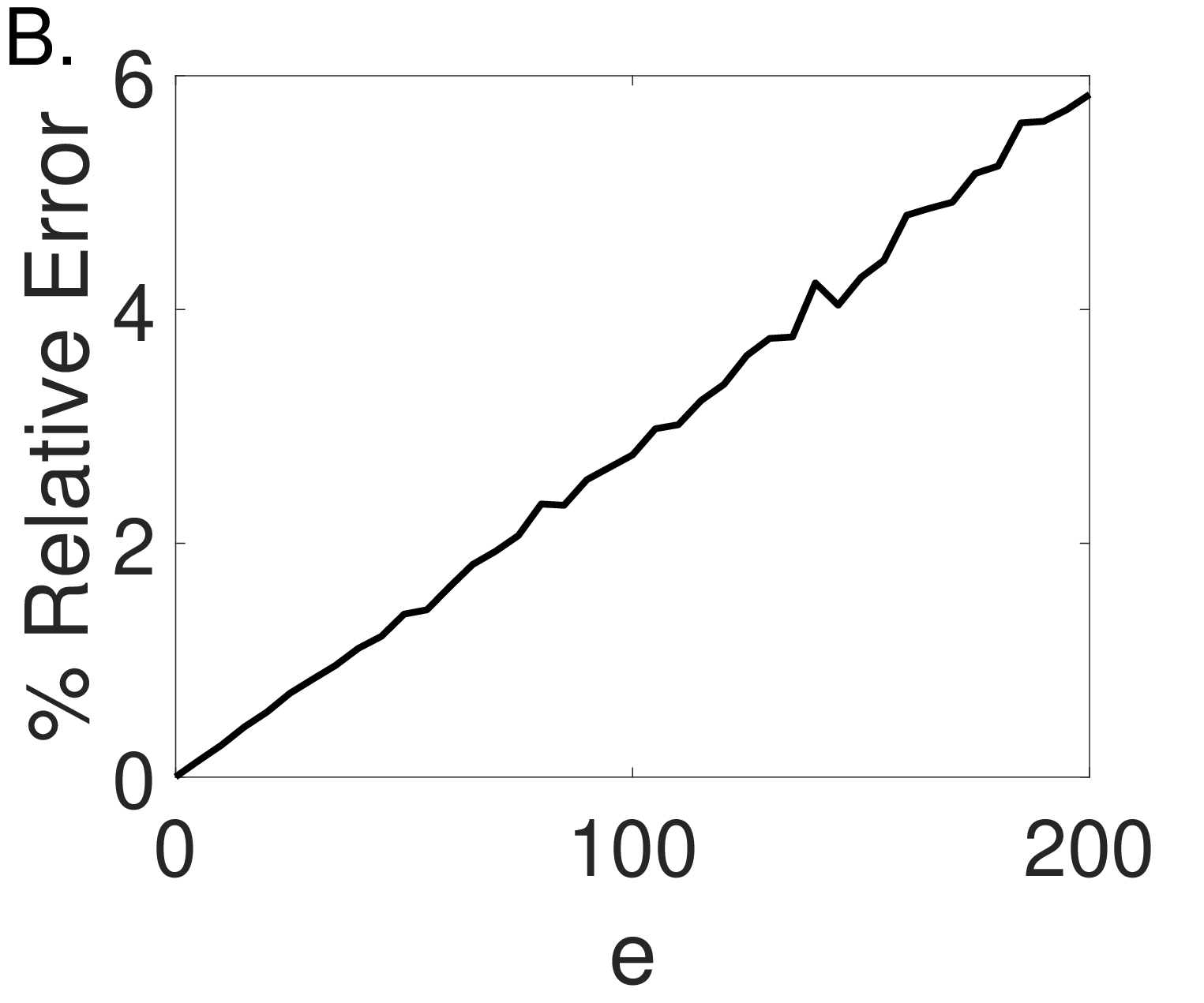}\\
\includegraphics[width=0.5\linewidth]{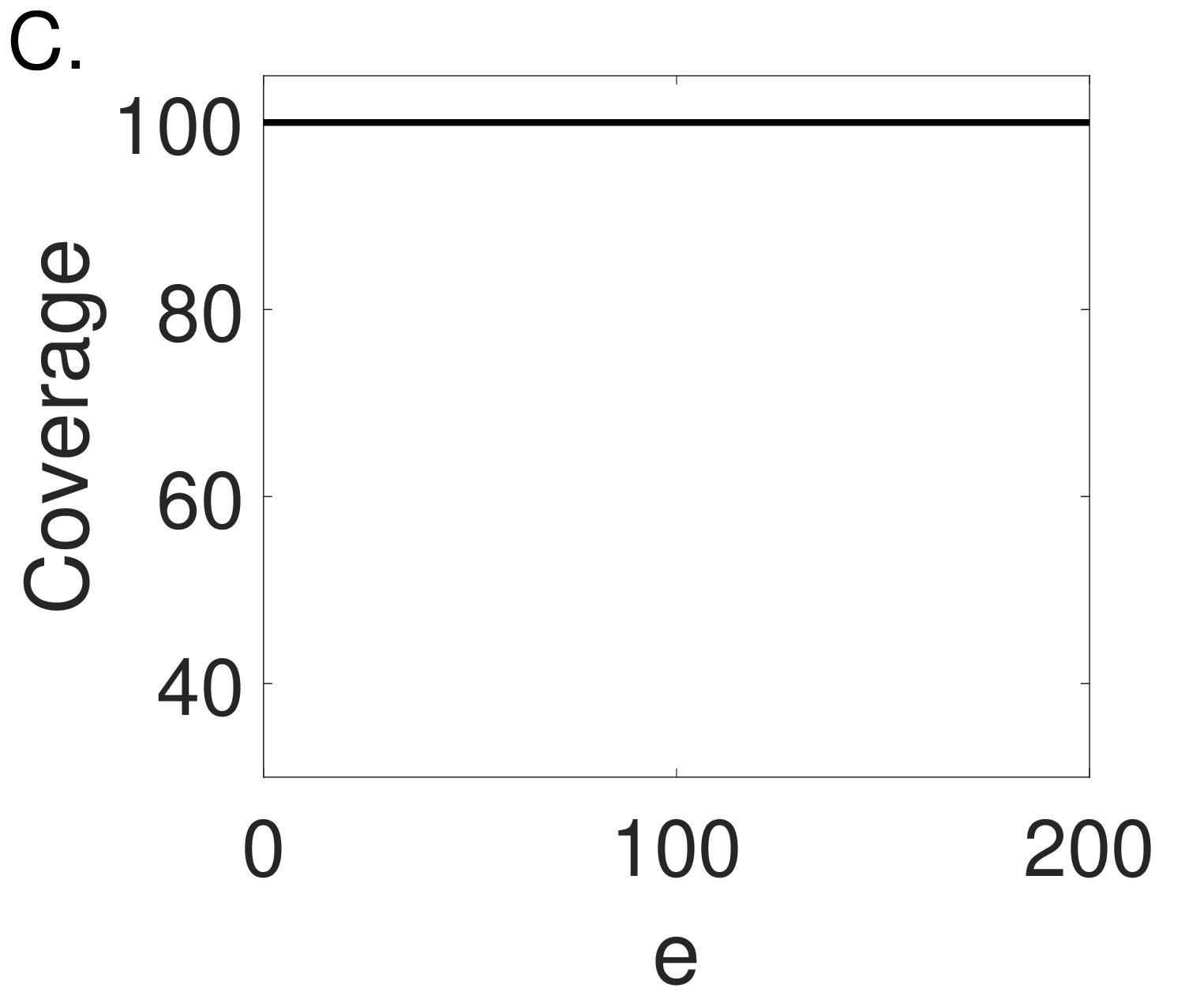}
    \caption{Using WENDy with equations \eqref{eq:wf_sir_conpop} we find the SIR model to be generally practically identifiable for additive noise between $[0\%,200\%]$. (A.) The area in blue denotes where the model is $(e,q)$-identifiable, and the area in white denotes where the model is not $(e,q)$-identifiable. The $(e,q)$-identifiability was determined from 1,000 simulations at each respective error level, e. The black $\pmb{\mathsf{X}}$ denotes the $(50,10)-$identifiability cutoff. Because this point falls within the blue region, we can say that at a $50\%$ additive error ratio in the data, the MSE of $\widehat{\beta}$ remains below the square of $10\%$ of the magnitude of $\beta$. \textcolor{black}{The red star \textcolor{red}{\textasteriskcentered } marks the $(10,20)$-identifiability criterion.} (B.) The relative error determined from 1,000 simulations at each respective error level, e. (C.) The proportion of estimated $95\%$ confidence intervals at $e=[0\%,200\%]\%$ that contain the true value of $\beta$.}
    \label{fig:sir_pract_ident}
\end{figure}

We additionally compare the $(e,q)$-identifiability of model equations \eqref{eq:wf_sir_conpop} to the average relative error and the coverage of the $95\%$ confidence intervals generated at each additive error ratio. The relative error generated for 1,000 simulated datasets at each additive error level is given in Figure \ref{fig:sir_pract_ident}B, and the coverage of the $95\%$ confidence intervals for $\beta$ is given in Figure \ref{fig:sir_pract_ident}C. Here, the $(e,q)$-identifiability and the relative error criteria are in agreement. The relative error remains below $6\%$ even at a high additive error ratio, $e=200\%$.  Unlike the blood concentration model, the coverage remains $100\%$ across all additive error ratios. However, it is important to note that while the parameter estimates remain very accurate at high noise, the uncertainty in the estimate still increases with increasing additive noise ($e$), similar to the blood concentration model.

\textcolor{black}{Additionally, we consider an example of $(e,q)$-identifiability where noise is multiplicatively lognormally distributed, i.e., $y(t)=\Omega(U(t),w)\epsilon$ and $\epsilon\sim\text{logNorm}(0,e).$ We estimate the parameter $\beta$ using WENDy for simulated data with log-normal \textcolor{black}{multiplicative observation error}. We vary the \textcolor{black}{multiplicative observation error} ratio in the simulated data between $e=0\%$ and $e=20\%$ such that the standard deviation of the error is $\sigma=e\log(\text{RMS}(\Omega(t)))$, and at each \textcolor{black}{multiplicative observation error} ratio, we generate 1,000 noisy datasets. The $(e,q)$-identifiability for $e=[0\%,20\%]$ and $q=[1\%,100\%]$ of model \eqref{eq:wf_sir_conpop} is given in Figure \ref{fig:sir_pract_ident_lognorm}A. Unlike in the case of additive noise, we find that the weak form SIR model is $(e,q)$-identifiable for all estimate error ratios above the $q=40\%$ level, but not the $q=10\%$ level. At error ratios above $e=16\%$, the SIR model is no longer (e,10)-identifiable. Multiplicative noise causes a greater impact near the peak in the SIR model, which we suspect accounts for the difference in parameter recovery between the additive and multiplicative noise cases. Note that the model is still $(10,20)-$identifiable. Therefore, we consider a simple SIR model with daily observations to be generally practically identifiable for both additive Gaussian and multiplicative lognormal noise.} 

\begin{figure}[ht]
    \centering
    \includegraphics[width=0.5\linewidth]{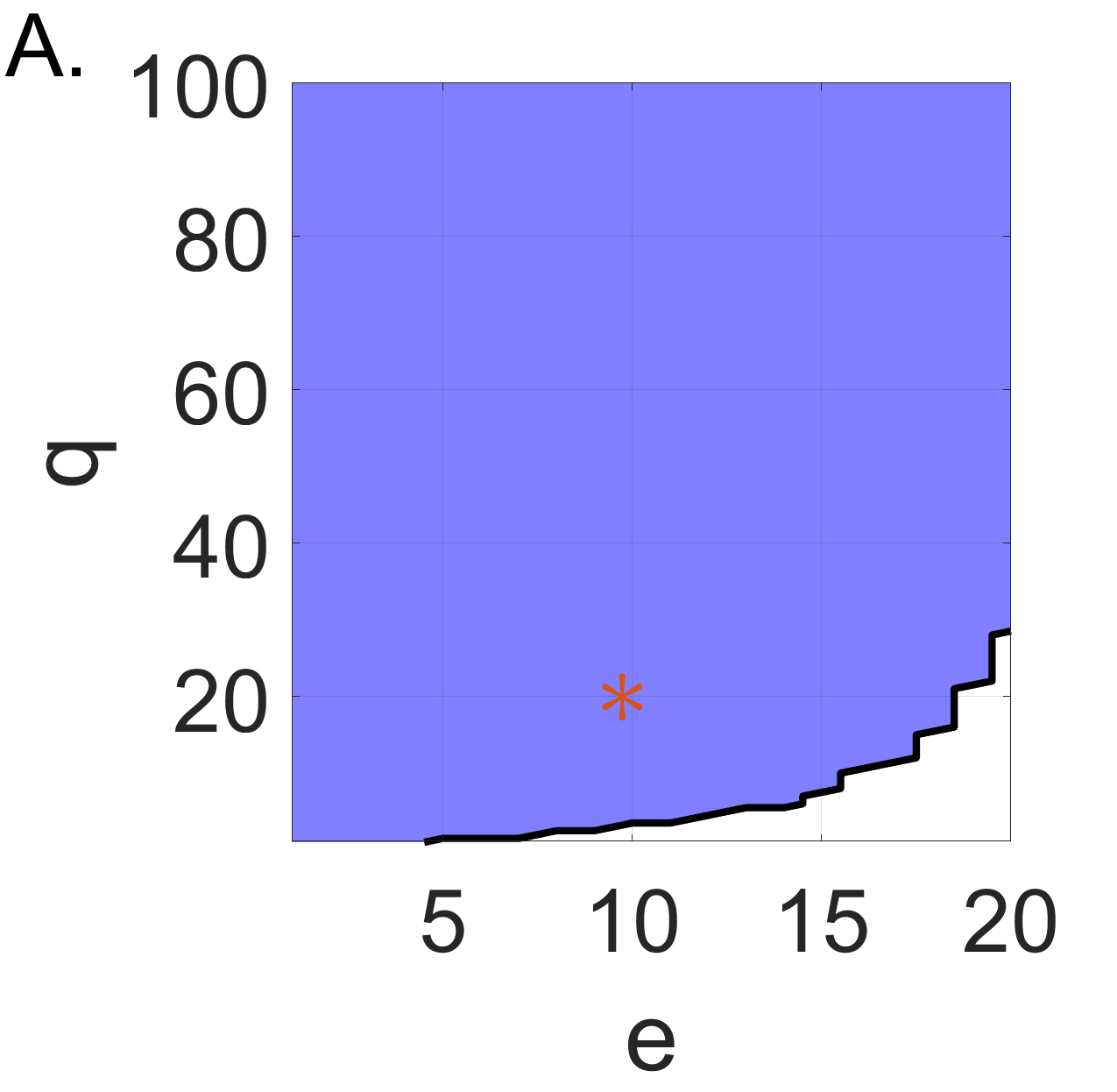}\includegraphics[width=0.5\linewidth]{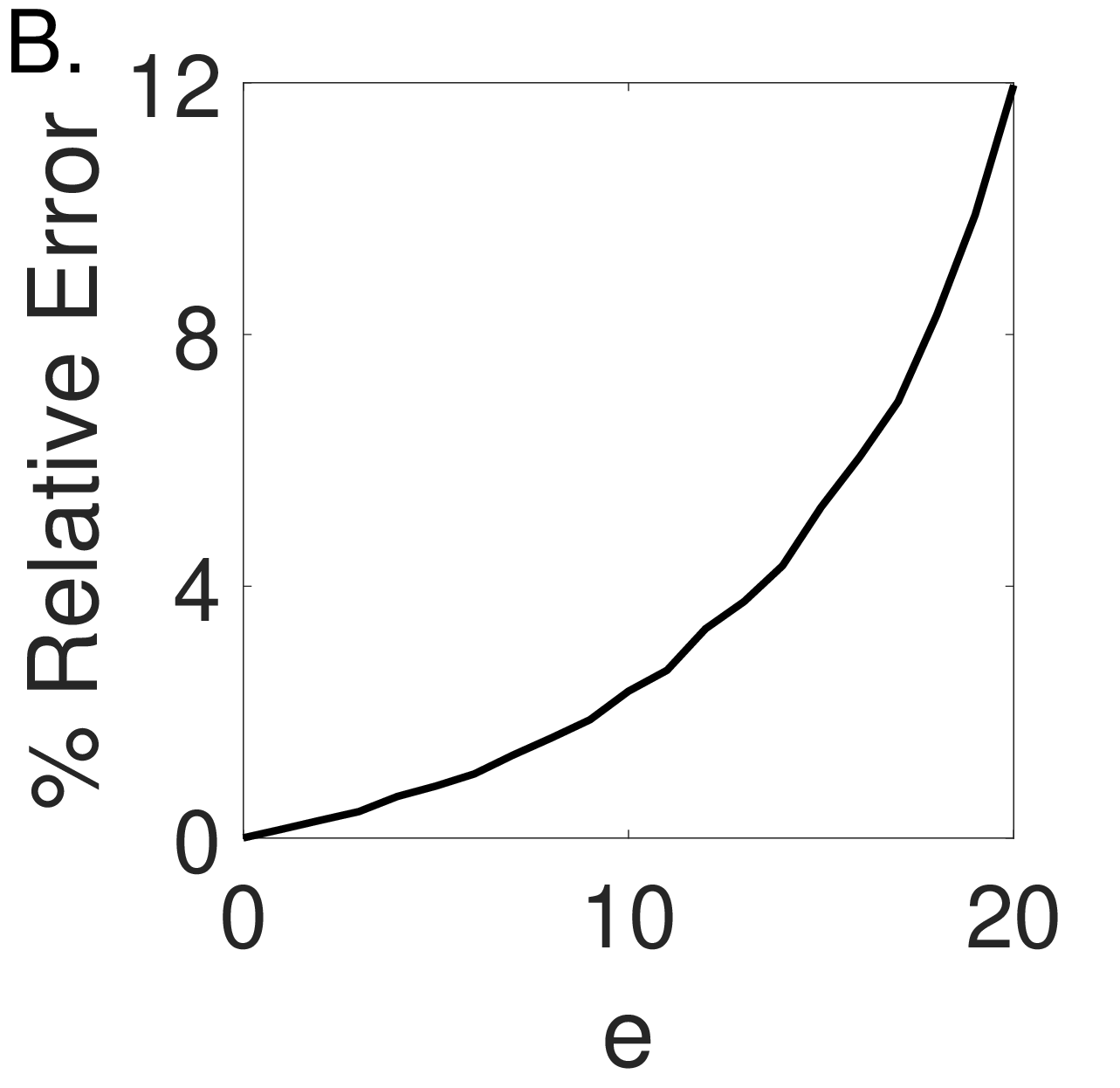}\\
\includegraphics[width=0.5\linewidth]{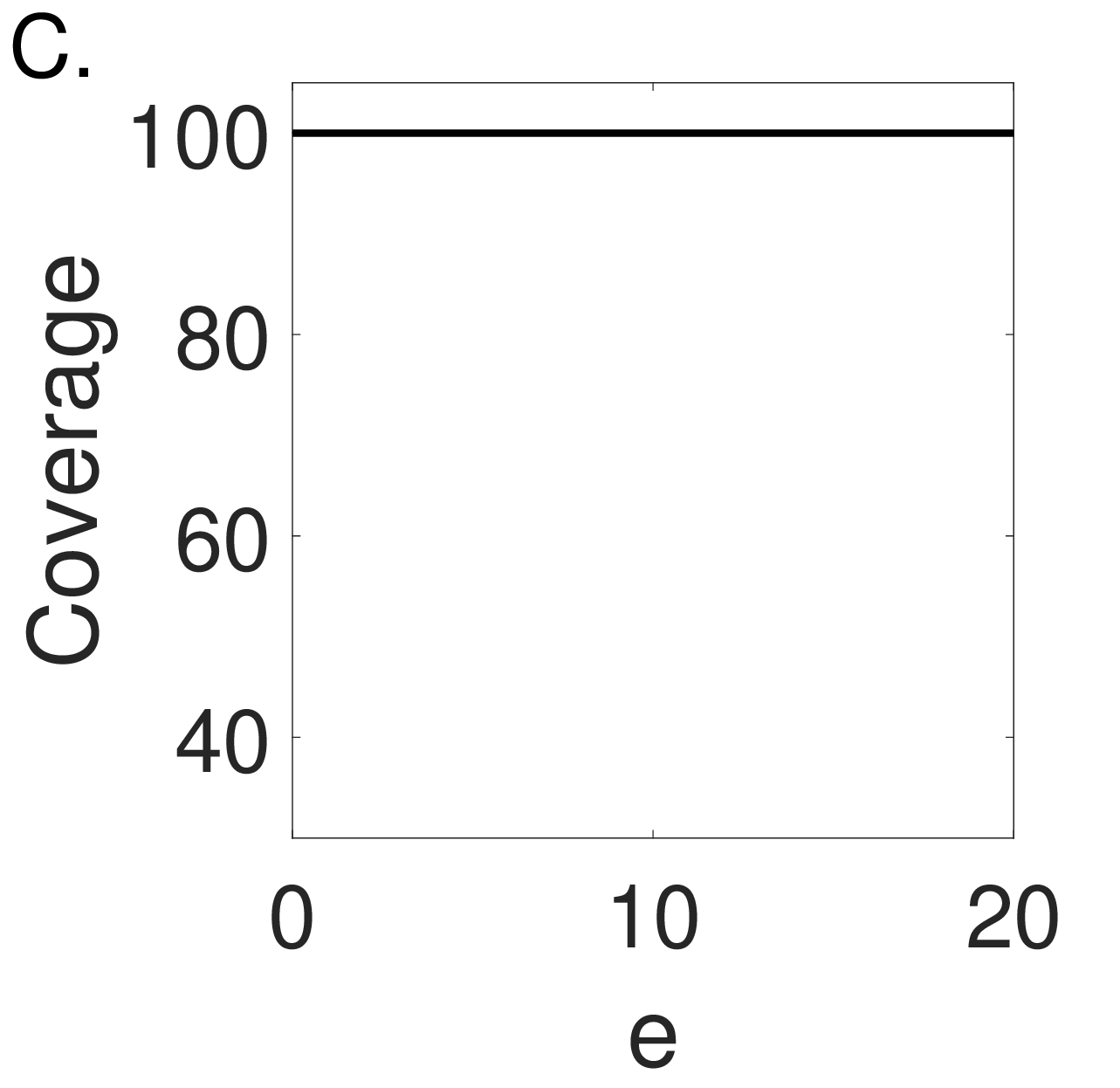}
    \caption{\textcolor{black}{Using WENDy with equations \eqref{eq:wf_sir_conpop} we find the SIR model to be generally practically identifiable for multiplicative noise in $[0\%,20\%]$. (A.) The area in blue denotes where the model is $(e,q)$-identifiable, and the area in white denotes where the model is not $(e,q)$-identifiable. The $(e,q)$-identifiability was determined from 1,000 simulations at each respective error level, e. \textcolor{black}{The red star \textcolor{red}{\textasteriskcentered } marks the $(10,20)$-identifiability criterion.} (B.) The relative error determined from 1,000 simulations at each respective error level, e. (C.) The proportion of estimated $95\%$ confidence intervals across $e=[0\%,20\%]$ that contain the true value of $\beta$.}}
\label{fig:sir_pract_ident_lognorm}
\end{figure}

\textcolor{black}{We additionally compare the $(e,q)$-identifiability of model equations \eqref{eq:wf_sir_conpop} to the average relative error and the coverage of the $95\%$ confidence intervals generated at each multiplicative error ratio. The relative error generated for 1,000 simulated datasets at each \textcolor{black}{multiplicative observation error} level is given in Figure \ref{fig:sir_pract_ident_lognorm}B, and the coverage of the $95\%$ confidence intervals for $\beta$ is given in Figure \ref{fig:sir_pract_ident_lognorm}C. Here, the $(e,q)$-identifiability and the relative error criteria are in agreement. The relative error remains below $12\%$ for \textcolor{black}{multiplicative observation error} ratios $e=[0\%,20\%]$.  Similarly to the case with Gaussian additive noise, the coverage remains $100\%$ across all additive error ratios. Again, it is important to note that while the parameter estimates remain accurate at high noise, the uncertainty in the estimate still increases with increasing additive noise ($e$).}
\end{exmp}

\begin{figure}[ht]
    \centering
    \includegraphics[width=\linewidth]{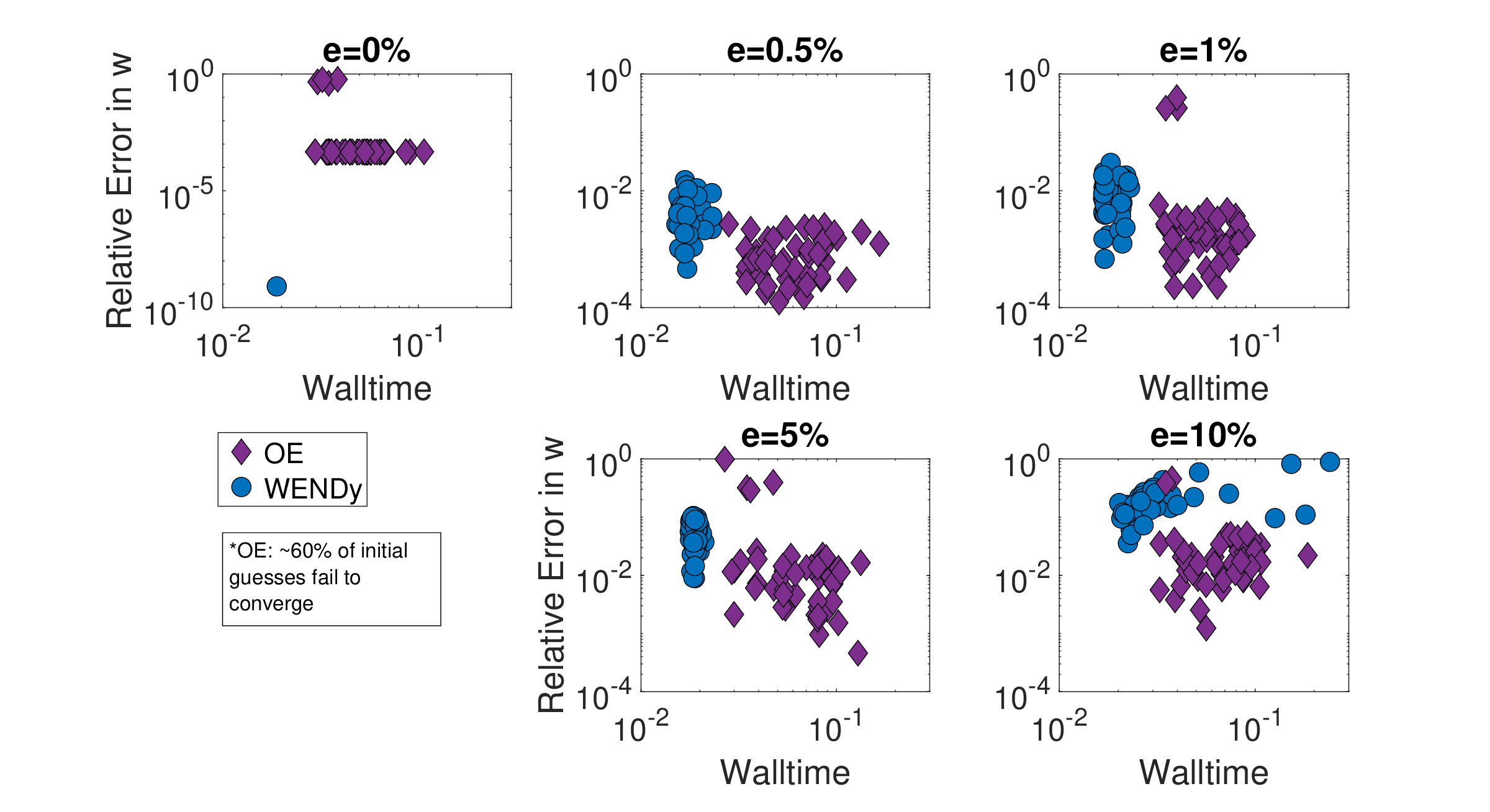}
    \caption{Using WENDy with equations \eqref{eq:weak_bld_cnc} we are able to recover the parameters $w=[w_1, w_2, w_3]$ for the blood-tissue diffusion model from noisy data both more quickly and more consistently than using an output error method. The relative error in $w$ versus walltime is given for 50 fittings of equations \eqref{eq:weak_bld_cnc} to $x_1$ compartment data of varying additive error ratios found using WENDy (blue dots) or 50 fittings of model equations \eqref{eq:bld_cnc} using output error (purple diamonds) for additive error ratio \textcolor{black}{in}  $e=[0\%,10\%]$. Note that for $\sim60\%$ of initial guesses, the OE method fails to converge, and that only successful optimizations for the OE method have been presented above. \textcolor{black}{In contrast, WENDY converges in all datasets.}}
\label{fig:bld_cnc_time}
\end{figure}

\medskip
\noindent
\textbf{Comparison to output error.} As demonstrated above, often the parameter estimation must be performed many times in order to gain an understanding of the practical identifiability of the system, meaning fast estimation methods lead to a more convenient assessment of identifiability. Table \ref{tab:pract_times} gives the median time to \textcolor{black}{compute parameter estimates over 1000 simulated datasets} for a $5\%$ additive noise ratio using WENDy on equation \eqref{eq:weak_bld_cnc} and using an output error method on \eqref{eq:bld_cnc}. \textcolor{black}{The WENDy method is over three times faster than the OE method.} Additonally, in figure \ref{fig:bld_cnc_time}, the relative errors in parameter \textcolor{black}{$\widehat{w}=[\widehat{w_1},\widehat{w_2},\widehat{w_3}]$} for 50 simulated datasets of additive error ratios $e=0\%,0.5\%,1\%,5\%,$ and $10\%$ are plotted
against the walltime. At all error levels, WENDy is \textcolor{black}{approximately five times } faster than the output error method. The median walltimes for the WENDy method fall between 1.27-2.53$\times 10^{-2}$ seconds, while the median output error walltimes fall between 5.4-7.3$\times 10^{-2}$ seconds. The median relative error is lower for the output error, between 0.064-1.7$\times 10^{-2}$ for $e=0.5\%,1\%,\%5,10\%$ versus between 0.36-2.7$\times 10^{-2}$ using WENDy. However, the output error is more likely to fail to converge and has a failure rate between 57.3$\%$ and $63.2\%$, while the WENDy method converges for all simulated datasets. 

\begin{table}[ht]
    \centering
    \begin{tabular}{c|c|c}
    Model&WENDy&OE\\
    \hline
    Blood-tissue diffusion     & $19$ sec & $70$ sec \\
    SIR     & $0.7$ sec & $140$ sec   
    \end{tabular}
    \caption{Median time to generate \textcolor{black}{parameter estimates for $D=1,000$ simulated datasets} using the WENDy method vs the OE methods at a $5\%$ additive noise ratio for the Blood-tissue diffusion and SIR models.}
\label{tab:pract_times}
\end{table}

\begin{figure}[ht]
    \centering
    \includegraphics[width=\linewidth]{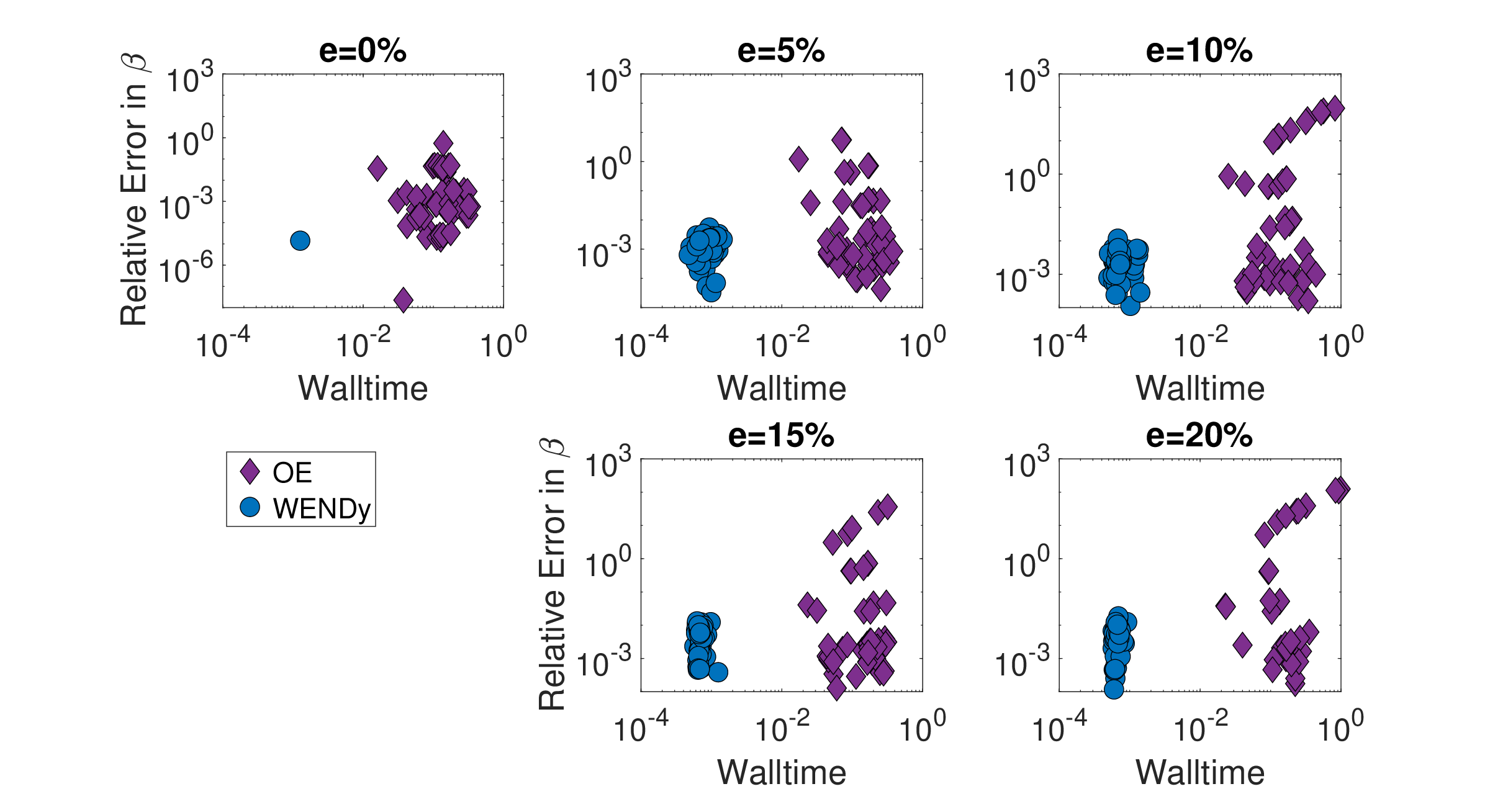}
    \caption{Using WENDy with equations \eqref{eq:wf_sir_conpop}, we are able to recover the transmission rate $\beta$ for the SIR model from noisy data, both more quickly than using an output error method. The relative error in $\beta$ versus walltime is given for 50 fittings of equations \eqref{eq:wf_sir_conpop} to infected compartment data of varying additive error ratio found using WENDy (blue dots) or 50 fittings of model equations \eqref{eq:SIR} using output error (purple diamonds) for additive error ratio \textcolor{black}{in} $e=[0\%,20\%$].}
    \label{fig:sir_time}
\end{figure}

In Figure \ref{fig:sir_time}, the relative errors \textcolor{black}{for the estimator of }  parameter $\beta$ for 50 simulated datasets of additive error ratios $e=0\%,5\%,10\%,15\%,$ and $20\%$ are plotted
against the walltime using WENDy on equations \eqref{eq:wf_sir_conpop} and using an output error method on \eqref{eq:SIR}. At all error levels, WENDy is multiple orders of magnitude faster than the output error method. The median walltimes for the WENDy method fall between 6.4-9.1$\times 10^{-4}$ seconds, while the median output error walltimes fall between 1.3-1.6$\times 10^{-1}$ seconds. Table \ref{tab:pract_times} shows that the WENDy method can be used in under a second to test practical identifiability on a single error ratio level. The median relative error is similar, within $2\times10^{-3}$, between the WENDy method and the OE method at error levels $e=5\%,10\%,15\%$. However, at the $e=20\%$ additive error ratio, the relative error using WENDy is lower, $4.6\times10^{-3},$ compared to the relative error using the output error, $1.6\times10^{-2}.$ Ultimately, the benefit of smaller walltimes for estimation of parameters in both the blood-tissue diffusion and SIR models allows for the calculations on thousands of simulated datasets, like those used to generate Figures \ref{fig:bld_cnc_pract_ident} and \ref{fig:sir_pract_ident}.

\section{Conclusion}

In this work, we present a weak-form framework for efficiently assessing \emph{a priori} model identifiability, summarized below:
\begin{enumerate}
    \item We apply differential elimination methods to generate an input-output equation. At this step, we can also assess the structural identifiability of the model, which is \textcolor{black}{a} prerequisite to practical identifiability.
    \item We cast the input-output equation into its weak form. This step involves identifying terms of the weak-form input-output equation where integration by parts can and cannot be applied. Terms where integration by parts is not possible may be managed by either a substitution of variables or an approximation of higher-order derivatives.
    \item We employ a \textcolor{black}{simulation-}based method of assessing practical identifiability. In this process, we generate simulated datasets for various levels of \textcolor{black}{observational} error ratios, $e$. We then apply WENDy to weak-form input-output equations for efficient parameter estimation. 
    \item We assess practical identifiability from these simulations using a newly defined criterion, $(e,q)-$identifiability.
\end{enumerate}

The $(e,q)$-identifiability criterion offers a method for \textcolor{black}{assessing} practical identifiability, which is based on the noise in observed variables $e$ and the mean-square error of the parameter estimator $q$. We demonstrate this $(e,q)$ criterion via computing the practical identifiability of these two examples in terms of mean-squared error and in terms of average relative error and coverage. This criterion is better able to capture changes in the quality of the parameter estimator due to increased noise in the data compared to criteria based on average relative errors. Additionally, while in this study we apply the criterion to ODE systems with \textcolor{black}{additive Gaussian and multiplicative lognormal measurement error}, it \textcolor{black}{has the potential to} be extended to a broad range of problems, including discrete-time systems, PDE systems, and problems with non-Gaussian noise. Assessing identifiability in PDE systems is a challenging problem, which has been addressed recently by \cite{ByrneHarringtonOvchinnikovEtAl2025Nonlinearity,CiocanelDingMastromatteoEtAl2024BullMathBiol,LiuSuhMainiEtAl2024JRSocInterface}, but it still remains difficult (if not impossible) to test structural identifiability for arbitrary PDE systems. Therefore, methods that can efficiently assess the practical identifiability of these systems are particularly important.

Critically, the practical identifiability analyses in this study rely on the efficiency of weak-form parameter estimation. We present a method for weak-form parameter estimation of ODE systems with unobserved compartments, utilizing differential elimination to generate weak-form input-output equations and subsequently applying WENDy. This method is efficient, accurate, and robust to noise. Additionally, applied to a blood-tissue diffusion system (Model \eqref{eq:bld_cnc}), a comparison with the results in \cite{BoulierKorporalLemaireEtAl2014ComputerAlgebrainScientificComputing} yields that the WENDy is better than the Hartley and polynomial test functions and strongly competitive with the interpolated polynomials as well as the pure integral approach. \textcolor{black}{Currently, the optimal choice in test functions for weak-form parameter estimation is an open question. \cite{TranBortz2025arXiv250703206} proposes a method for constructing a set of test functions that minimizes the error introduced by quadrature.}

Regarding method performance, we note that in all the examples presented above (including at all error levels), WENDy is substantially faster than OE methods. In particular, WENDy is faster than the output error method for both the blood-tissue diffusion and SIR models, and is multiple orders of magnitude faster in the SIR case. Ultimately, the benefit of smaller walltimes for estimation of parameters in both the blood-tissue diffusion and SIR models allows for the calculations on thousands of simulated datasets, like those used to generate Figures \ref{fig:bld_cnc_pract_ident} and \ref{fig:sir_pract_ident}. Table \ref{tab:pract_times} demonstrates that for a single error ratio level, the WENDy method is able to generate \textcolor{black}{parameter estimates} faster than OE methods, with a particularly large advantage (seconds vs minutes \textcolor{black}{over 1,000 simulated datasets}) in the SIR model case. For the SIR model, the accuracy of the weak-form method is the same as the OE method for additive noise ratios between \textcolor{black}{$0\%$ and $15\%$}, and more accurate at high noise ratios of $e=20\%$. We further observe that the relative error increases linearly with the additive error ratio for this system. Linear error has also been observed for some systems in \cite{Chen2023Calcolo,MessengerBortz2021MultiscaleModelSimul,MessengerBortz2025IMAJNumerAnal} when performing \textcolor{black}{equation error} regression-based optimization. \textcolor{black}{While the median accuracy of the weak-form method is generally significantly higher than the accuracy of the output error method \cite{BortzMessengerDukic2023BullMathBiol}, as observed in the SIR example, we did find that it was slightly lower for the blood-tissue diffusion model. However, in the blood-tissue diffusion example, the weak-form method did not fail to converge, while the OE method (which relies on inspired initial parameter estimates) failed in over half of the simulated datasets. More specifically,} for Example \ref{ex:1}, $\sim60\%$ of the OE optimizations failed to converge, even with reasonable initial parameter guesses (see Figure \ref{fig:bld_cnc_time}). \textcolor{black}{While in this study, we compare the computational efficiency with respect to a single instance of parameter estimation, these comparisons are important and scalable to more statistically sophisticated modeling frameworks, which often require repeated applications of a parameter estimation method.}

\begin{figure}[ht]
    \centering
\includegraphics[width=0.45\linewidth]{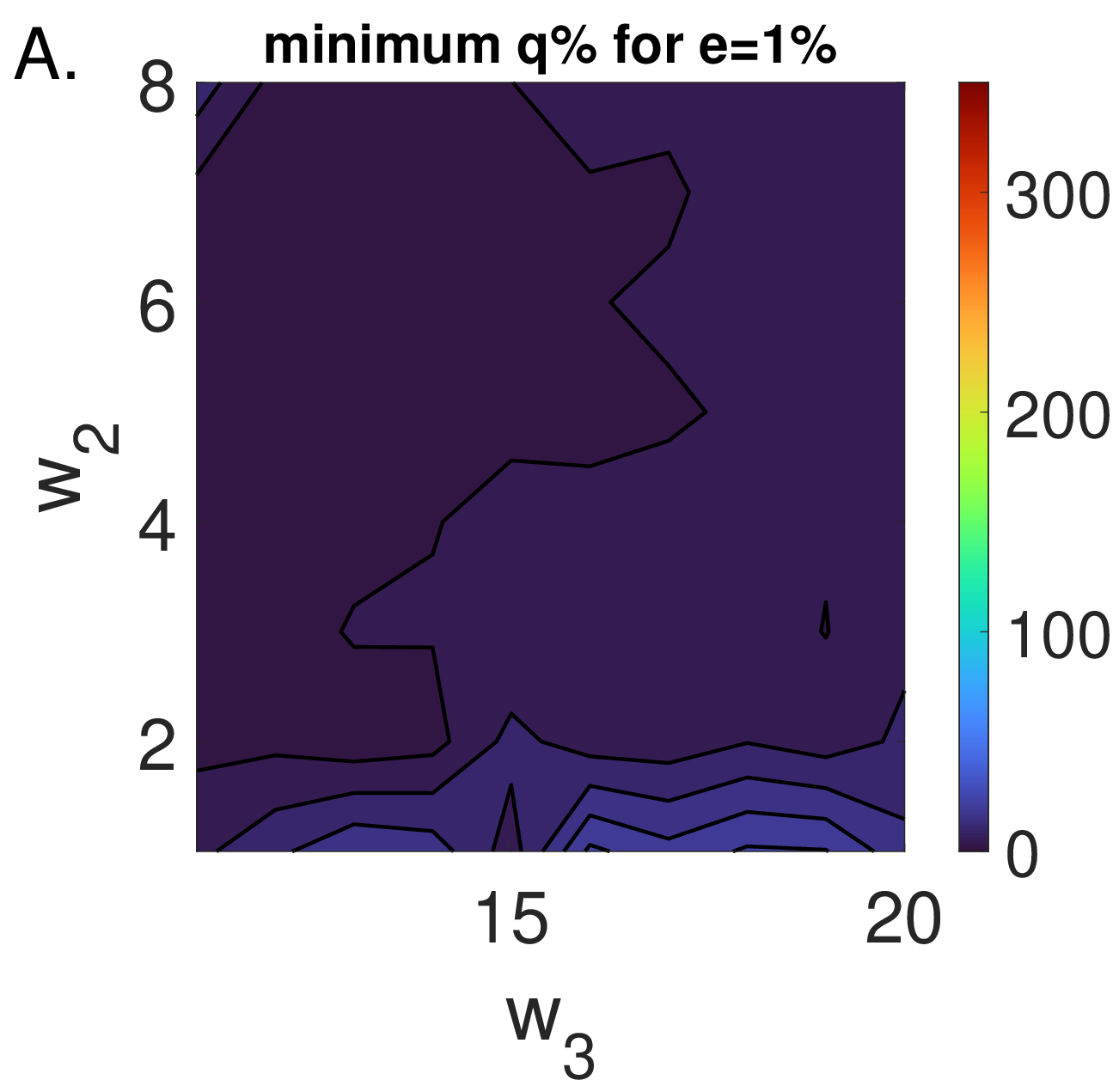}\includegraphics[width=0.45\linewidth]{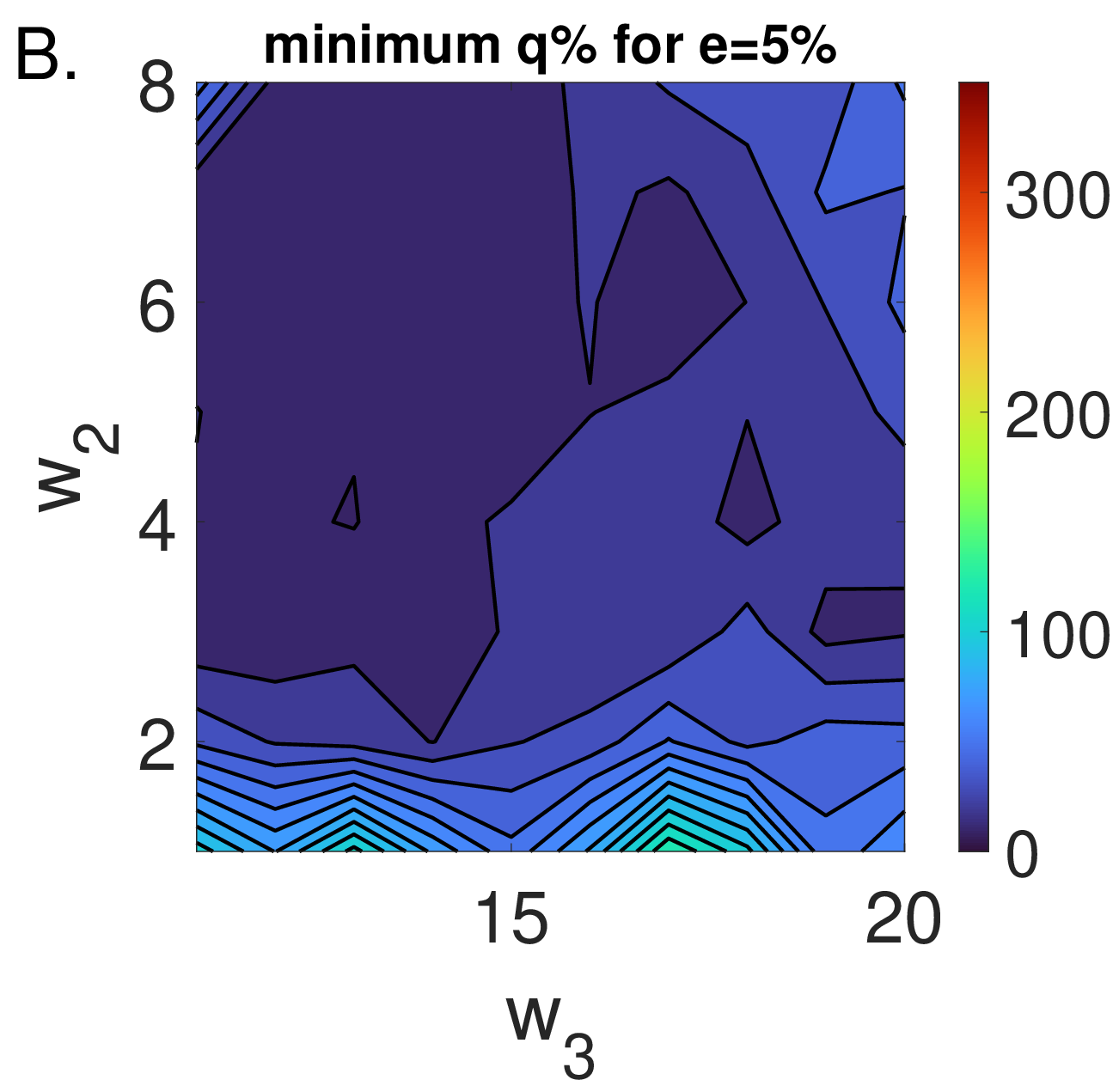}\\
\includegraphics[width=0.45\linewidth]{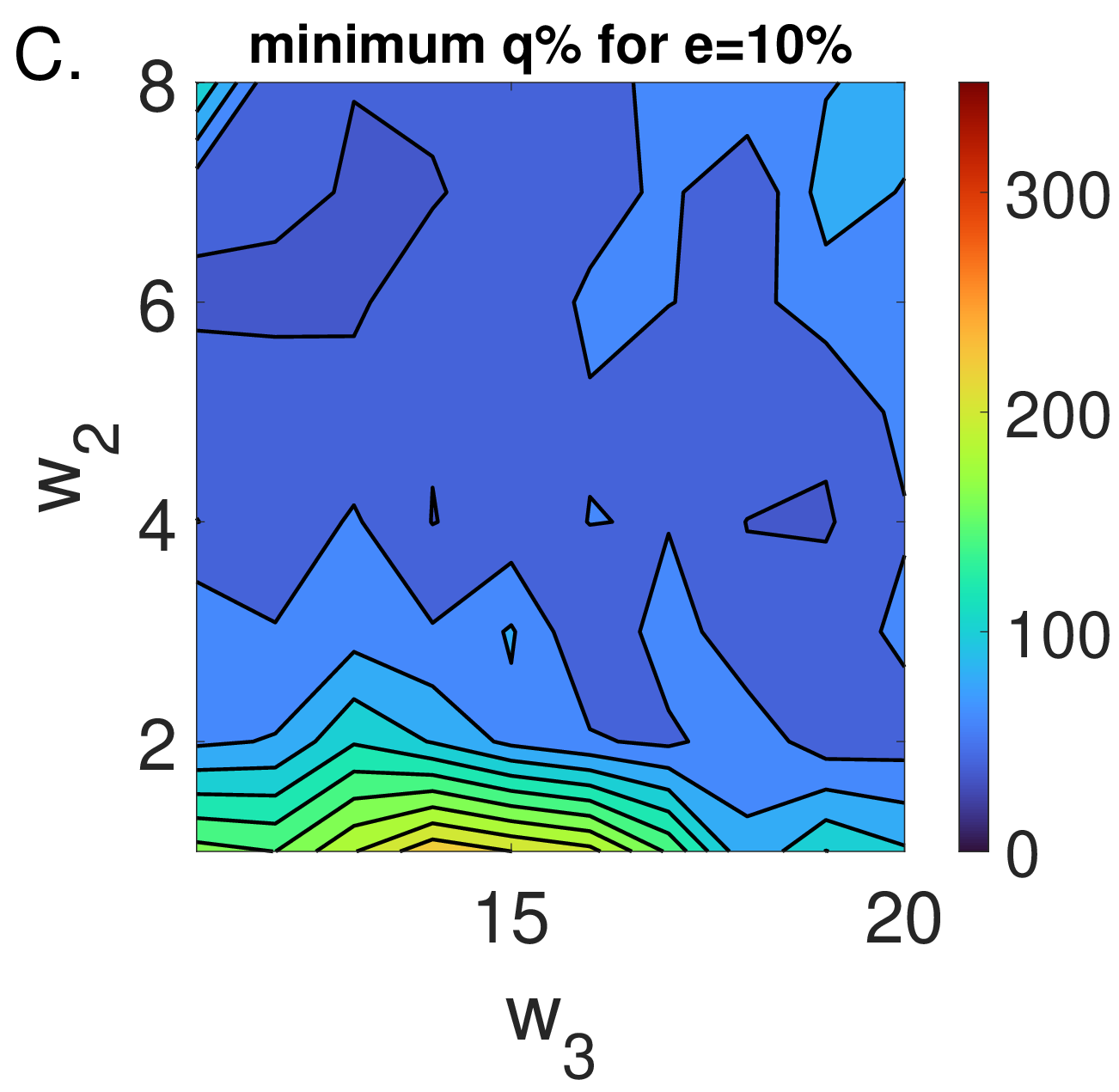}\includegraphics[width=0.45\linewidth]{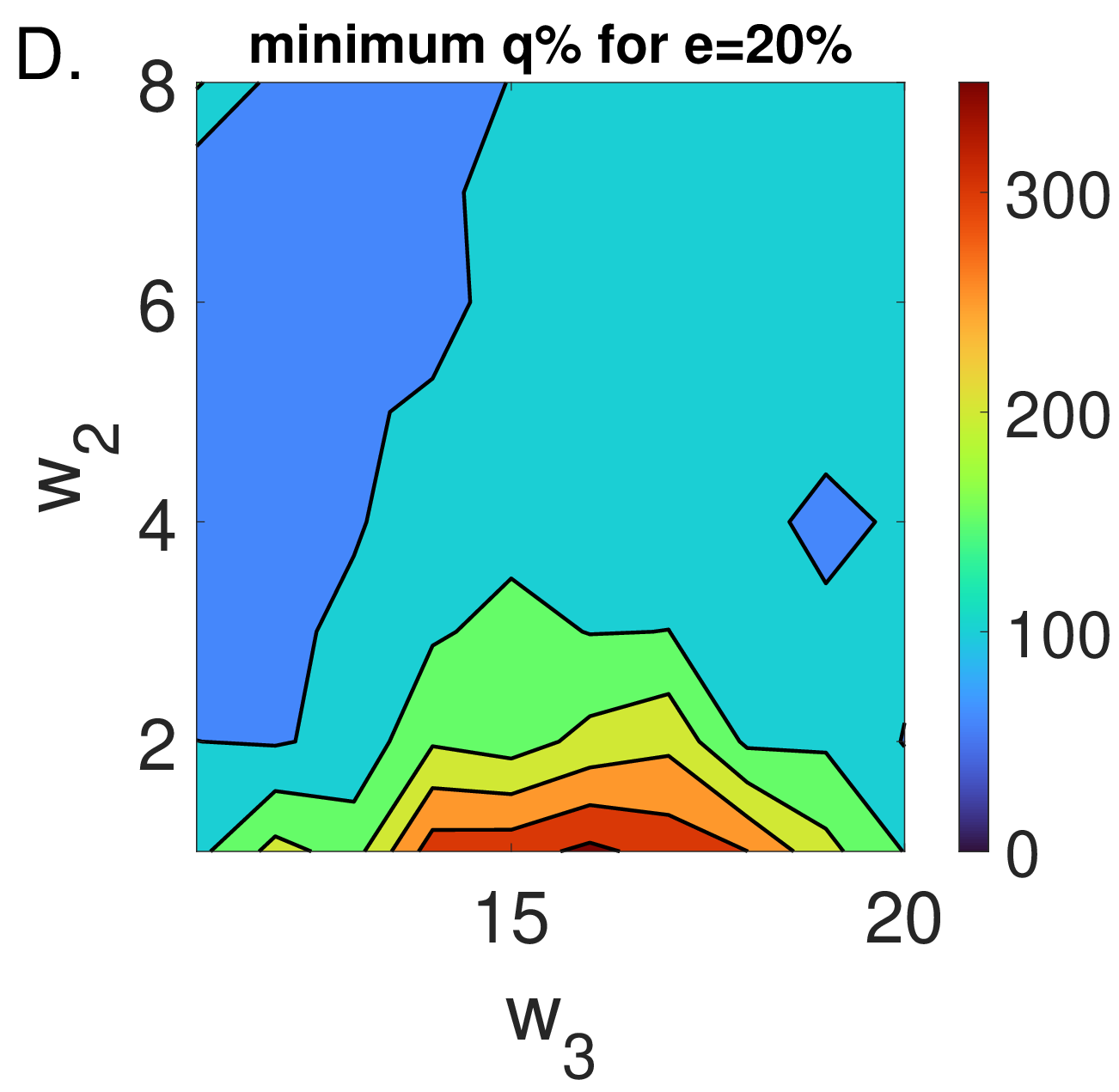}\\
    \caption{\textcolor{black}{The (e,q)-identifiability criterion holds locally in the parameter space. The minimum estimator error ratio, $q$, satisfied by the blood-tissue diffusion model for $w_1=30$ and varying parameters $w_2=[1,9]$ and $w_3=[10,20]$ using observational error ratios (A.) $e=1\%$, (B.) $e=5\%$, (C.) $e=10\%$, and (D.) $e=20\%$. The $(e,q)-$identifiability was determined from 1,000 simulations at each respective error level, $e$.}}
    \label{fig:eq_heatmap}
\end{figure}

\textcolor{black}{While global practical identifiability analysis is an emerging area of research \cite{BholaDuraisamy2023SciRep,DobreBastogneRichard2010IFACProceedingsVolumes,ElWajehJungBongartzEtAl2022BullMathBiol}  our $(e,q)$ criterion, like profile likelihood analysis and the FIM method, the $(e,q)$  method is local. This implies that, following best modeling practices, one may need to test a range of values in the parameter space rather than near a single parameterization of a system. For example, as we demonstrate in Appendix \ref{app:profile-likelihood},  depending on the dynamics of the blood concentration, the parameter $w_3$ of the blood diffusion model may or may not be practically identifiable at a single observational noise level. \textcolor{black}{However, we can also choose to assess $(e,q)$-identifiability on a range of parameters, rather than individual parameterizations of a specific model. In Figure \ref{fig:eq_heatmap}, we show, using the blood-tissue diffusion model, for a range of parameter choices in $w_2$ and $w_3$, that if one parameterization yields an  $(e,q)-$identifiable model, then a range of nearby parameterizations will also be $(e,q)-$identifiable. More generally, $(e,q)$-identifiability can be used in this way to explore a biologically informed parameter space. }
}

One limitation of this study \textcolor{black}{is the reliance on the generation of the weak-form input-output equation. The generation of the full input-output equation of systems of higher dimension can quickly become computationally difficult. Future work will address weak-form parameter practical identifiability for systems where input-output equations cannot be explicitly found. Additionally, information about the initial conditions of the system is lost in the weak-form transformation.} There are examples of ODE systems where initial conditions can play an important role in the identifiability of key model parameters \cite{ChowellDahalLiyanageEtAl2023JMathBiol,MiaoXiaPerelsonEtAl2011SIAMRev,VaseyMessengerBortzEtAl2025JComputPhys}. A possible direction to address this limitation is to work with an integral representation of the equation near the end points of the domain to recover information about the initial condition lost in the transformation to the weak form. Future exploration of the impact of initial conditions and methods to incorporate this lost information in our weak-form framework is needed. 

Lastly, in this study, we approach practical identifiability through simulated data, and our $(e,q)-$criterion provides an \emph{a priori} method of studying model identifiability and robustness to noise. \textcolor{black}{The $(e,q)$-identifiability method assesses uncertainty in the parameter estimates through repeated simulation by leveraging the fast computational time of the WENDy method.  Our method is comparable to the alternative methods such as the FIM and profile likelihood (see Table \ref{tab:pla_CI_compare} in Appendix \ref{app:profile-likelihood} for a comparison of confidence intervals generated from the WENDy method versus from the profile likelihood method with output error optimization), albeit often several orders of magnitude faster.}


\section{Acknowledgments}
Research reported in this publication was supported in part by the NIGMS Division of Biophysics, Biomedical Technology and Computational Biosciences grant R35GM149335, in part by the NSF Division Of Environmental Biology (EEID grant DEB-2109774), and in part by the NIFA Biological Sciences (grant 2019-67014-29919). \textcolor{black}{We thank Rainey Lyons (University of Colorado) for insightful discussions on test function formulation. We also thank the reviewers for feedback on the input-output definition of structural identifiability.}

\section{Data and Code Availability}
The code for this paper will be provided on Github at \url{https://github.com/MathBioCU/weak-form-practical-identifiability}.


\bibliography{WIdentifiability}             

\begin{appendices}
\section{Test Function Selection}\label{app:testfunc}

It is well known that the choice of test function has a significant impact on the performance of weak-form methods. There have been several efforts to wisely select test functions, including spectral matching \cite{MessengerBortz2021JComputPhys} and quadrature error minimization \cite{BortzMessengerDukic2023BullMathBiol}.  And, more recently, a method has been proposed to construct test functions for accurate parameter estimation \cite{TranBortz2025arXiv250703206}.  However, optimal selection of test functions remains an open problem.  In this work, we choose the test function optimally, based on knowing the true parameter. This ensures that the focus of this work can remain on the impact of constructing weak-form Input/Output equations.

\begin{figure}[H]
    \centering
    \includegraphics[width=0.5\linewidth]{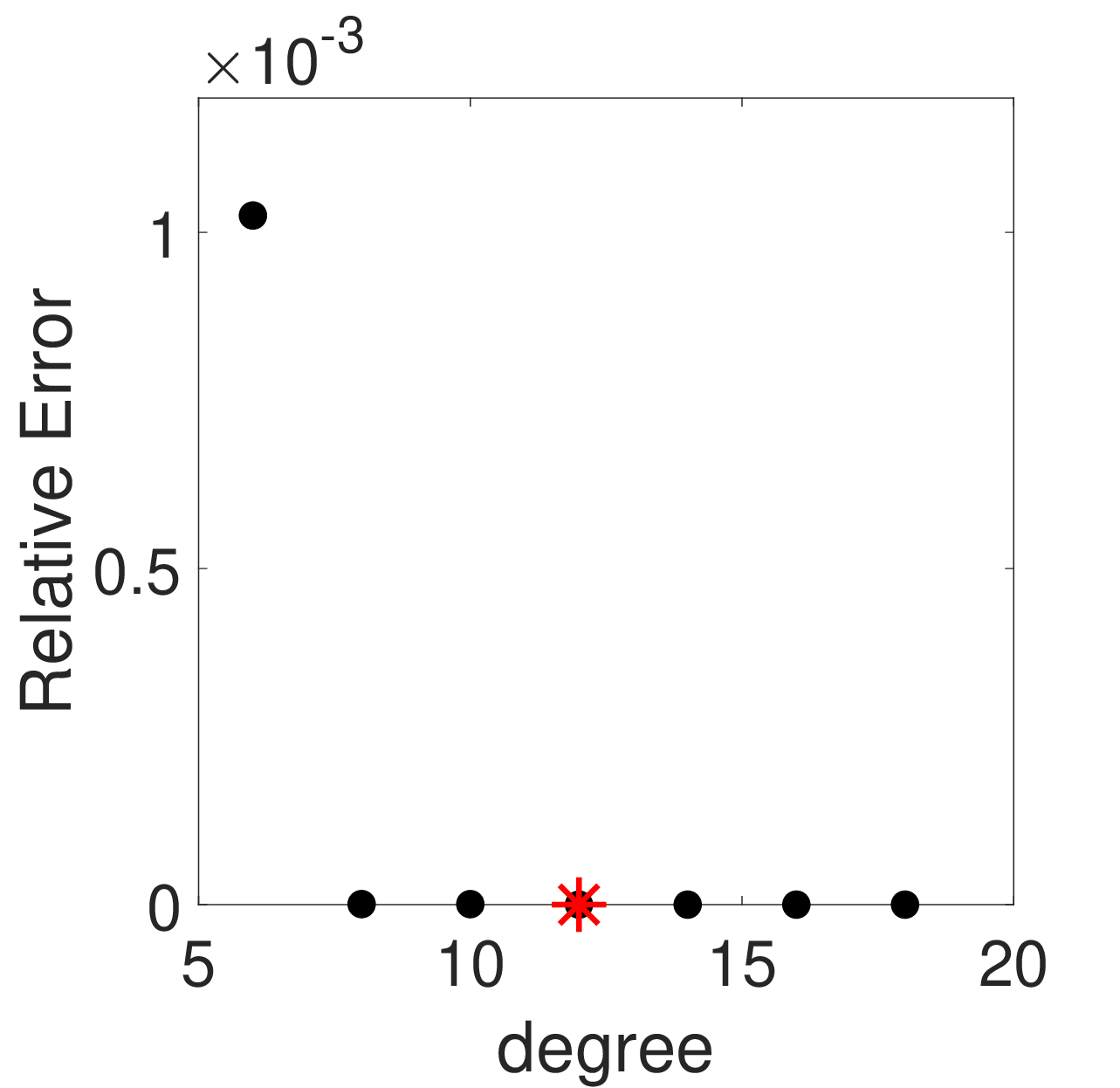}
    \caption{The relative error in parameters $w$ when fitting model \eqref{eq:weak_bld_cnc} to 400 observations using WENDy for polynomial test function choices of degree  $ = [6,18]$. The red star \textcolor{red}{\textasteriskcentered } indicates the  $12^\text{th}$ order polynomial used in this study.}
    \label{fig:appA_deg_bld_cnc}
\end{figure}

\begin{figure}[H]
    \centering
    \includegraphics[width=0.5\linewidth]{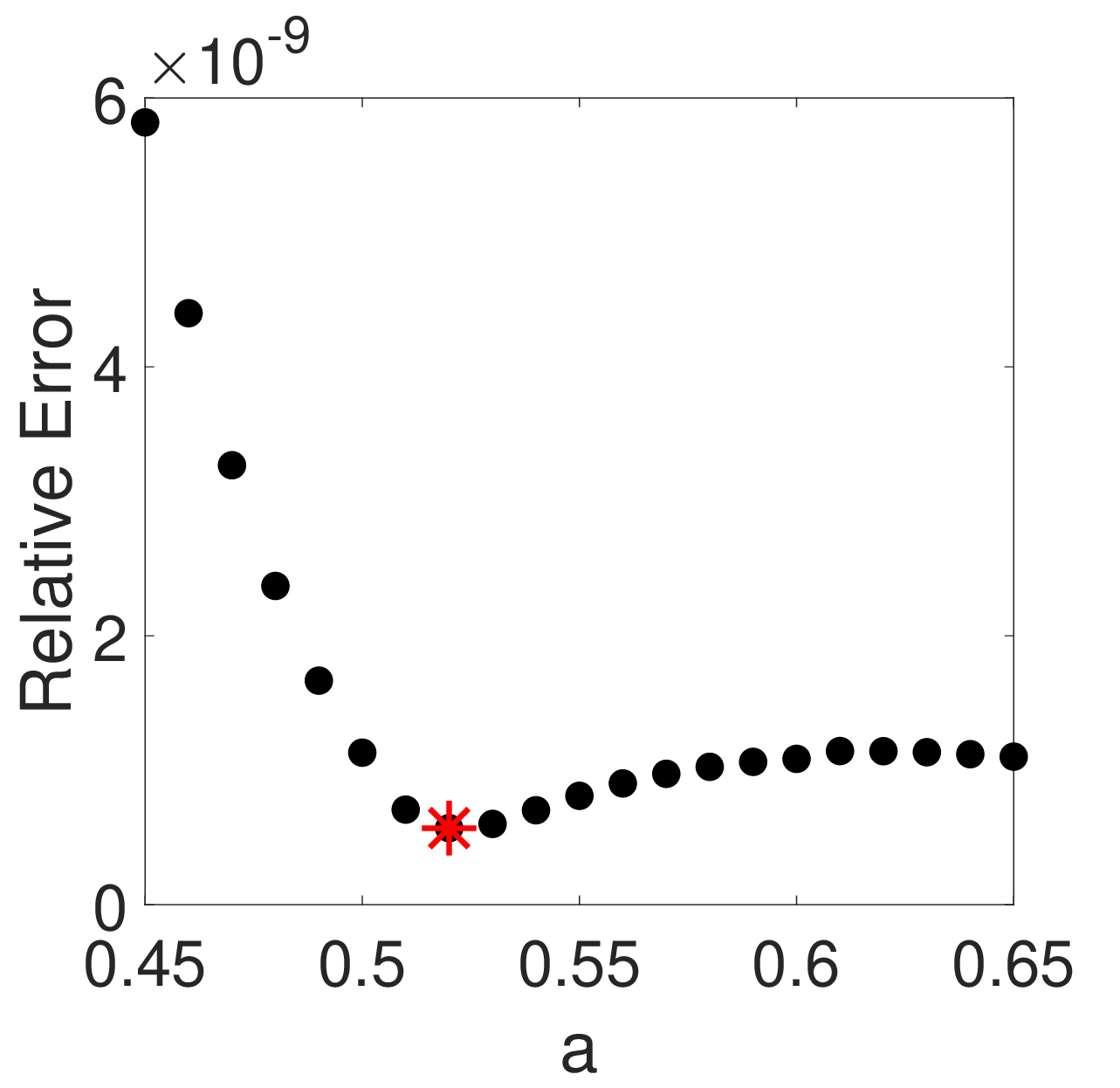}
    \caption{The relative error in parameters $w$ when fitting model \eqref{eq:weak_bld_cnc} to 400 observations using WENDy for polynomial test function radii of support $a=[0.45,0.65]$. The red star \textcolor{red}{\textasteriskcentered } indicates the choice of $a=0.52$ used in this study.}
    \label{fig:appA_rad_bld_cnc}
\end{figure}

We select the degree of the polynomial test function and the radius of compact support for each example presented in this paper by observing the relative error in parameter estimates. For model \eqref{eq:weak_bld_cnc}, we select our test functions to be $12^\text{th}$ order polynomials with a radius of support $a=0.52$. Figure \ref{fig:appA_deg_bld_cnc} shows the relative error in $w$ for polynomial test function choices of order 6 through 18. We observed a significant improvement in accuracy when moving to a $12^\text{th}$ order polynomial, but did not gain accuracy by increasing the order further. Figure \ref{fig:appA_rad_bld_cnc} shows the relative error in $w$ for compact support for radii from $a=[0.45,0.65]$.

\begin{figure}[H]
    \centering
    \includegraphics[width=0.5\linewidth]{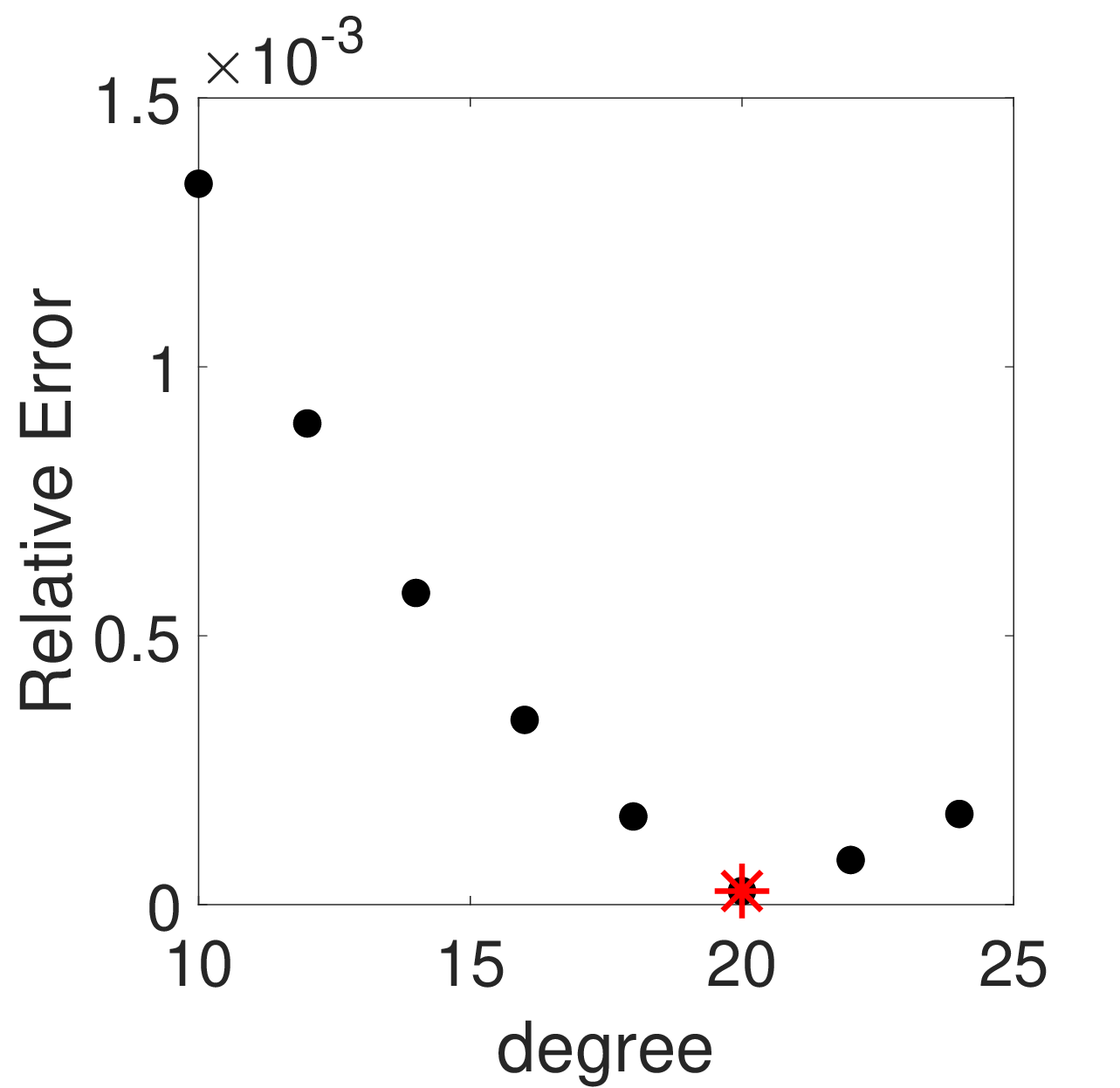}
    \caption{The relative error in parameters $\beta$ when fitting model \eqref{eq:wf_sir_conpop} to 31 observations using WENDy for polynomial test function choices of degree  $ = [10,26]$. The red star \textcolor{red}{\textasteriskcentered } indicates the  $20^\text{th}$ order polynomial used in this study.}
    \label{fig:appA_deg_sir}
\end{figure}

\begin{figure}[H]
    \centering
    \includegraphics[width=0.5\linewidth]{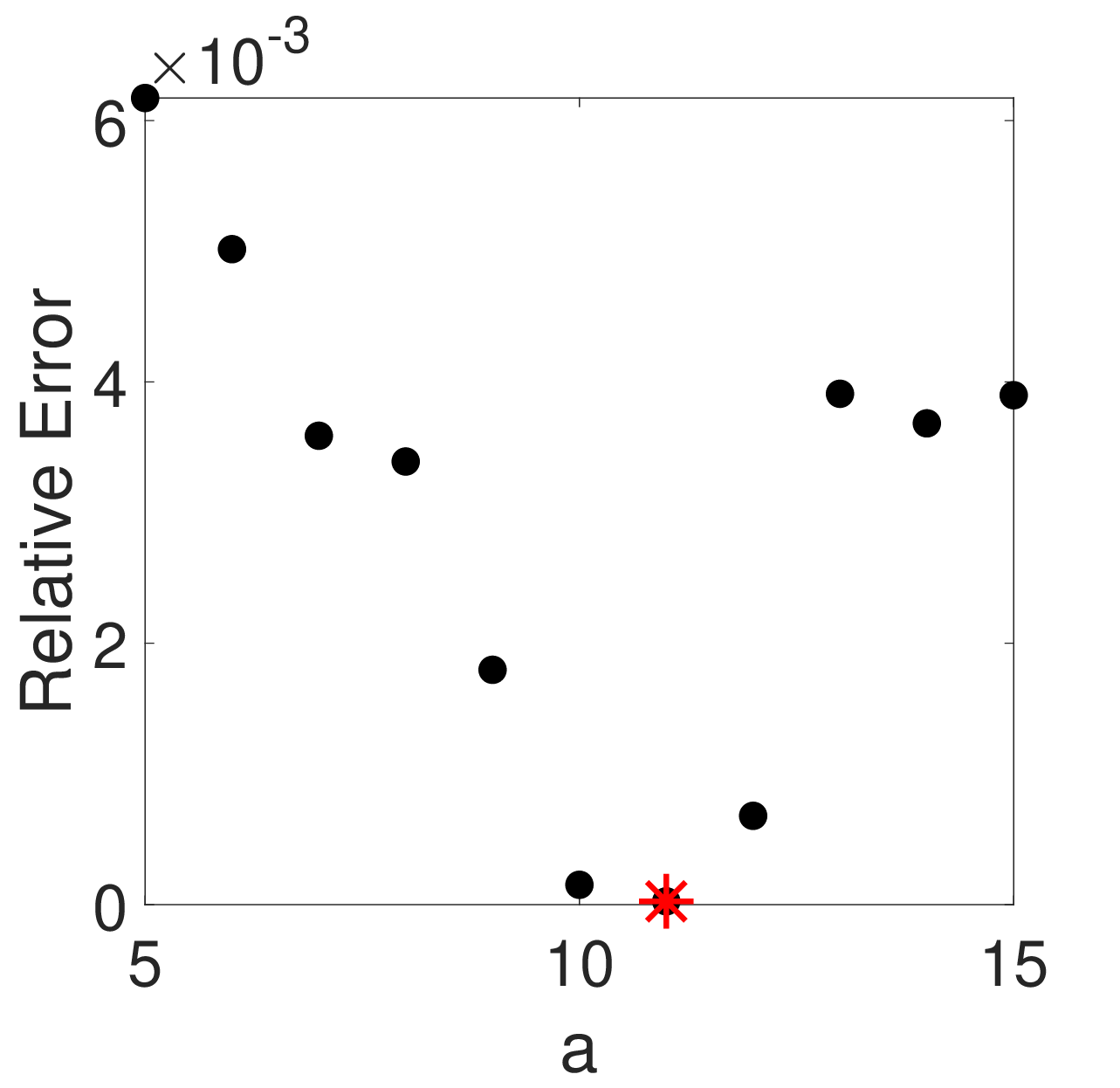}
    \caption{The relative error in parameters $\beta$ when fitting model \eqref{eq:weak_bld_cnc} to 31 observations using WENDy for polynomial test function radii of support $a=[5,15]$. The red star \textcolor{red}{\textasteriskcentered } indicates the choice of $a=11$ used in this study.}
    \label{fig:appA_rad_sir}
\end{figure}

For model \eqref{eq:wf_sir_conpop}, we select our test functions to be $20^\text{th}$ order polynomials with a radius of support $a=11$. Figure \ref{fig:appA_deg_sir} shows the relative error in $\beta$ for polynomial test functions of degree  $ = [10,26]$. Figure \ref{fig:appA_rad_sir} shows the relative error in $\beta$ for compact support for radii from $a=[5,15]$.

\section{Derivations of weak-form input-output equations for SIR model}\label{app:derv_wkIO}
\noindent
\textcolor{black}{\textbf{Derivation of Equation \eqref{eq:wf_sir_conpop}.}}\\
\textcolor{black}{Consider the input-out representation of the SIR model given by equation \eqref{eq:io_sir}. We divide by $I$ to obtain the following,}
\textcolor{black}{
\begin{equation}
    0=\frac{\ddot{I}I-\dot{I}^2}{I^2}+\beta\alpha I+\beta \dot{I}.\label{eq-app:alt_wk_step1}
\end{equation}
}
\textcolor{black}{Notice that,
\begin{equation*}
    \frac{d}{d\textsf{t}}(\frac{\dot{I}}{I})=\frac{\ddot{I}I-\dot{I}^2}{I^2}.
\end{equation*}}
\textcolor{black}{ Additionally, the equation \eqref{eq:SIR} we know,
\begin{equation*}\begin{split}
    \frac{dR}{d\textsf{t}}=\alpha I \Rightarrow
    R(t)=\int_0^t\alpha I(s)\textsf{d}s.\end{split}
\end{equation*}
Note, here we assume a naive population, i.e., $R(0)=0$ and $S(0)=N-I(0)$. Given that we observe the infected compartment, we are able to compute $R(t)$ through numeric integration.}
\textcolor{black}{We are now able to generate the following equation from equation \eqref{eq-app:alt_wk_step1},
\begin{equation}
    \begin{split}
        \frac{d}{d\textsf{t}}(\frac{\dot{I}}{I})&=-\beta\frac{d}{d\textsf{t}}\left(\int_0^t\alpha I(s)d\textsf{s}+I\right)\\
        &=-\beta\frac{d}{d\textsf{t}}(R(t)+I).
    \end{split}\label{eq-app:alt_wk_step2}
\end{equation}}
\textcolor{black}{Note, 
\begin{equation*}
    \frac{\dot{I}}{I}=\frac{\beta SI-\alpha I}{I}=\beta S_0-\alpha.
\end{equation*}}
\textcolor{black}{Next, we integrate equation \eqref{eq-app:alt_wk_step2} with respect to $t$ to obtain an alternative form of equation \eqref{eq:io_sir},
\begin{equation*}
    \frac{\dot{I}}{I}=\beta(R(t)+I)+\beta S_0-\alpha.
\end{equation*}}
\textcolor{black}{We then multiply by $I$,
\begin{equation}
    \dot{I}+\alpha I=-\beta(R(t)+I-S_0)I.\label{eq-app:alt_sir_IO}
\end{equation}}
\textcolor{black}{We obtain the weak-form input-output equation \eqref{eq:wf_sir_conpop} by convoling equation \eqref{eq-app:alt_sir_IO} with a known test fucntion $\phi$.}

\noindent
\textcolor{black}{\textbf{An alternative form of Equation \eqref{eq:wf_sir_conpop}.}}\\

Consider the model equation, which is presented again below,
\begin{equation}\begin{split}
    \dot{S}&=-\beta SI,\\
    \dot{I}&=\beta SI-\alpha I,\\
    \dot{R}&=\alpha I.
\end{split}\label{eq-app:SIR}
\end{equation}
By solving for $S$ from the $\dot{S}$ equation, we obtain equation the following,
\begin{equation}
    S=\frac{\dot{S}}{-\beta I}.\label{eq-app:Ssub}
\end{equation}
From the $\dot{S}$ equation, we can also obtain the following equation by taking the derivative, 
\begin{equation}
    \ddot{S}=-\beta \dot{S}I-\beta S\dot{I}.\label{eq-app:Sdotsub}
\end{equation}
From the $\dot{S}$ and $\dot{I}$ equations we can obtain,
\begin{equation}
    \dot{S}=-\beta SI +\alpha I -\alpha I = -(\beta SI -\alpha I) -\alpha I= -\dot{I}-\alpha I.\label{eq-app:Sdotsub2}
\end{equation}
From the $\dot{S}$ and $\dot{I}$ equations, we can obtain  the following equation by taking the derivative, 
\begin{equation}
    \ddot{I}=\frac{d}{\textsf{d}t}(\beta SI-\alpha I)=\frac{d}{\textsf{d}t}(-\dot{S}-\alpha I)=-\ddot{S}-\alpha \dot{I}.\label{eq-app:Idotdotsub}
\end{equation}
Substituting equations \eqref{eq-app:Ssub} and \eqref{eq-app:Sdotsub2} in equation \eqref{eq-app:Sdotsub} we obtain equation below,
\begin{equation}\begin{split}
    \ddot{S}&=-\beta \dot{S}I-\beta S\dot{I}=-\beta(-\alpha I-\dot{I})I-\beta \frac{\dot{S}}{-\beta I}\dot{I}\\&=\beta\alpha I^2+\beta I\dot{I}+(-\alpha I-\dot{I})\frac{\dot{I}}{I}=\beta\alpha I^2+\beta I\dot{I}-\alpha \dot{I}-\frac{\dot{I}^2}{I}. \end{split}\label{eq-app:Sdotdot}
\end{equation}
Next, we substitute equation \eqref{eq-app:Sdotdot} in equation \eqref{eq-app:Idotdotsub} to obtain equation \eqref{eq:io_sir}, an input-output equation for the SIR model for observed state variable $I,$ which we present again below,
\begin{equation*}
    \begin{split}
        \ddot{I}&=-\ddot{S}-\alpha I\\
        \ddot{I}&=-\beta\alpha I^2-\beta I\dot{I}+\alpha \dot{I}+\frac{\dot{I}^2}{I}-\alpha \dot{I}\\
        0&=-\frac{\dot{I}^2}{I}+\beta I\dot{I}+\beta\alpha I^2.
    \end{split}
\end{equation*}
Notice that three of the terms on the right-hand side of \eqref{eq:io_sir} can be written in the form \eqref{eq:diff_form1}, or in other words, can be written a derivative of a function of only $I$. However, the term $\frac{\dot{I}^2}{I}$ cannot be put in this form. We re-write equation \eqref{eq:io_sir} in the form \eqref{eq:diff_form2} by converting as many terms as possible to form \eqref{eq:diff_form1} to obtain equation the following,
\begin{equation}
    0=-\frac{\dot{I}^2}{I}+\beta\alpha I^2+\frac{d}{\textsf{d}t}\frac{\beta}{2}I^2+\frac{d^2}{\textsf{d}t^2}I.\label{eq-app:io-form2}
\end{equation}
Next, we multiply by test function $\phi$ and integrate over domain 0 to $T$ to obtain equation \eqref{eq:wf_sir}, the weak-form SIR input-output equation for observed state $I$, which is presented again below,
\begin{equation*}
\int_0^T\ddot{\phi}I\textsf{d}t-\int_0^T\phi\frac{\dot{I}^2}{I}\textsf{d}t=\beta\int_0^T\dot{\phi}\frac{I^2}{2}\textsf{d}t-\beta\alpha\int_0^T\phi I^2\textsf{d}t.
\end{equation*}
From equation \eqref{eq:SIR} we know,
\begin{equation*}
    \dot{I}^2=(\beta SI)^2-2\beta\alpha SI^2+\alpha^2I^2.
\end{equation*}
Thus, using equation \eqref{eq-app:Sdotsub2}, we can then rewrite the term which is not integrable by parts as follows,
\begin{equation}
    \frac{\dot{I}^2}{I}=\beta^2 S^2I-2\beta\alpha SI+\alpha^2=\beta S\dot{S} -2\beta\alpha SI+\alpha^2I.\label{eq:sqdotI}
\end{equation}
Substituting equation \eqref{eq:sqdotI} into equation \eqref{eq:wf_sir} we obtain the following
\begin{equation}
\begin{split}
    \int_0^T\phi\frac{\dot{I}^2}{I}\textsf{d}t&=\beta\int_0^T\phi S\dot{S}\textsf{d}t-\beta\alpha\int_0^T\phi 2SI\textsf{d}t+\alpha^2\int_0^T\phi \textsf{d}t,\\
&=-\beta\int_0^T\dot{\phi}\frac{S^2}{2}\textsf{d}t-\beta\alpha\int_0^T\phi 2SI\textsf{d}t+\alpha^2\int_0^T\phi I\textsf{d}t.
\end{split}\label{eq:wf_sqdot_int}
\end{equation}
Note that on the right-hand side of equation \eqref{eq:wf_sqdot_int} is written in the terms of the state variables, $S$ and $I$, and the test function and its derivative, $\phi$ and $\dot{\phi}$, and does not include derivatives of the state variables (as desired). 
We assume to only observe the state variable, $I$, and therefore, we cannot use equation \eqref{eq:wf_sqdot_int} directly. From the equation \eqref{eq:SIR} we know,
\begin{equation*}\begin{split}
    \frac{dR}{dt}=\alpha I \Rightarrow
    R(t)=\int_0^t\alpha I(s)\textsf{d}s.\end{split}
\end{equation*}
Note, here we assume a naive population, i.e., $R(0)=0$. Given that we observe the infected compartment, we are able to compute $R(t)$ through numeric integration.

Let the total population $N=S(t)+I(t)+R(t)$ be constant and known. Then, $S(t)=N-I(t)-R(t)$ can be computed from only observations of the state variable $I(t)$. Then, using our assumption of a known constant population, we are able to generate the following equation from equation \eqref{eq:wf_sqdot_int}, which is given again below,
\begin{equation*}
\begin{split}
    \int_0^T\phi\frac{\dot{I}^2}{I}\textsf{d}t
&=\beta\int_0^T\dot{\phi}\frac{(N-R(t)-I)^2}{2}\textsf{d}t-\beta\alpha\int_0^T\phi2(N-R(t)-I)I\textsf{d}t+\alpha^2\int_0^T\phi I\textsf{d}t.
\end{split}
\end{equation*}
Finally, we can write equation \eqref{eq-app:wk_sir_alt}, a weak-form SIR model that reveals parameter $\beta$ using only observations of state $I$ and a known population size, which is given again below,
\begin{equation}
\begin{split}
    \int_0^T\ddot{\phi}I\textsf{d}t-\alpha^2\int_0^T\phi I \textsf{d}t&=\beta\int_0^T\dot{\phi}(\frac{I^2-(N-R(t)-I)^2}{2}\textsf{d}t\\&-\beta\alpha\int_0^T\phi(I^2+2(N-R(t)-I)I)\textsf{d}t,\\
    R(t)&=\int_0^t\alpha I(s)\textsf{d}s.\label{eq-app:wk_sir_alt}
\end{split}
\end{equation}
\textcolor{black}{Using WENDy with equation \ref{eq-app:wk_sir_alt} to estimate parameter $\beta$ yields identical results to using WENDy with equation \ref{eq:wf_sir_conpop}}.

\section{Structural Identifiability of Example Models}\label{app:weak-struct}

\textbf{Example 1}

Below, we present again equation \eqref{eq:bldcnc_io_og}, the input-output equation for model \ref{eq:bld_cnc} in the strong form, 

\begin{equation*}
\begin{split}
\ddot{x}_1(t)x_1(t)^2+2\ddot{x}_1(t)x_1(t)+\ddot{x}_1(t)+w_1\left(x_1(t)^2+x_1(t)\right)\\+w_2\left(\dot{x}_1(t)x_1(t)^2+2\dot{x}_1(t)x_1(t)\right)+w_3\dot{x}_1(t)=0,
\end{split}
\end{equation*}
where $w_1=k_{21}V_e$, $w_2=k_{12}+k_{21}$, and $w_3=k_{12}+k_{21}+V_e$. In order to verify the structural identifiability from this input-output representation, we can consider Definition 2. However, for this to be a sufficient condition for the structural identifiability, the rank of the Wronskian of the differential monomials of the input-output representation must be checked \cite{OvchinnikovPogudinThompson2023AAECC}. Recently, \cite{DongGoodbrakeHarringtonEtAl2023SIAMJApplAlgebraGeometry} have developed a projection-based approach analogous to the differential elimination approach that takes into consideration conditions for the equivalence of multi and single experiment identifiability. We confirm both the form of equation \eqref{eq:bldcnc_io_og} and that the model parameters are globally structurally identifiable using \texttt{find\_ioequations} and \texttt{assess\_identifiability} commands from the \texttt{StructuralIdentifiability} package in Julia \cite{DongGoodbrakeHarringtonEtAl2023SIAMJApplAlgebraGeometry}. While \texttt{StructuralIdentifiability.jl} also produces the above strong form input-output equation, we chose to use the \texttt{Differential Algebra v4} package to generate equation \eqref{eq:bldcnc_io_og} in order to perform the integration by parts procedure that is needed to generate the weak-form in this study.

\begin{figure}[H]
    \centering
    \includegraphics[width=0.5\linewidth]{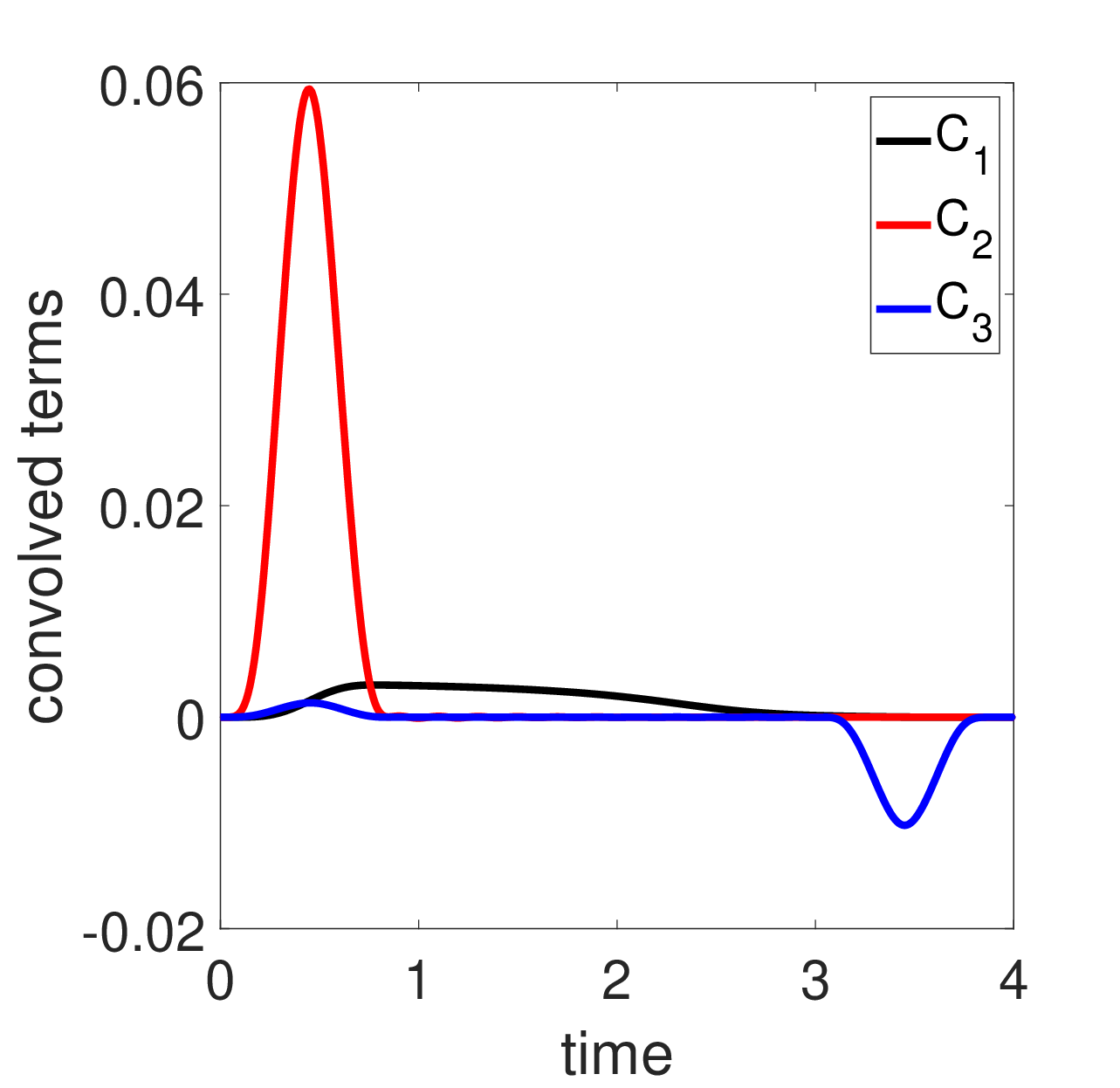}
    \caption{Dynamics of terms $C_1,$ $C_2$, and $C_3$ from the weak-form input-output equation \eqref{eq:weak_bld_cnc} for parameters $k_{12}=5,$ $k_{21}=1,$ and $V_e=6$.}
    \label{fig:appC_bld_cnc_Cterms}
\end{figure}

While we cannot directly test the identifiability of the weak-form input-output equation \eqref{eq:weak_bld_cnc}, we can confirm the unique recoverability of the parameters in the discretized WENDy equation $\mathbf{G}\mathbf{w}^*=\mathbf{b}$. In the case of no noise, 
\begin{equation*}
    \mathbf{G}=\begin{bmatrix}
        \mathbf{\Phi}\frac{\mathbf{y}}{\mathbf{y}+1}&-\mathbf{\dot{\Phi}}\frac{\mathbf{y}^2}{\mathbf{y}+1}&\mathbf{\dot{\Phi}}\frac{1}{\mathbf{y}+1}
        \end{bmatrix}
\end{equation*}
For the chosen test function matrix, 400 noise-free observations, and assuming model parameters $k_{12}=5,$ $k_{21}=1,$ and $V_e=6$, we verify G is full rank using the SVD algorithm in \texttt{Matlab} with a tolerance of $1e-12$. This result supports that, at least local to the parameter choices used in this example, the transformation from the strong to weak-form input-output system did not result in a loss of parameter identifiability. We also verify that $\text{rank}(\mathbf{\Phi})=\text{rank}(\dot{\mathbf{\Phi}})=K$ with a tolerance of $1e-12$, for this example $K=15$. To further support this claim, we graph the terms $C_1,$ $C_2$, and $C_3$ from the weak-form input-output equation \eqref{eq:weak_bld_cnc} in Figure \ref{fig:appC_bld_cnc_Cterms}. Notice that the convolved terms each provide a unique signal.

\noindent
\textbf{Example 2}

Similarly, for the SIR example, we confirm both the form of equation \eqref{eq:io_sir} and that the model parameters are globally structurally identifiable using \texttt{find\_ioequations} and \texttt{assess\_identifiability} commands from the \texttt{StructuralIdentifiability} package in Julia \cite{DongGoodbrakeHarringtonEtAl2023SIAMJApplAlgebraGeometry}.

While we cannot directly test the identifiability of the weak-form input-output equation \eqref{eq:wf_sir_conpop}, we can confirm the unique recoverability of the parameters in the discretized WENDy equation $\mathbf{G}\mathbf{w}^*=\mathbf{b}$. In the case of no noise, 
\begin{equation*}
    \mathbf{G}=\begin{bmatrix}
        \mathbf{\Phi}\left(\mathbf{y}^2-(N-\mathbf{R}-\mathbf{y})^2\right)&\mathbf{\dot{\Phi}}\left(\mathbf{y}^2+2(N-\mathbf{R}-\mathbf{y})\mathbf{y}\right)
        \end{bmatrix}
\end{equation*}
For the chosen test function matrix, 31 noise-free observations, and assuming model parameters N=10,000, $S_0=N-1$, $I_0=1$, $R_0=0$, $\beta=\frac{5.5}{N}$, and $\gamma=5$, we verify G is full rank using the SVD algorithm in \texttt{Matlab} with a tolerance of $1e-12$. This result supports that, at least local to the parameter choices used in this example, the transformation from the strong to weak-form input-output system did not result in a loss of parameter identifiability. We also verify that $\text{rank}(\mathbf{\Phi})=\text{rank}(\dot{\mathbf{\Phi}})=K$ with a tolerance of $1e-12$, for this example $K=4$. 

\section{Example 1 $(e,q)-$identifiability with lognormal noise}\label{app:exp1_lognormal}

\begin{figure}[ht]
    \centering
    \includegraphics[width=0.5\linewidth]{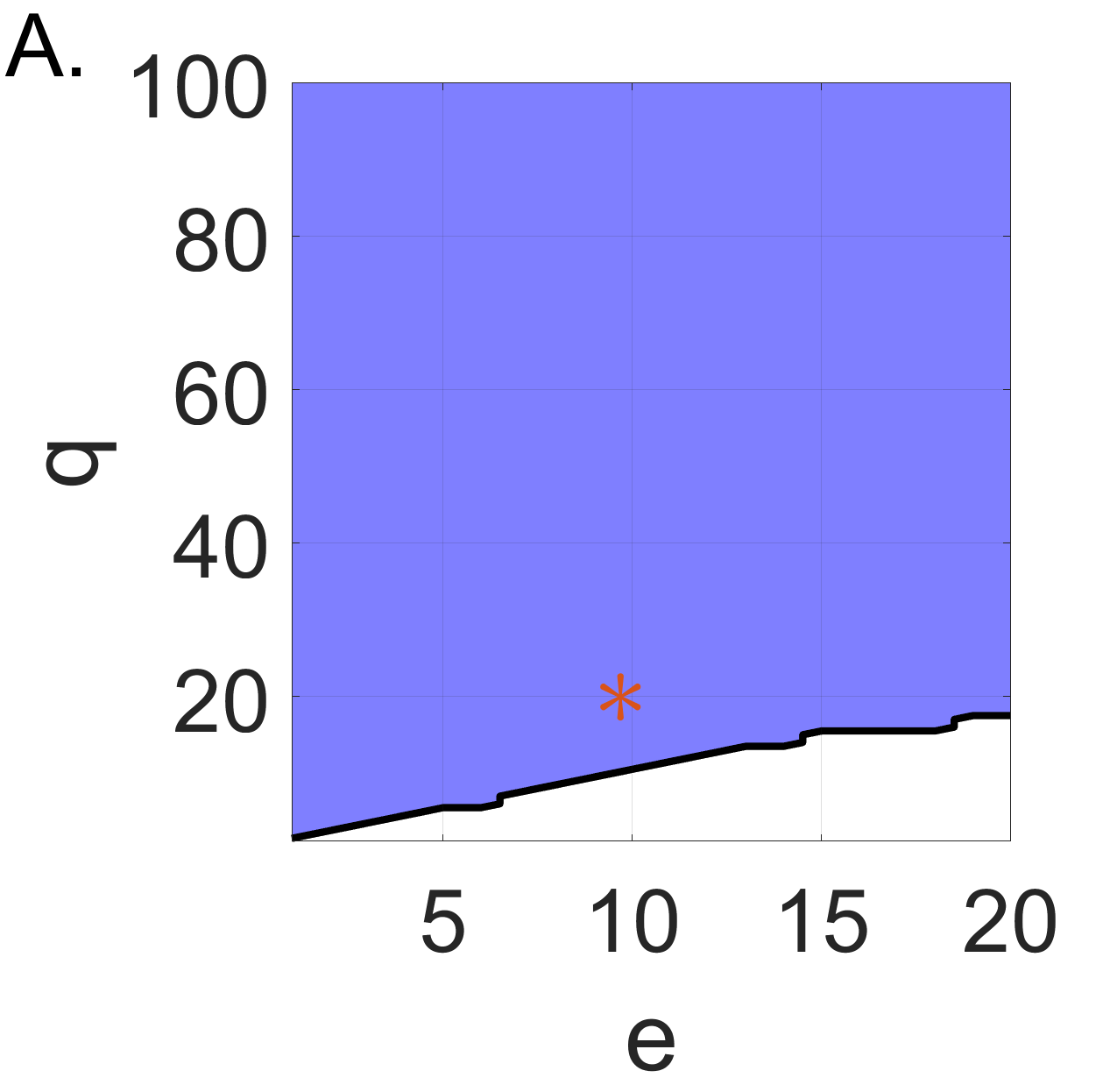}\includegraphics[width=0.5\linewidth]{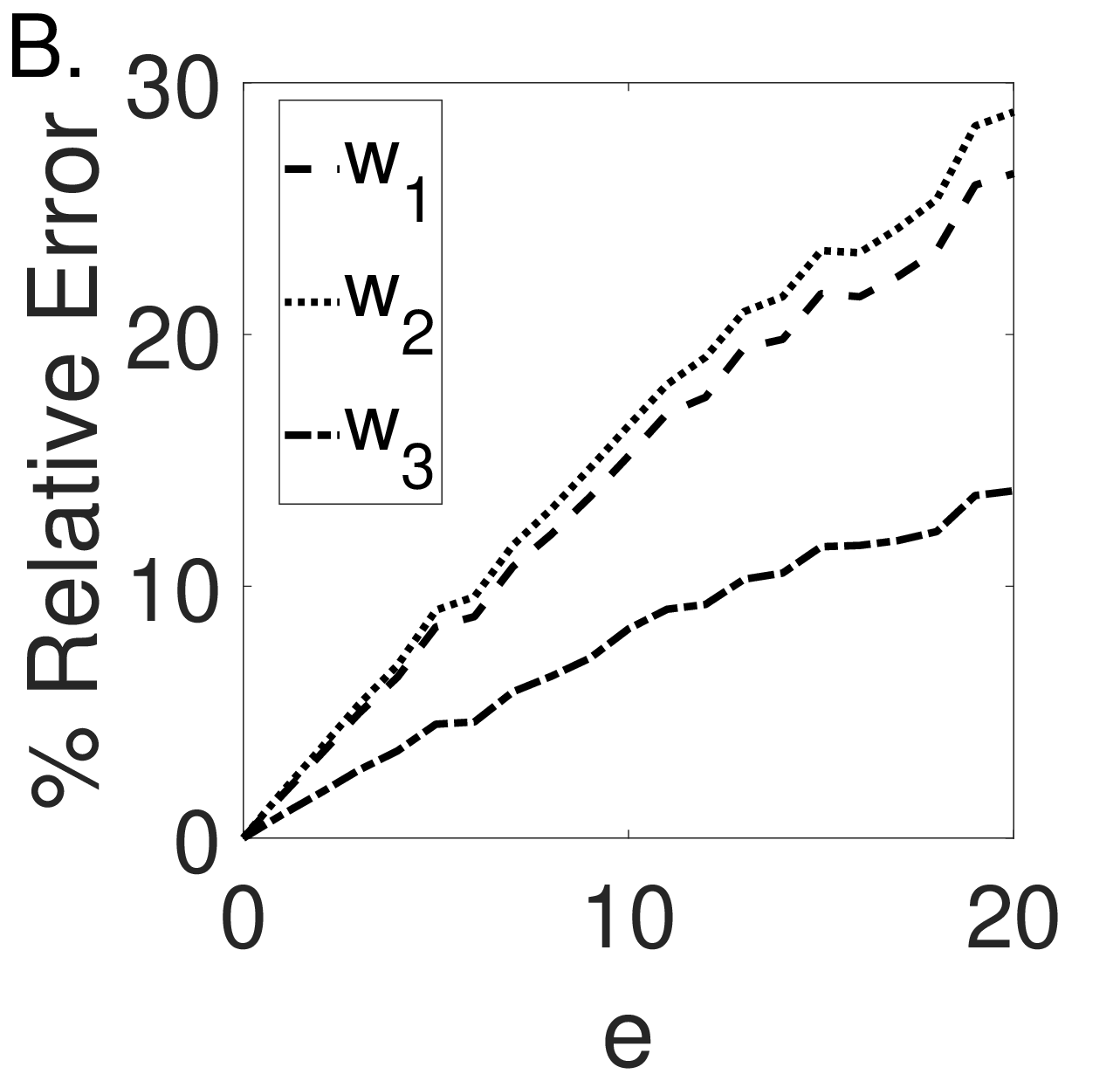}\\
\includegraphics[width=0.5\linewidth]{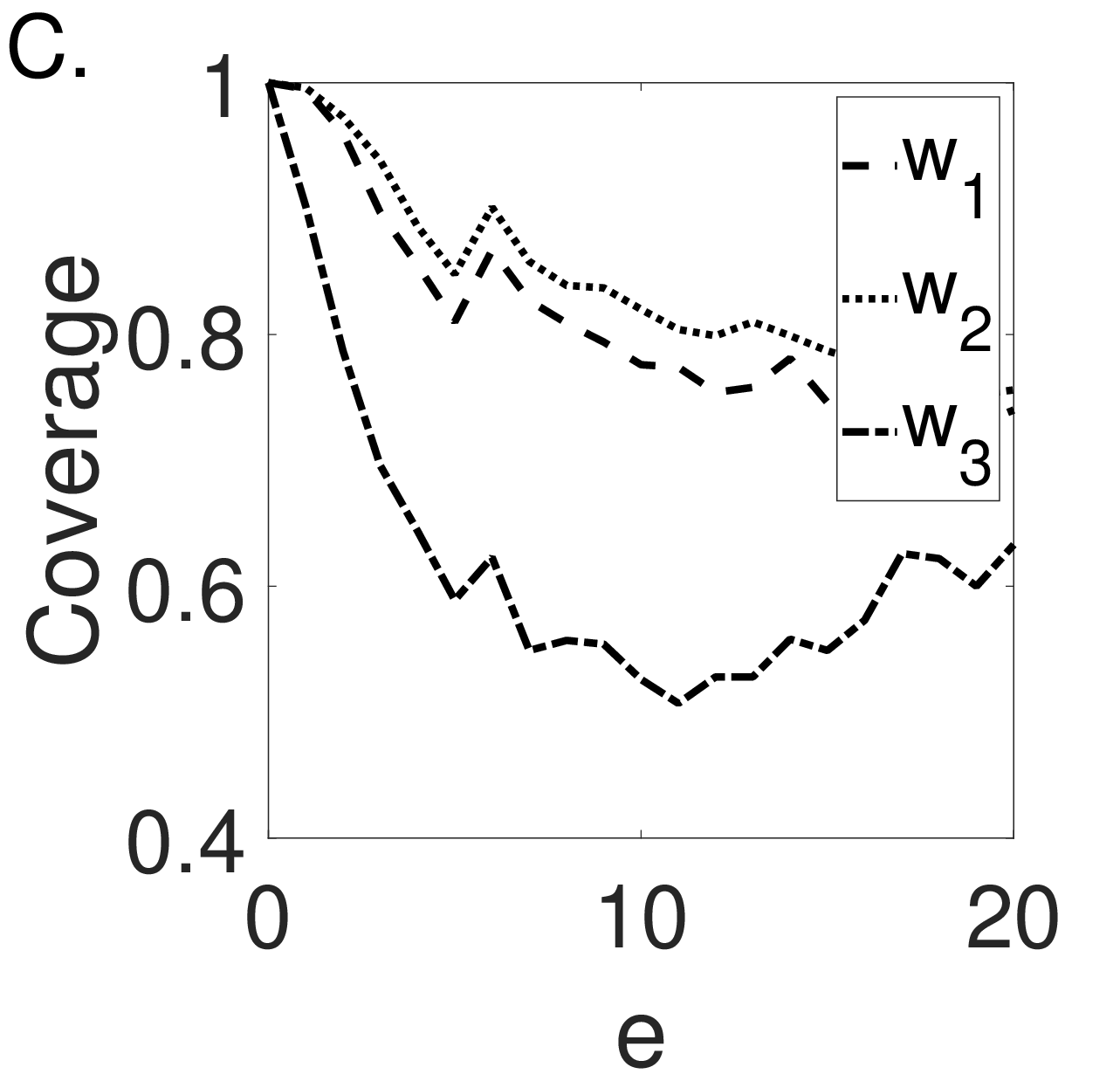}
    \caption{Using WENDy with equations \eqref{eq:weak_bld_cnc} we find the SIR model to be generally practically identifiable for multiplicative noise between $[0\%,20\%]$. (A.) The area in blue denotes where the model is $(e,q)$-identifiable, and the area in white denotes where the model is not $(e,q)$-identifiable. The $(e,q)$-identifiability was determined from 1,000 simulations at each respective error level, e. The red star \textcolor{red}{\textasteriskcentered } marks the $(10,20)$-identifiability criterion. (B.) The relative error determined from 1,000 simulations at each respective error level, e. (C.) The proportion of estimated $95\%$ confidence intervals at $e=[0\%,20\%]\%$ that contain the true value of $w_i$.}
\label{fig:bldcnc_pract_ident_lognorm}
\end{figure}

 We estimate the parameters $w_1$, $w_2$, and $w_3$ using WENDy for simulated data with log-normal \textcolor{black}{multiplicative observation error}. We vary the \textcolor{black}{multiplicative observation error} ratio in the simulated data between $e=[0\%,20\%]$ such that the standard deviation of the error is $\sigma=e\log(\text{RMS}(\Omega(t)))$, and at each \textcolor{black}{multiplicative observation error} ratio, we generate 1,000 noisy datasets. The $(e,q)$-identifiability for $e=[0\%,20\%]$ and $q=[1\%,100\%]$ of model \eqref{eq:weak_bld_cnc} is given in Figure \ref{fig:bldcnc_pract_ident_lognorm}A. Unlike in the case of additive noise, we find that the weak form model is (10,20)-identifiable, or generally practically identifiable.

We additionally compare the $(e,q)$-identifiability of model equations \eqref{eq:weak_bld_cnc} to the average relative error and the coverage of the $95\%$ confidence intervals generated at each additive error ratio. The relative error generated for 1,000 simulated datasets at each \textcolor{black}{multiplicative observation error} level is given in Figure \ref{fig:bldcnc_pract_ident_lognorm}B, and the coverage of the $95\%$ confidence intervals for $w_1$, $w_2$, and $w_3$ is given in Figure \ref{fig:bldcnc_pract_ident_lognorm}C. Here, the $(e,q)$-identifiability and the relative error criteria are in agreement. 

\section{\textcolor{black}{Comparison to Profile Likelihood}}\label{app:profile-likelihood}

\begin{figure}[ht]
    \centering
    \includegraphics[width=0.5\linewidth]{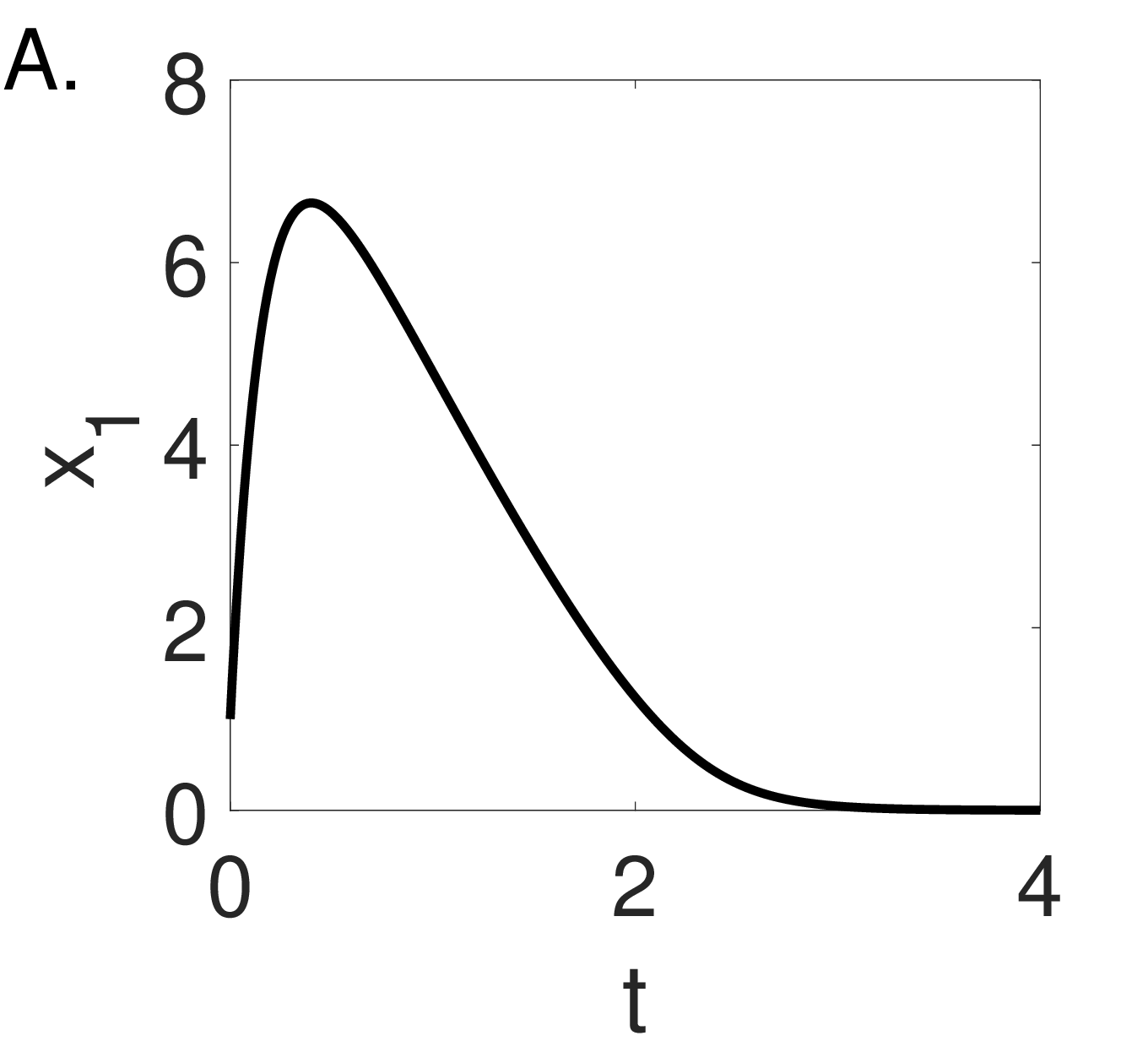}\includegraphics[width=0.5\linewidth]{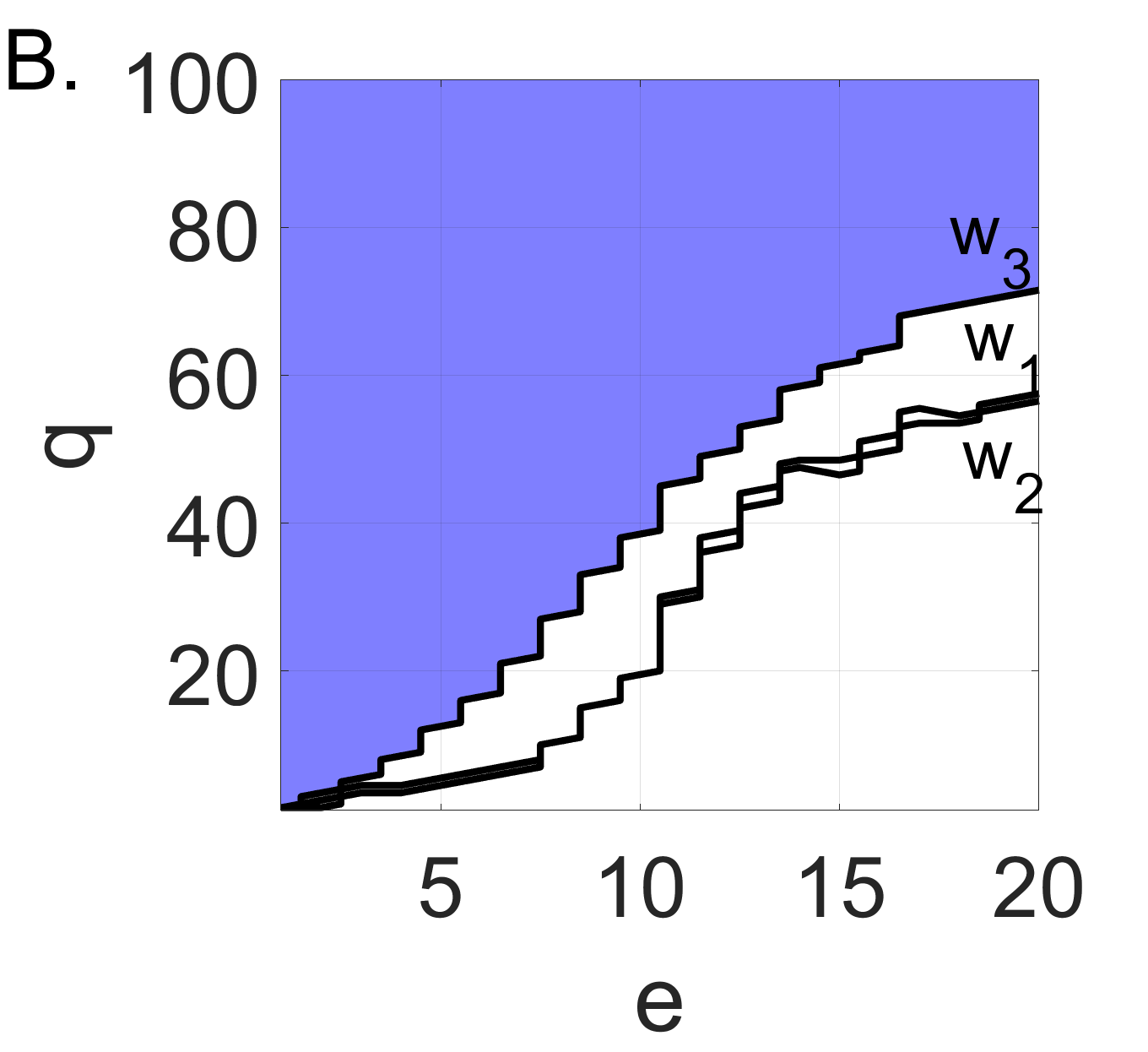}\\
    \includegraphics[width=0.8\textwidth]{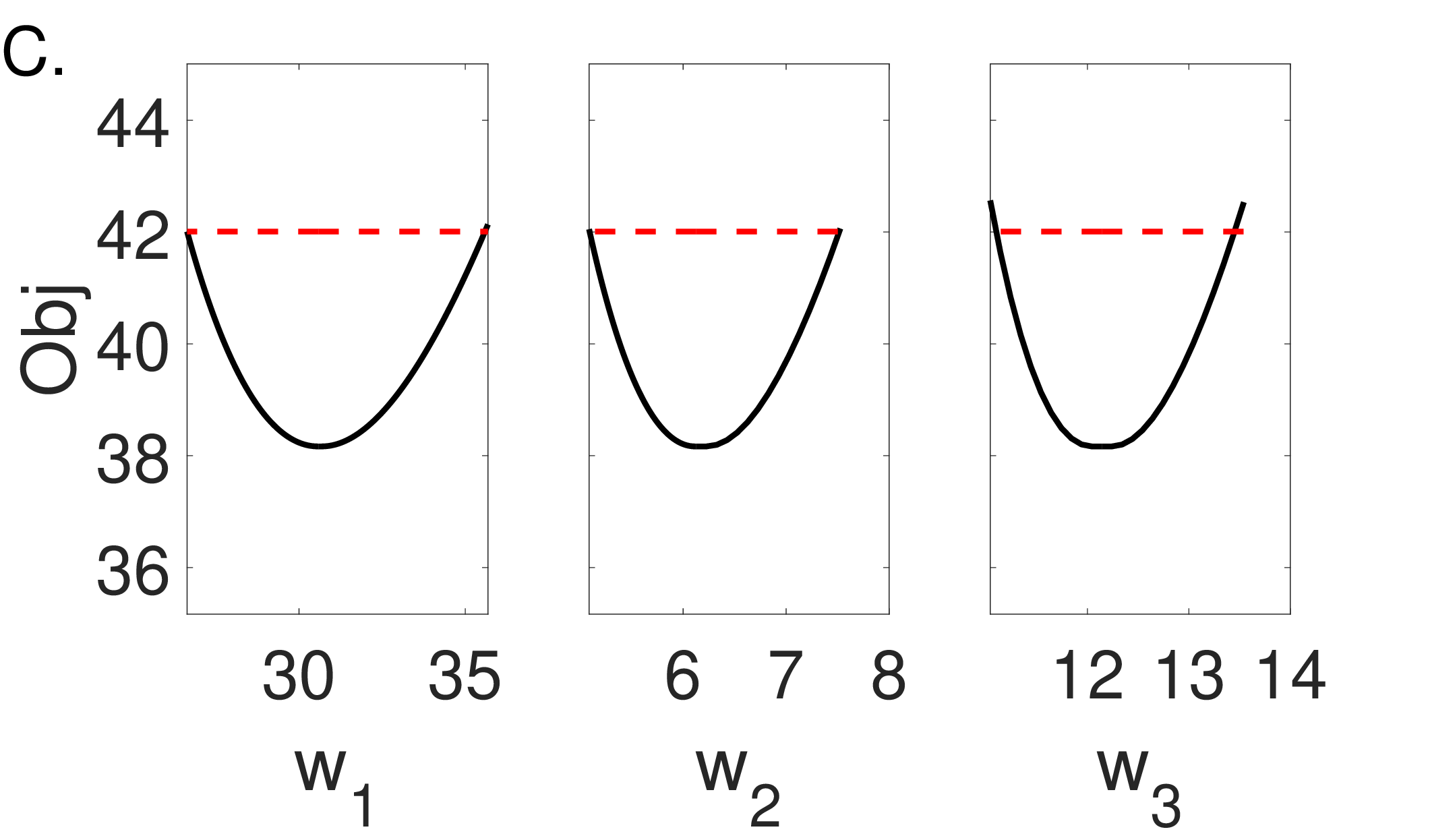}
    \caption{\textcolor{black}{At $e=10\%$, the blood-diffusion model given by equation \eqref{eq:bld_cnc} with parameters $w=[30, 6, 12]$ is $(10, 39)-$identifiable and practically identifiable by profile likelihood analysis when using 40 observations. (A.) Continuous dynamics of the $x_1$ compartment of model given by equation \eqref{eq:bld_cnc} with parameters $w=[30, 6, 12]$. (B.) The area in blue denotes where the model is $(e,q)$-identifiable, and the area in white denotes where the model is not $(e,q)$-identifiable. The $(e,q)$-identifiability was determined from 1,000 simulations at each respective error level, $e$. (C.) Profile likelihood (black line) for parameter $w_1,$ $w_2$, $w_3$ where the red dashed line represents the $95\%$ confidence level threshold in output error objective function value.}}
    \label{fig:PLA_compare_identifiable}
\end{figure}

\textcolor{black}{We compare the $(e,q)-$identifiability for the blood-diffusion model given by equation \eqref{eq:bld_cnc}, found using the weak-form equation error method, to the profile likelihood (PLA), found using an output error method, for two different choices of parameters $w$.  As can be seen in Figure \ref{fig:PLA_compare_identifiable}, we $(10,q)$-identifiability of the blood diffusion model parameters and profile likelihood identifiability with $10\%$ observational error are in agreement. Further, for observational error under $10\%$ we find agreement between the WENDy confidence intervals and the PLA confidence intervals, which are given in Table \ref{tab:pla_CI_compare}. In a different region of the parameter space, we find the blood-diffusion model is not $(e,q)-$identifiable for observational error greater $0.5\%$ and that this practical non-identifiability is driven by uncertainty in parameter $w_3$. We find agreement for these results using the profile likelihood method at a $3\%$ observational error level. \textcolor{black}{However, we note that a WENDy-based confidence interval is predicted for parameter $w_3$ despite both $(e,q)$ and PLA predicted non-identifiability. We hypothesize this discrepancy is driven by limitations in our approximation of the covariance matrix, which is described in Appendix \ref{app:wendy_covar}.} It is important to note that for the figures and confidence intervals generated using the profile likelihood method, the true parameters were used as an initial guess for the first output error optimization step, \textcolor{black}{to prevent the \texttt{lsqnonlin} optimization from failing to converge at higher levels of noise.}}

\begin{table}[ht]
    \centering
    \begin{tabular}{c|c|c|c|c}
       $e$  & Method & $\text{CI}_{w_1}$ &$\text{CI}_{w_2}$ & $\text{CI}_{w_3}$\\
       \hline
        $1\%$ & WENDy & $[28.57,31.43]$ & $[5.67,6.33]$ &$[11.78,12.22]$\\
        & PLA & $[26.63,34.96]$ & $[5.09,7.34]$ & $[11.07,13.31]$\\
        \hline
        $5\%$ & WENDy & $[26.57,33.43]$ & $[5.21,6.79]$ &$[11.40,12.59]$\\
        & PLA & $[26.38,35.29]$ & $[5.03,7.42]$ & $[11.04,13.37]$\\
        \hline
        $10\%$ & WENDy & $[24.78,35.22]$ & $[4.79,7.21]$ &$[10.24,13.76]$\\
        & PLA & $[26.24,35.58]$ & $[4.99,7.53]$ & $[11.01,13.44]$\\
    \end{tabular}
    \caption{\textcolor{black}{WENDy and Profile likelihood (PLA) $95\%$ confidence intervals for the blood-diffusion model given by equation \eqref{eq:bld_cnc} with true parameters $w=[30, 6, 12]$.}}
    \label{tab:pla_CI_compare}
\end{table}

\begin{figure}[ht]
    \centering
    \includegraphics[width=0.5\linewidth]{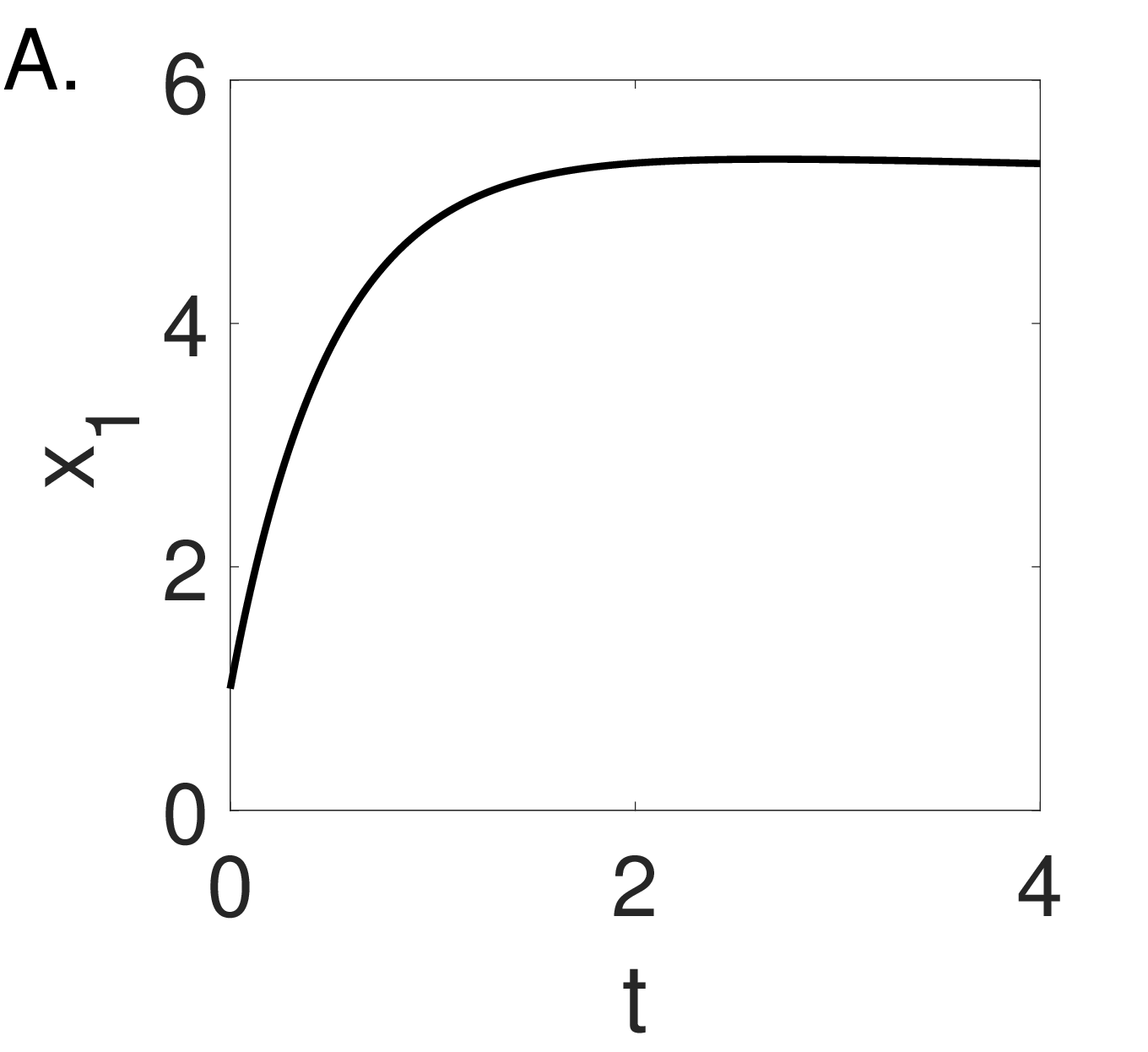}\includegraphics[width=0.5\linewidth]{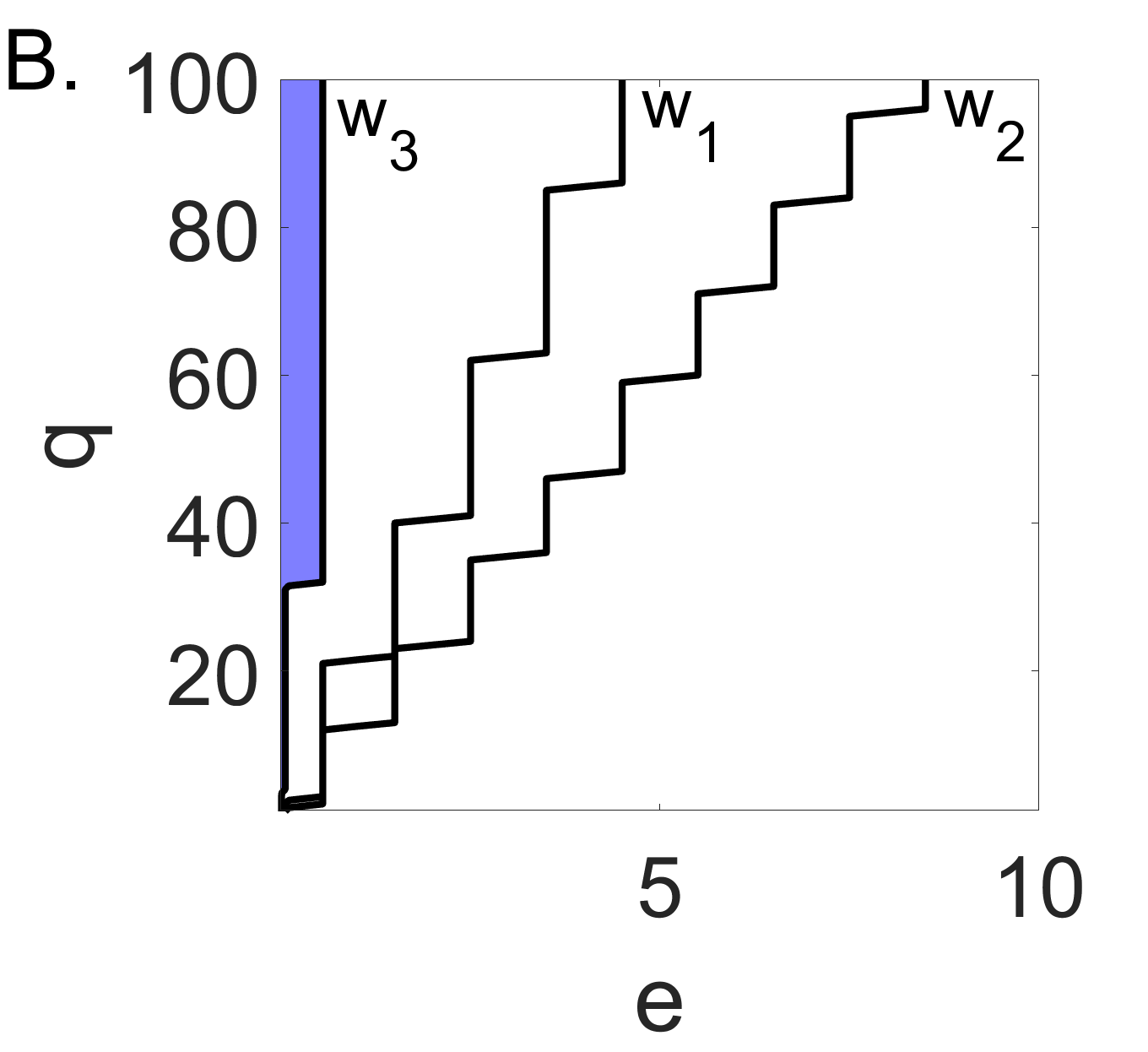}\\
    \includegraphics[width=0.8\textwidth]{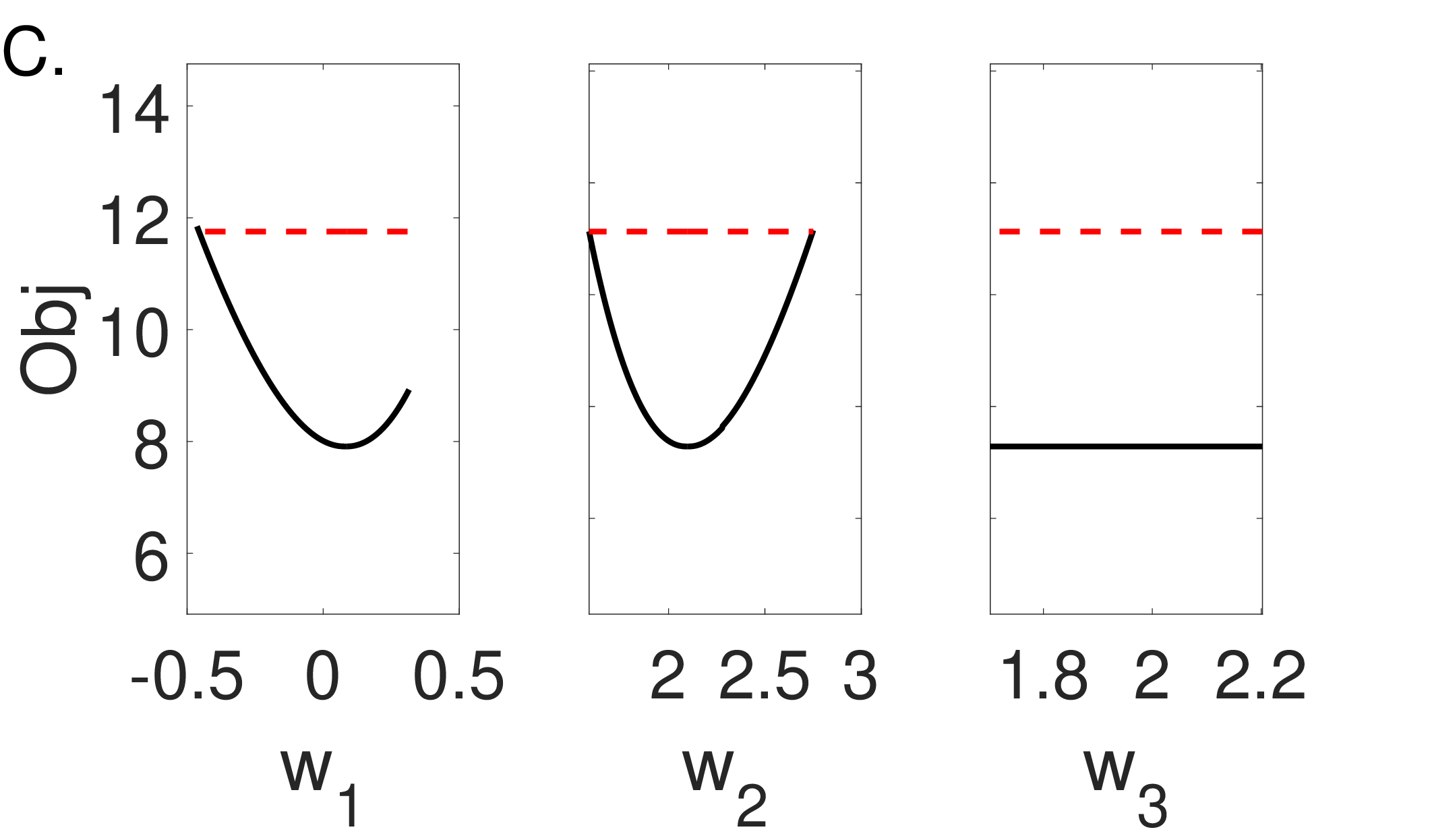}
    \caption{\textcolor{black}{At $e=3\%$, the blood-diffusion model given by equation \eqref{eq:bld_cnc} with parameters $w=[0.1,2,2.1]$ is not $(e, q)-$identifiable and not practically identifiable by profile likelihood analysis when using 40 observations. (A.) Continuous dynamics of the $x_1$ compartment of the model given by equation \eqref{eq:bld_cnc} with parameters $w=[0.1,2,2.1]$. (B.) The area in blue denotes where the model is $(e,q)$-identifiable, and the area in white denotes where the model is not $(e,q)$-identifiable. The $(e,q)$-identifiability was determined from 1,000 simulations at each respective error level, $e$. (C.) Profile likelihood (black line) for parameter $w_1,$ $w_2$, $w_3$ where the red dashed line represents the $95\%$ confidence level threshold in output error objective function value.}}
    \label{fig:PLA_compare_nonidentifiable}
\end{figure}

\begin{table}[ht]
    \centering
    \begin{tabular}{c|c|c|c|c}
       $e$  & Method & $\text{CI}_{w_1}$ &$\text{CI}_{w_2}$ & $\text{CI}_{w_3}$\\
       \hline
        $3\%$ & WENDy & $[0.044,0.19]$ & $[0.57,2.79]$ &$[-18.74,39.72]$\\
        & PLA & $[-0.44,-]$ & $[1.59,2.75]$ & $[-,-]$\\
    \end{tabular}
    \caption{\textcolor{black}{Profile likelihood (PLA) $95\%$ confidence intervals for the blood-diffusion model given by equation \eqref{eq:bld_cnc} with true parameters $w=[0.1,2,2.1]$.}}
    \label{tab:pla_CI_compare_nonident}
\end{table}
\FloatBarrier
\section{\textcolor{black}{WENDy: Covariance Matrix}}\label{app:wendy_covar}

\textcolor{black}{In this Appendix, we provide a brief description of the WENDy method approximation of the covariance matrix and derived parameter confidence intervals. A more detailed derivation of the covariance is provided in reference \cite{BortzMessengerDukic2023BullMathBiol}.}

\textcolor{black}{Consider the case of additive normal noise, i.e. $\mathbf{y}=\mathbf{y}^*+\varepsilon$ where $\varepsilon\sim \mathcal{N}(0,\sigma^2).$ The residual of equation $\mathbf{Gw^*=b}$ as defined in Section \ref{subsec:wendy1} is given below,}
\begin{equation*}
    \textcolor{black}{\mathbf{r}(\mathbf{y,w})=\mathbf{G}(\mathbf{y})\mathbf{w}-\mathbf{b}(\mathbf{y}).}
\end{equation*}

\textcolor{black}{Both $\mathbf{G}$ and $\mathbf{b}$ act as nonlinear transformations of the noise present in the observational data $\mathbf{y}$. By taking the first-order Taylor expansion around the true output $\mathbf{y^*}$, we can obtain the following approximation of the residual,}
\begin{equation}
    \textcolor{black}{\mathbf{r}(\mathbf{y,w})\approx \mathbf{r}(\mathbf{y^*,w})+J_{\mathbf{y}}(\mathbf{w})\varepsilon,}
\end{equation}
\textcolor{black}{where $J_{\mathbf{y}}$ is the Jacobian of $\mathbf{r}$ with respect to observational data. The covariance of the residuals is then approximated as}
\begin{equation}
    \textcolor{black}{\textbf{S}_R=\text{Cov}(\mathbf{r}(\mathbf{y,w}))\approx J_{\mathbf{y}}(\mathbf{w})\text{Cov}(\varepsilon)J_{\mathbf{y}}(\mathbf{w})^T,}
\end{equation}
\textcolor{black}{where $\text{Cov}(\varepsilon)=\sigma^2\mathbf{I}_M.$}

\textcolor{black}{Using this covariance structure, WENDy applies an iteratively reweighted least squares algorithm to solve the following generalized least-squares problem}
\begin{equation}
   \textcolor{black}{ \min_{\mathbf{w}} (\mathbf{Gw-b})^T\textbf{S}_R^{-1}(\mathbf{Gw-b}).}
\end{equation}

\textcolor{black}{From this relationship we can write the approximate parameter covariance matrix}
\begin{equation}
    \textcolor{black}{\textbf{S}_w\approx(\mathbf{G}^T\mathbf{G})^{-1}\mathbf{G}^T\textbf{S}_R\mathbf{G}(\mathbf{G}^T\mathbf{G})^{-1}.}
\end{equation}
\textcolor{black}{And, the $(1-\alpha)$ confidence interval around the parameter estimate $\hat{w}_i$}
\begin{equation}
   \textcolor{black}{ [\hat{w}_i\pm F_{\textbf{S}_{w_{ii}}}(1-\alpha/2)],}
\end{equation}
\textcolor{black}{where $F$ is the CDF of the normal distribution with mean zero and variance $\textbf{S}_{w_{ii}}$.}

\textcolor{black}{In the case that we consider multiplicative lognormal noise, i.e. $\mathbf{y=y^*\varepsilon}$ where $\log(\varepsilon)\sim\mathcal{N}(0,\sigma^2)$, we assume $\text{Cov}(\varepsilon)\approx\sigma^2\mathbf{diag}(y_1,..,y_M),$ where $\mathbf{diag(x)}$ is a diagonal matrix with entries of $\mathbf{x}.$}

\end{appendices}

\end{document}